\documentclass[11pt, letterpaper]{article}
\usepackage{amsfonts,amsmath,amsopn,amssymb,amsthm,bbm,latexsym,mathrsfs,pstricks,verbatim,caption,  pst-node,stmaryrd}
\let\n\noindent

\def\A{{\cal{A}}}

\let\d\partial

\let\s\sigma

\let\R\rangle
\let\l\left
\let\r\right

\def\z{{\cal Z}}
\let\lf\lfloor
\let\rf\rfloor
\let\lc\lceil
\let\rc\rceil

\let\La\Lambda
\let\la\lambda

\let\d=\partial

\let\la\lambda
\let\La\Lambda

\def\y{{\infty}}

\let\Rw\Rightarrow
\def\l{{\left}}
\def\r{{\right}}
\def\rw{\rightarrow}

\def\R{\rangle}

\let\d=\partial

\def\w{{\tilde w}}

\let\ka\kappa

\begin{document}

\let\Rw\Rightarrow
\let\rw\rightarrow
\let\l\left
\let\r\right
\let\s\sigma
\let\ka\kappa
\let\de\delta

\def\M{{\cal M}}

\def\n{{\tilde n}}

\title{\vskip60pt {\bf Paths and partitions: combinatorial descriptions of the parafermionic states}}




\smallskip
\author{ \bf{ Pierre
Mathieu} \\ 
\\
D\'epartement de physique, de g\'enie physique et d'optique,\\
Universit\'e Laval,
Qu\'ebec, Canada, G1K 7P4.\\
pmathieu@phy.ulaval.ca}

\vskip .2in
\bigskip

\maketitle
\let\a\alpha
\def\ah{\hat{\alpha}}
\def\bh{\hat{\beta}}
\let\b\beta
\let\n\noindent

\def\B{{\text{B}_{k-1}}}

\def\F{\text{ABF}_{k}}
\def\Fl{\text{ABF}_{k,\ell}}

\def\Ft{{\text{ABF}}^*_{k}}
\def\Ba{\text{B}_{k-1,a}}

\def\wf{w^{\text{FB}}}
\def\ws{w^{\text{sc}}}

\def\N{{\mathcal N}^{[k-1]}}
\def\Nk{{\tilde {\mathcal N}}^{[k]}}
\def\nt{{\tilde n}}
 \def\exrw{\xrightarrow{{\rm ex} }}
\def\rrw{\xrightarrow{{\rm r} }}

\def\x{{\tilde x}}
\def\m{{\tilde m}}

\def\Z{\mathcal {Z}}

\vskip0.3cm
\centerline{{\bf ABSTRACT}}
\vskip18pt

The $\Z_k$ parafermionic conformal field theories, despite the relative complexity of their modes algebra, offer the simplest context for the study of the bases of states and their different combinatorial representations. Three bases are known. The classic one is given by strings of the fundamental parafermionic operators whose sequences of modes are in correspondence with restricted  partitions with parts at distance $k-1$ differing at least by  2.   Another basis is expressed in terms of the ordered modes of the $k-1$ different parafermionic fields, which are in correspondence with the so-called multiple partitions. Both types of partitions have a natural (Bressoud) path representation. Finally, a third basis, formulated in terms of different  paths, is inherited from the solution of the restricted solid-on-solid model of Andrews-Baxter-Forrester. The aim  of this work is to review, in a unified and pedagogical exposition, these four different combinatorial representations of the states of the $\Z_k$ parafermionic models.

The first part of this article presents the different paths and partitions and their bijective relations; it is purely combinatorial, self-contained and elementary;  it can be read independently of  the conformal-field-theory applications. The second part links this combinatorial analysis with the bases of states of the $\Z_k$ parafermionic theories. With the prototypical example of the parafermionic models worked out in detail, this analysis contributes to fix some foundations for the combinatorial study of more complicated theories. Indeed, as we briefly indicate in ending, generalized versions of both the Bressoud and the Andrews-Baxter-Forrester paths emerge naturally in the description of the minimal models.


\newpage

{\tableofcontents}
\newpage

\section{Introduction}

\subsection{Quasi-particles and fermionic character formula}

The modern emergence of  conformal field theory \cite{BPZ} rests heavily 
 on the representation theory of the Virasoro algebra \cite{FF82,Kac}, via the field-state correspondence.
  This has certainly clarified the classification of the fields.
Each primary field is associated with a highest-weight state and its conformal family is represented by the full highest-weight  module. 
But more importantly, the representation theory has lead to a neat characterization  of those models (dubbed minimal) that can be solved exactly. The minimal models are singled out by the completely degenerate nature of their representations, which contain an infinite number of singular vectors.
 These singular vectors are at the root of the models solvability: they are associated to null fields that lead to differential equations for the correlation functions. 

The (holomorphic) Hilbert space resulting from the removal of the singular vectors in the space of states reduces to
a direct sum  of
irreducible modules.
 The distribution of states in these modules, or more concisely said, their character, is then easily obtained in closed form, via an exclusion-inclusion process \cite{Roc}. 

Albeit very general,
 the description of states resulting from the subtractions of singular submodules does not capture much of the particular physics of each model. 
By contrast, such would be the case for a quasi-particle description.
That would amount to represent each state by a sequence of creation operators, specific to each model, acting on a highest-weight state. The full module would thereby be generated by a simple filling process
subject to some exclusion rules.
 Non-uniqueness in such a quasi-particle descriptions is even expected \cite{KKMMa}: different realizations would signal different integrable perturbations \cite{Zam}. In that vein, a correspondence is anticipated between the quasi-particles and the massive particles of the off-critical theory (and more generally, between the conformal  basis states and the states of the off-critical massive theory \cite{DN}). 

To a large extend, working out  such a quasi-particle description of rational conformal field theories is still at the level of a program, despite impressive recent progress \cite{F(12a),F(12b),F(13)} (briefly placed in context in the concluding section). However, clear and distinctive imprints of this yet-to-be-shaped quasi-particle representation theory are visible: these are the fermionic character formulae \cite{KKMMa}. The  fermionic qualitative refers to a manifestly positive multiple-sum expression: the manifest positivity is taken as a signature of  the packing of quasi-particles, while the different summations refer to the different quasi-particle types.

%

The simplest context for studying quasi-particle bases is, somewhat paradoxically, 
the $\Z_k$ parafermionic conformal field theories \cite{ZFa}. These are defined by an extension of the conformal algebra generated by $k-1$ different fields with fractional dimension and realizing a $\Z_k$ cyclic symmetry. The fractional dimension of the chiral generators brings severe complications, the most noticeable one being that the parafermionic operators (the modes of the parafermionic fields) satisfy generalized commutation relations expressed as infinite sums. In spite of this, these models display important simplifying features. For instance, all singular vectors can be found in closed form \cite{JMa}. In addition, the spectrum of primary fields is completely determined by the commutation relations \cite{ZFa}.  And, of more immediate interest in the present context, two (quasi-particle) bases of states are known.

\subsection{Parafermionic bases of states, paths and partitions: their origins}

The first result pertaining to the description of the  states of the irreducible modules
of the $\Z_k$ parafermionic models (and actually: of a conformal theory)
goes back to the seminal work of Lepowsky and Primc \cite{LP}.
 The basis of states they obtained is formulated in terms of restricted partitions, namely partitions satisfying the condition: $\la_j\geq \la_{j+k-1}+2$ 
-- the restriction being interpreted as a kind of generalized exclusion principle \cite{LW}. This is a pioneer work in another aspect: the generating function for this basis of parafermionic states provides the first example of a fermionic character formula
(a topic that  has been
launched many years later  by the Stony-Brook group 
\cite{KKMMa,KKMMb,KMM}). The key step in the construction of these $\Z_k$ fermionic characters uses  a result of Andrews for the generating function of restricted partitions
\cite{An}, which is nothing but  the sum-side of the generalized (Andrews-Gordon) version of the famous Rogers-Ramanujan identities.

These restricted partitions have a natural path representation, whose origin drags us into a little detour.
Restricted partitions were known to be related to partitions with prescribed successive ranks. To explain this link, recall that
  a partition $\mu= (\mu_1,\mu_2,\cdots)$ can be represented by a Ferrers (or Young)  diagram  with rows of length $\mu_j$. 
  Let $\mu'_j$ be the length of its columns. Via this diagram, the partition has a Frobenius representation:
  \begin{equation}
 \mu=  \begin{pmatrix} s_1&s_2&\cdots &s_d\\ t_1&t_2&\cdots &t_d \end{pmatrix}\;,
    \end{equation}
    with $s_j= \mu_j-j$, $t_j= \mu'_j-j $ and $d$ is the largest integer such that $\mu_d\geq d$.
The successive  ranks are defined as $SR(j)= s_j-t_j$. Fix an integer $i$ such that $1\leq i\leq k$. The prescription on the successive ranks is then $SR(j)\in [-i+2, 2k-i-1] $ for all $j$. Partitions satisfying this condition are equinumerous to those partitions with
$\la_j\geq \la_{j+k-1}+2$  and at most $i-1$ parts equal to 1 \cite{AnS} (see also \cite{Andr} chap. 9). Burge \cite{Bu} has displayed a nice bijection between those two types of partitions. It relies on the association of a partition of  each kind with a common binary word. This binary word has then been interpreted as a path in \cite{AnB}. The definition of these paths has subsequently been refined in \cite{AgB,BreL}. 
These are what we call here the Bressoud  paths. 
Roughly, a Bressoud path is composed of North-East (NE) and South-East (SE) edges in the first quadrant, with the constraint that the peaks have height at most $k-1$. In addition, portions of the paths can be disconnected, i.e., separated by segments of the $x$-axis.

By construction, Bressoud paths are in one-to-one correspondence with partitions satisfying $\la_j\geq \la_{j+k-1}+2$  and at most $i-1$ parts equal to 1. For a path, the number $i$ is related to the value of its initial vertical point, its final point being forced to lie on the $x$-axis. 

Now given that the states of the parafermionic theories are related to restricted partitions
and that these are related to Bressoud paths, parafermionic states can thus be represented by these paths. In this context, the number $i$ selects the highest-weight module. This is our first path description of parafermionic states. 

The basis of parafermionic states in \cite{LP,JMb} is formulated in terms of the modes of a single parafermionic field. However, it  is possible to set up a basis that involves the modes of the $k-1$ non-trivial parafermionic fields \cite{Geo,JM.A}. A multi-parafermionic basis element is naturally associated to a so-called multiple partition, an ordered set of $k-1$ partitions. Each partition in this set is composed of parts that are related to the modes of a given type of parafermionic field. Since multiple partitions and restricted partitions are two different expressions of the $\Z_k$ basis elements, they must be in one-to-one correspondence. This is indeed so \cite{JM.A,Mult}. In particular, 
this connection explains in which sense a restricted partition can be viewed as being composed of `particles' of charge running form 1 to $k-1$. 
The correspondence between these two types of partitions also implies that multiple partitions are bijectively related to Bressoud paths.

Notice, en passant, that although the Bressoud paths do describe the parafermionic states via their (purely combinatorial) correspondence with restricted or multiple partitions, they turn out to have a nice particle interpretation. 
In a path context, particles refer to the basic blocks in terms  of which the paths are built. Here these are basic triangles.
 But this pictorial particle interpretation happens to match perfectly 
the parafermionic quasi-particle description in that the different particles are precisely the different types of parafermionic operators. Indeed, a path is a sequence of particles that can be put in correspondence with a sequence of parafermionic operators. The energy and the charge of a particle in the path correspond respectively to  the mode value and the charge of the correlative parafermionic operator. A path is thus associated with a sequence of parafermionic operators, albeit not in the ordering characteristic of the multiple partitions.

We have so far recounted the origin of three of our four combinatorial representations of the parafermionic states: the restricted and  multiple partitions, and the Bressoud paths.

The second path description has its origin in the solution of the restricted-solid-on-solid (RSOS) model of Andrews, Baxter and Forrester \cite{ABF}. By means of the corner-transfer matrix,  the one-point probability of the order variable is expressed there in terms of a configuration sum. A path is interpreted as the contour of a particular configuration.
 The model has different regimes in which the paths are weighted differently. It has been noticed that the parafermionic states are described by the paths pertaining to regime II (an observation which seems to go back to \cite{Kyoto}). These paths are called here ABF paths.

ABF paths are similar to Bressoud paths in that they are composed of sequences of NE and SE edges lying in the first quadrant and they both terminate on the $x$-axis. But they differ in two essential ways. At first,  the ABF paths are necessarily connected: there are no zero-height horizontal  segments within these paths. They thus embody a definite notion of length. (The parafermionic states are recovered in the infinite length limit.) A second difference is that the peaks in the ABF paths can have height up to $k$, as opposed to $k-1$ for Bressoud paths.

Again, since they both describe the parafermionic states, a precise correspondence between these
two path descriptions is forecasted. Roughly, such a correspondence should eliminate their differences. As just indicated, these variances are that ABF paths can have peaks of height $k$ while Bressoud paths can have horizontal portions on the $x$-axis. Hence, heuristically, a ABF path gets transformed into a Bressoud path by crushing its height-$k$ peaks and inversely, a Bressoud path is transformed into a ABF path by uplifting to height $k$ its flat portions.  This is the crude idea behind the correspondence. Its precise formulation 
relies on the common representation  of a path of each type by the same multiple partition \cite{JMpath}. In this context, a multiple partition arises as a canonical reordering of the path data as specified by the position and the charge (which is related to the height) of its peaks. 
The key tool underlying  the reordering process is an exchange relation that is abstracted from the generalized commutation relations of the parafermionic modes. Actually, the exchange relation is the central technical device in this work. 


\subsection{The organization of the article}
The article is divided into two parts. The first one, composed of  Sections  \ref{SBres}-\ref{SABF}, is devoted to the presentation of the
four different combinatorial representations of the states of the $\Z_k$ parafermionic models. In order of their appearance, these are: 
the Bressoud paths, the restricted partitions, the multiple partitions and the  ABF paths. 

 In Section \ref{SBgf}, a detailed and pedagogical derivation of the generating function for the Bressoud paths (defined in Section \ref{SBdef}) is presented, in the spirit of the fermi-gas constructive method of  \cite{OleJS}.
This actually  represents the simplest example for the application of this method. Its detailed study, 
from a point of view that embodies a particle-like interpretation, is further motivated by being the starting point of the analysis for a large number of conformal field theories, as indicated in Section \ref{Sgb}.

The restricted partitions are defined in Section \ref{Sdefrp}. A simplified form of the Burge correspondence between these and the Bressoud paths is given in Section \ref{Sbur}. 
Multiple partitions are introduced in Section \ref{Sdefmp}; their bijective relation with paths and restricted partitions are  presented in Sections \ref{SBmp} and \ref{Srp+mp} respectively. Section \ref{Sgfmp} presents a remarkably simple derivation of their generating function. In perspective, this is the core of the simplest (known) derivation of the fermionic expression of the parafermionic characters. 
 Section \ref{SABF} is concerned with ABF paths; their link with the other three objects passes through their connexion with multiple partitions, which  is worked out in Section \ref{SABFmp}. The key result is presented  in Section \ref{SBvsAFB}, where the precise relation between Bressoud and ABF paths is displayed.

\begin{figure}[ht]
\caption{{\footnotesize The links between the partitions and paths.}}
\label{fig0}
\begin{center}
\begin{pspicture}(-4,-.5)(13,1.8)

\rput[tr](0,0){\rnode{A}{\psframebox{Bressoud paths}}}
\rput[tr](6,2){\rnode{B}{\psframebox{Restricted Partitions }}}
\rput[tr](11,0){\rnode{C}{\psframebox{ABF paths}}}
\rput[tr](6,0){\rnode{E}{\psframebox{Multiple ~Partitions~~}}}
\ncline[nodesep=3pt]{<->}{A}{E}
\ncline[nodesep=3pt]{<->}{E}{C}
\ncline[nodesep=3pt]{<->}{E}{B}

\ncline[nodesep=3pt]{<->}{A}{B}

\rput(1,1.){\scriptsize{\S \ref{Sbur}}}
\rput(1,0){\scriptsize{\S \ref{SBmp} }}
\rput(7.4,0){\scriptsize{\S \ref{SABFmp} - \S \ref{SBvsAFB}}}
\rput(4.8,1.){\scriptsize{\S \ref{Srp+mp}}}

\end{pspicture}
\end{center}
\end{figure}
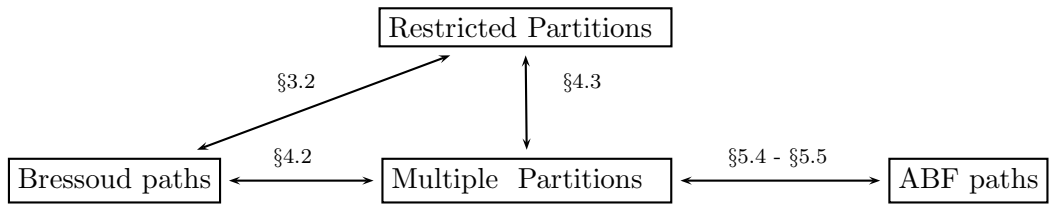

A sketch of the various bijective relations is pictured in Fig. \ref{fig0}. 
Note that although the presentation of these correspondences is to a large extend a review work, their exposition has been unified,
somewhat simplified and, in some instances, rephrased  in purely combinatorial terms, independently of the parafermionic interpretation. 
In consequence, this first part can be read independently of  the conformal-field-theory applications. 

The connection with the bases of states for the $\Z_k$ parafermionic models is presented in Section \ref{SpvsZ}. Finally, some results  concerning similar path descriptions for another parafermionic theory and  the minimal models, are summarized in Section \ref{Sother}.

A more detailed presentation of each section is contained in their respective opening paragraph.

\section{Bressoud paths}\label{SBres}

We introduce our first type of paths: the Bressoud paths, defined in Section \ref{SBdef}. The subsequent section is devoted to the construction of their generating function. Given the bijection with the parafermionic basis of states to be demonstrated later, this amounts to an explicit construction of the fermionic form of the parafermionic character  (up to a correction to be explained in  Section \ref{Sbas1}). This constructive method has also the interesting particularity of revealing the `particle content' of a path. Here `particles' are understood as the basic constituents in terms of which all the paths can be described.

\subsection{Defining the Bressoud paths}\label{SBdef}

{\bf  Bressoud paths}, or $\B$ paths for short, are defined as follows  \cite{BreL}:

\n {\bf Shape}: A $\B$ path is a sequence of integral points $(x,y)$ within the strip $x\geq 0$ and $0\leq y\leq k-1$, with adjacent points linked by NE, SE or H (horizontal) edges, the latter being allowed only if they lie on the $x$-axis.

\n In this work, it is understood that NE and SE edges link neighboring points (usually called vertices) whose heights differ by 1, i.e., they link $(x,y)$ and $(x+1,y\pm 1)$ respectively.

\n {\bf Initial and end points:} For a fixed value of $k$, the set of all  paths can be subdivided into classes  specified by the  initial point $a$, with $0\leq a\leq k-1$.
The final point is forced to lie on the $x$ axis.  There is no definite notion of length for a $\B$  path:  a path can always  be considered to be completed by an infinite tail  of H edges. 

\begin{figure}[ht]\
\caption{{\footnotesize A typical B$_{4,2}$ path (where the first index is the value of $k-1$, the maximal weight, and the second one indicates the initial point $a=2$). The charge of each peak  is given in parenthesis.}} 
\label{fig1}
\begin{center}
\begin{pspicture}(2,0)(13.5,3)
{\psset{yunit=35pt,xunit=35pt,linewidth=.8pt}
\psline{-}(0.3,0.3)(0.3,1.5) \psline{->}(0.3,0.3)(13.0,0.3)
\psset{linestyle=dotted}
\psline{<->}(0.6,0.6)(1.2,0.6)
\psline{<->}(01.2,0.6)(4.2,0.6)
\psline{<->}(4.2,0.6)(4.8,0.6)
\psline{<->}(4.8,0.6)(6.6,0.6)
\psline{<->}(2.7,0.9)(3.9,0.9)
\psline{<->}(8.7,0.9)(9.3,0.9)
\psset{linestyle=solid}
\psline{-}(0.3,0.3)(0.3,0.4) \psline{-}(0.6,0.3)(0.6,0.4)
\psline{-}(0.9,0.3)(0.9,0.4) \psline{-}(1.2,0.3)(1.2,0.4)
\psline{-}(1.5,0.3)(1.5,0.4) \psline{-}(1.8,0.3)(1.8,0.4)
\psline{-}(2.1,0.3)(2.1,0.4) \psline{-}(2.4,0.3)(2.4,0.4)
\psline{-}(2.7,0.3)(2.7,0.4) \psline{-}(3.0,0.3)(3.0,0.4)
\psline{-}(3.3,0.3)(3.3,0.4) \psline{-}(3.6,0.3)(3.6,0.4)
\psline{-}(3.9,0.3)(3.9,0.4) \psline{-}(4.2,0.3)(4.2,0.4)
\psline{-}(4.5,0.3)(4.5,0.4) \psline{-}(4.8,0.3)(4.8,0.4)
\psline{-}(5.1,0.3)(5.1,0.4) \psline{-}(5.4,0.3)(5.4,0.4)
\psline{-}(5.7,0.3)(5.7,0.4) \psline{-}(6.0,0.3)(6.0,0.4)
\psline{-}(6.3,0.3)(6.3,0.4) \psline{-}(6.6,0.3)(6.6,0.4)
\psline{-}(6.9,0.3)(6.9,0.4) \psline{-}(7.2,0.3)(7.2,0.4)
\psline{-}(7.5,0.3)(7.5,0.4) \psline{-}(7.8,0.3)(7.8,0.4)
\psline{-}(8.1,0.3)(8.1,0.4) \psline{-}(8.4,0.3)(8.4,0.4)
\psline{-}(8.7,0.3)(8.7,0.4)
\psline{-}(9.0,0.3)(9.0,0.4)
\psline{-}(9.3,0.3)(9.3,0.4) \psline{-}(9.6,0.3)(9.6,0.4)
\psline{-}(9.9,0.3)(9.9,0.4) \psline{-}(10.2,0.3)(10.2,0.4)
\psline{-}(10.5,0.3)(10.5,0.4) \psline{-}(10.8,0.3)(10.8,0.4)
\psline{-}(11.1,0.3)(11.1,0.4) \psline{-}(11.4,0.3)(11.4,0.4)
\psline{-}(11.7,0.3)(11.7,0.4) \psline{-}(12,0.3)(12,0.4)
\psline{-}(12.3,0.3)(12.3,0.4) \psline{-}(12.6,0.3)(12.6,0.4)

\rput(0.9,-0.05){{\scriptsize $2$}}
\rput(2.1,-0.05){{\scriptsize $6$}} \rput(3.3,-0.05){{\scriptsize $10$}}
\rput(4.5,-0.05){{\scriptsize $14$}}\rput(5.7,-0.05){{\scriptsize $18$}}
\rput(7.5,-0.05){{\scriptsize $24$}}
\rput(9,-0.05){{\scriptsize $29$}}\rput(9.9,-0.05){{\scriptsize $32$}}
\rput(12,-0.05){{\scriptsize $39$}}
 \psline{-}(0.3,0.6)(0.4,0.6)
\psline{-}(0.3,0.9)(0.4,0.9) \psline{-}(0.3,1.2)(0.4,1.2)
\psline{-}(0.3,1.5)(0.4,1.5) 

\rput(0.05,0.9){{\scriptsize $2$}} \rput(0.05,1.5){{\scriptsize $4$}}
\rput(0.05,0.6){{\scriptsize $1$}} \rput(0.05,1.2){{\scriptsize $3$}}
\psline{-}(0.3,0.9)(0.6,0.6) \psline{-}(0.6,0.6)(0.9,0.9)
\psline{-}(0.9,0.9)(1.2,0.6) \psline{-}(1.2,0.6)(1.5,0.9)
\psline{-}(1.5,0.9)(1.8,1.2) \psline{-}(1.8,1.2)(2.1,1.5)
\psline{-}(2.1,1.5)(2.4,1.2) \psline{-}(2.4,1.2)(2.7,0.9)
\psline{-}(2.7,0.9)(3.0,1.2) \psline{-}(3.0,1.2)(3.3,1.5)
\psline{-}(3.3,1.5)(3.6,1.2) \psline{-}(3.6,1.2)(3.9,0.9)
\psline{-}(3.9,0.9)(4.2,0.6) \psline{-}(4.2,0.6)(4.5,0.9)
\psline{-}(4.5,0.9)(4.8,0.6) \psline{-}(4.8,0.6)(5.1,0.9)
\psline{-}(5.1,0.9)(5.4,1.2) \psline{-}(5.4,1.2)(5.7,1.5)
\psline{-}(5.7,1.5)(6.0,1.2)
 \psline{-}(6.0,1.2)(6.3,0.9)\psline{-}(6.3,0.9)(6.6,0.6)
\psline{-}(6.6,0.6)(6.9,0.3) \psline{-}(6.9,0.3)(7.2,0.6)
\psline{-}(7.2,0.6)(7.5,0.9) \psline{-}(7.5,0.9)(7.8,0.6)
\psline{-}(7.8,0.6)(8.1,0.3) \psline{-}(8.1,0.3)(9,1.2)
\psline{-}(9.0,1.2)(9.3,0.9) \psline{-}(9.3,0.9)(9.6,1.2)
\psline{-}(9.6,1.2)(9.9,1.5) \psline{-}(9.9,1.5)(10.2,1.2)
\psline{-}(10.2,1.2)(10.5,0.9) \psline{-}(10.5,0.9)(10.8,0.6)
\psline{-}(10.8,0.6)(11.1,0.3)
\psline{-}(11.4,0.3)(12,0.9)\psline{-}(12.0,0.9)(12.6,0.3)

\rput(0.9,1.15){{\scriptsize $(1)$}}
\rput(2.1,1.85){{\scriptsize$(3)$}}
\rput(3.3,1.85){{\scriptsize $(2)$}}
\rput(4.5,1.15){{\scriptsize $(1)$}}
\rput(5.7,1.85){{\scriptsize $(3)$}}
\rput(7.5,1.15){{\scriptsize$(2)$}}
\rput(9,1.45){{\scriptsize $(1)$}}
\rput(9.9,1.85){{\scriptsize $(4)$}}
\rput(12,1.15){{\scriptsize  $(2)$}}
}
\end{pspicture}
\end{center}
\end{figure}
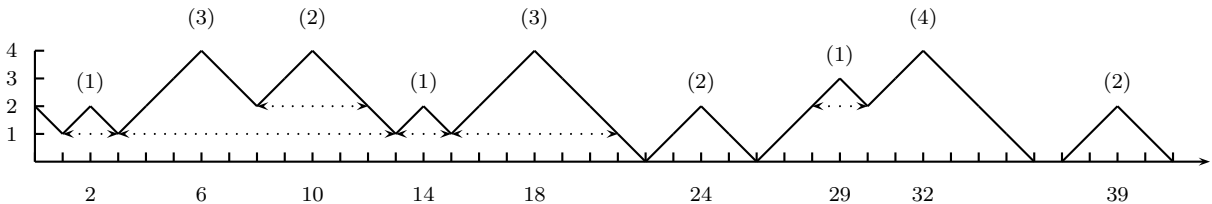

\n $\B$ paths with initial point $a$ will be denoted $\Ba$. An example of B$_{4,2}$ path is given in Fig. \ref{fig1}.

\n {\bf Weight}: The weight $w$ of a lattice path is  the sum of the $x$-coordinate of all its  peaks:
\begin{equation} \label{defwpath} w= \sum_{x\geq1} w(x)\qquad {\rm where} \qquad w(x)=
\begin{cases}
x & \text{if $x$ is the position of a peak} ,\\
 0 &\text{otherwise}\;. \end{cases}
\end{equation}

\n The weight of the path of Fig. \ref{fig1} is thus $2+6+10+14+18+24+29+32+39$.

Peaks are characterized by their height or more precisely, their relative height \cite{BreL}, which we will call their charge \cite{OleJS}.  This is a key concept for the combinatorial interpretation of the paths.
For an isolated peak described by a triangle starting and ending on the $x$ axis,  the charge is equal to the height. This is not as simple within a {\bf charge complex} \cite{OleJS} (or a mountain  \cite{BreL}), that is, within portions of the paths delimited by two points on the horizontal axis (or the $y$ axis and the first point on the $x$ axis) and containing more than one peak.  This situation requires a criterion more  precise that  the mere height and it is formulated as follows.
  
  \n {\bf Charge:} The charge of 
a peak with coordinates $(x,y)$ is the largest integer $c$ such that we can find two points $(x',y-c)$ and $(x'',y-c)$ on the path  with $x'<x<x''$ and such that between these two points there are no peaks of height larger than $y$ and every peak of height equal to $y$ has weight larger than $x$ \cite{BP}.  

In other words, starting from a peak, we go down in both directions and from the first point (that closest to the peak) at which the path changes its direction, on one side or the other, we draw a baseline from which the  height is read: this height is the charge.
Note that if two peaks in a complex have the same height, 
it is the leftmost peak which is attributed the largest charge (a convention that is captured by the above precise definition).

For the path pictured in Fig. \ref{fig1},  the value of the charge is given above each peak. A dotted line indicates the line from which the height must be measured to give the charge. In this example, there are four charge complexes.

The charge content of a path is the specification of the numbers  $m_j$ of peaks of  charge $j$, for $1\leq j\leq k-1$. For instance, the path of Fig. \ref{fig1} has $m_1=m_2=3, m_3=2$ and $ m_4=1$. 

\n {\bf Total charge:} The total charge $m$ of a path is the sum of the charges of all its peaks: 
\begin{equation}\label{toch}
m=\sum_{j=1}^{k-1} jm_j.
\end{equation}

\n {\bf Particles}: Peaks of charge $j$ are also called particles of charge $j$ (or type-$j$ particles).
From now on, we will  use both terms (peaks and particles) interchangeably and refer to a set  of values of $\{m_j\}$ either as the charge or the particle content of the path.

\subsection{The generating function for paths: direct method }\label{SBgf}

There are various ways of computing the generating function for $\B$ paths.  Here, we use a direct  fermi-gas-type method pioneered by Warnaar \cite{OleJS} in a different context.
Although rather elementary, this example has apparently not been worked out along this line  before (see \cite{BreL} for a different derivation based on a recursive method and \cite{W97} for a derivation in the same spirit but for related paths).

\subsubsection{The strategy in few steps}

The problem is to enumerate all the $\B$ paths with fixed initial position $a$, taking into account their weight. The sought-for generating function is thus
\begin{equation}G_a(q)= \sum_{\text{$\Ba$ paths}} q^{w} = \sum_{n\geq 0} d(n;a)\, q^n,
\end{equation}
where $d(n;a)$ is the number of paths of weight $n$ and initial vertical position $a$.

The problem is broken into two parts: 

\begin{enumerate}

\item Enumerate and weight all the paths with a fixed charge content  $\{m_j\}$.

This part is itself worked out in different steps as follows:
\begin{enumerate}
\item For a fixed   $\{m_j\}$, identity the configuration (that is, the distribution of the particles along the path) that minimizes the weight. Call this configuration the {\bf minimal-weight configuration} (mwc). Evaluate its weight $w_{\text{mwc}}(a)$.

\item Enumerate all possible configurations that can be obtained from this minimal-weight configuration by the  displacements of the different particles and compute the corresponding weight relative to $w_{\text{mwc}}(a)$.

These displacements are subject to two rules:

\begin{enumerate}

\item {Identical particles are impenetrable. }

\item {Particles of different charges can penetrate under  the condition that the individual  charges  must be preserved. }

\end{enumerate}
\end{enumerate}

\item Sum over all values of $m_j$.

\end{enumerate}

The impenetrability of identical particles can be viewed as a hard-core repulsion and this property is the rationale for tagging this as a {\it fermi-gas method}. This  impenetrability criterion prevents displacements that would otherwise produce identical paths.

Before plunging into  the details of this enumerative problem, let us make some remarks that will render its analysis less formidable than a first sight evaluation might suggest.

\begin{enumerate}

\item The ordering of the peaks that characterize the minimal-weight configuration is easily determined by comparing the weight of the different configurations of {\it two peaks with different charges}. 

\item The $a$ dependence in $w_{\text{mwc}}(a)$
 is simply obtained by determining how the choice of $a$ affects the weight of a path composed of {\it a single peak}, as a function of its charge.

\item  The analysis of the different configurations obtained from the various interpenetration arrangements is again analyzed first by considering only {\it two particles of different charges}. The preservation of  the individual  charges, say $i,\,j$, means that no particle of charge $>\; \text{max}\, (i,j)$ should be generated in the penetration process.

\end{enumerate}

\subsubsection{The minimal-weight configuration}\label{Bmwc}

As just remarked, to identify the  minimal-weight configuration for a fixed charge content, one only has to investigate the case of two peaks of different charges, say 1 and 3. Since the weight of each peak is given by its $x$-position, the total weight is manifestly minimized when the two peaks are as close as possible to the origin. Various such configurations are displayed in Fig. \ref{fig5} (whose detailed structure will be further discussed below) for the case where $a=0$. This makes clear that  in order to minimize the weight, the peak 1 must be at the left of the peak 3. 


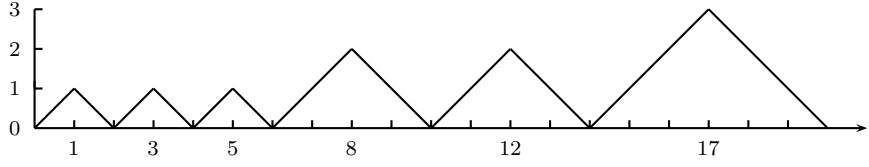
\begin{figure}[ht]
\caption{The minimal-weight configuration of a B$_{k-1,0}$ path for $k\geq 4$ with particle content: $m_1=3,\,m_2=2,\, m_3=1$. }
\vskip-1cm
\label{fig2}
\begin{center}
\begin{pspicture}(0,-0.5)(11,2.5)

{\psset{yunit=15pt,xunit=15pt,linewidth=.8pt}

\psline(0,1)(0,1.2)
\psline(1,0)(1,.2)     \psline(2,0)(2,.2)        \psline(3,0)(3,.2)
\psline(4,0)(4,.2)     \psline(5,0)(5,.2)        \psline(6,0)(6,.2)
\psline(7,0)(7,.2)     \psline(8,0)(8,.2)        \psline(9,0)(9,.2)
\psline(10,0)(10,.2)     \psline(11,0)(11,.2)        \psline(12,0)(12,.2)
\psline(13,0)(13,.2)     \psline(14,0)(14,.2)        \psline(15,0)(15,.2)
\psline(16,0)(16,.2)     \psline(17,0)(17,.2)        \psline(18,0)(18,.2)
\psline(19,0)(19,.2)

 \psline{-}(0,3.0)(0.2,3)
\psline{-}(0,2)(0.2,2)
\psline{-}(0,1)(0.2,1.)


\psline{-}(0,0)(0,3)
\psline{->}(0,0)(21,0)


\rput(-.5,0){\scriptsize 0}
\rput(-.5,1){\scriptsize 1}
\rput(-.5,2){\scriptsize 2}
\rput(-.5,3){\scriptsize 3}

\rput(1,-0.5){\scriptsize 1}
\rput(3,-0.5){\scriptsize 3}
\rput(5,-0.5){\scriptsize 5}
\rput(8,-0.5){\scriptsize 8}
\rput(12,-0.5){\scriptsize 12}
\rput(17,-0.5){\scriptsize 17}


\psline(0,0)(1,1)
\psline(1,1)(2,0)
\psline(2,0)(3,1)
\psline(3,1)(4,0)
\psline(4,0)(5,1)
\psline(5,1)(6,0)
\psline(6,0)(8,2)
\psline(8,2)(10,0)
\psline(10,0)(12,2)
\psline(12,2)(14,0)
\psline(14,0)(17,3)
\psline(17,3)(20,0)

\psset{dotsize=2pt}\psset{dotstyle=*}
}

\end{pspicture}
\end{center}
\end{figure}

This is generic for $\B$ paths: the minimal-weight configuration is obtained by ordering the peaks in increasing charge and packing them the closest to the origin. The example where  $m_1=3,\, m_2=2, \, m_3=1$ is displayed in Fig. {\ref{fig2}.
Its weight is $w_{\text {mwc}}=1+3+5+8+12+17$.
For a generic charge content, with $a=0$, one has:
\begin{equation}\label{mwc0}
w_{\text {mwc}}(0)= \sum_{i,j=1}^{k-1} \text{min}(i,j)\,m_i\, m_j.
\end{equation}
This is proved as follows. The weight of a sequence of $m_j$ peaks of charge $j$, at positions $j,3j,\ldots , j(2 m_j-1)$ is
\begin{equation}\label{aau}
\sum_{s=0}^{m_j-1} j(2s+1) = jm_j^2.
\end{equation}
This result holds for a sequence that  starts at the origin. But if  it starts instead at $x_0$, the weight at each peak is augmented by $x_0$; the total weight is now:
\begin{equation}\label{aav}
 jm_j^2+m_j x_0.
 \end{equation}
In the minimal-weight configuration, the  value of $x_0$ is simply the total diameter of all the particles of lower charge:
\begin{equation}
x_0 = 2\sum_{i=1}^{j-1} im_i.
\end{equation}
Substituting this value into (\ref{aau}) and summing over $j$ yields precisely (\ref{mwc0}).

Now let us see how $w_{\text {mwc}}(0)$ is affected by choosing the initial vertical position to be $a>0$. Figure \ref{fig3} illustrates the minimal-weight configuration of those paths composed of a single peak of different charge when $a=2$ and $k=5$. One sees that the increase of the initial point from 0 to 2 does not affect the weight of the peaks of charge 1 and 2, but it shifts that of charge 3 by 1 and that of charge 4 by 2. Note that it is the bound on the height $y\leq k-1=4$ that causes this shift, in that it forces  the path representing a single peak of charge $j>k-1-a$ to go down before moving upward to reach the maximal allowed height. In this example, the shift of the minimal weight of the particle $j$ caused by the displacement of the initial point from 0 to 
$a$, is  $\text{max}(j-2,0)$. Here the origin of the term $j-2$ is $j+a-(k-1)$, as we now demonstrate. 

If $x_{\text{min}}(j;a)$ stands for the minimal $x$-position of a peak of charge $j$ for a the path starting at the point $(0,a)$, then
\begin{equation}\label{mxpos}
x_{\text{min}}(j;a) = j+\text{max}(j+a-k+1,0),
\end{equation}
since when $j+a>k-1$, there must be $j+a-(k-1)$ SE edges before the path starts to go upward with $j$ NE edges. The peak position, hence the weight, is thus shifted by $j+a-(k-1)$.
With $m_j$ peaks, the shift in weight is $\text{max}(j+a-k+1,0)$ times $m_j$. The expression for $w_{\text {mwc}}(a)$ is thus
\begin{equation}\label{mwca}
w_{\text {mwc}}(a)= w_{\text {mwc}}(0)+ \sum_{j=0}^{k-1} \text{max}(j+a-k+1,0)\, m_j.
\end{equation}

\begin{figure}[ht]
\caption{The minimal-weight configuration for B$_{4,2}$ paths composed of a  single peak of charge ranging from 1 to 4.}
\label{fig3}
\begin{center}
\begin{pspicture}(-2,0)(16,2)

{\psset{yunit=12pt,xunit=12pt,linewidth=.8pt}

\psline(0,1)(0,1.2)
\psline(1,0)(1,.2)     \psline(2,0)(2,.2)        \psline(3,0)(3,.2)
\psline(4,0)(4,.2)     \psline(5,0)(5,.2)        \psline(6,0)(6,.2)
\psline(7,0)(7,.2)     \psline(8,0)(8,.2)        \psline(9,0)(9,.2)
\psline(10,0)(10,.2)     \psline(11,0)(11,.2)        \psline(12,0)(12,.2)
\psline(13,0)(13,.2)     \psline(14,0)(14,.2)        \psline(15,0)(15,.2)
\psline(16,0)(16,.2)     \psline(17,0)(17,.2)        \psline(18,0)(18,.2)
\psline(19,0)(19,.2)   
\psline(20,0)(20,.2)     \psline(21,0)(21,.2)        \psline(22,0)(22,.2)
\psline(23,0)(23,.2)     \psline(24,0)(24,.2)        \psline(25,0)(25,.2)

\psline(26,0)(26,.2)     \psline(27,0)(27,.2)        \psline(28,0)(28,.2)
\psline(29,0)(29,.2)     \psline(30,0)(30,.2)        \psline(31,0)(31,.2)

 \psline{-}(0,3.0)(0.2,3)
\psline{-}(0,2)(0.2,2)
\psline{-}(0,1)(0.2,1.) 
\psline{-}(0,4)(0.2,4.) 

\rput(1,4.8){\scriptsize (1)}
\rput(7,4.8){\scriptsize (2)}
\rput(16,4.8){\scriptsize (3)}
\rput(27,4.8){\scriptsize (4)}

\rput(1,-0.5){\scriptsize 1}
\rput(7,-0.5){\scriptsize 2}
\rput(16,-0.5){\scriptsize 4}
\rput(27,-0.5){\scriptsize 6}


\rput(-.5,0){\scriptsize 0}
\rput(-.5,1){\scriptsize 1}
\rput(-.5,2){\scriptsize 2}
\rput(-.5,3){\scriptsize 3}
\rput(-.5,4){\scriptsize 4}
\psline(0,2)(1,3)
\psline(1,3)(4,0)

\psline(5,2)(7,4)
\psline(7,4)(11,0)

\psline(12,2)(13,1)
\psline(13,1)(16,4)
\psline(16,4)(20,0)

\psline(21,2)(23,0)
\psline(23,0)(27,4)
\psline(27,4)(31,0)

\psline{-}(0,0)(0,4.)
\psline{-}(5,0)(5,4.)
\psline{-}(12,0)(12,4.)
\psline{-}(21,0)(21,4.)

\psline{-}(0,0)(4.,0)
\psline{-}(5,0)(11.5,0)
\psline{-}(12,0)(20.5,0)
\psline{-}(21,0)(31.5,0)

 \psline{-}(0,3.0)(0.2,3)
\psline{-}(0,2)(0.2,2)
\psline{-}(0,1)(0.2,1.) 
\psline{-}(0,4)(0.2,4.) 

 \psline{-}(5,3.0)(5.2,3)
\psline{-}(5,2)(5.2,2)
\psline{-}(5,1)(5.2,1.) 
\psline{-}(5,4)(5.2,4.) 

 \psline{-}(12,3.0)(12.2,3)
\psline{-}(12,2)(12.2,2)
\psline{-}(12,1)(12.2,1.) 
\psline{-}(12,4)(12.2,4.) 

 \psline{-}(21,3.0)(21.2,3)
\psline{-}(21,2)(21.2,2)
\psline{-}(21,1)(21.2,1.) 
\psline{-}(21,4)(21.2,4.) 

}

\end{pspicture}
\end{center}
\end{figure}
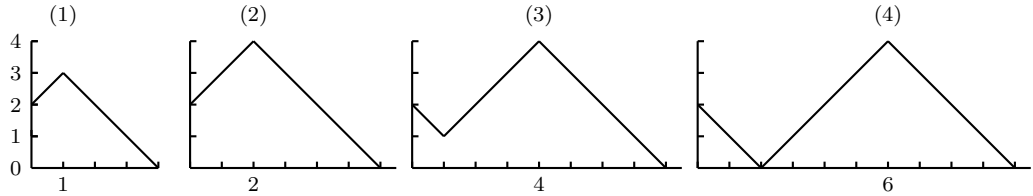

\subsubsection{Deforming the minimal-weight configuration}\label{Sdefo}

We now have to determine all the possible configurations with fixed particle content. These can all be obtained  from the minimal-weight configuration by a succession of simple displacements. Consider first the case where we have only identical particles, say  of charge 1. The starting configuration for $m_1=3$ is displayed in Fig. \ref{fig4}. We now have to determine the allowed displacements of the peaks and weight them. The rightmost peak can be displaced by any integer, say $\mu_1$.
Since the particles with identical charge cannot penetrate each other, the displacements of the other particles is limited: the second one can be displaced by any integer $\mu_2$ that is $\leq \mu_1$. Similarly, the leftmost one can be displaced by $\mu_3\leq \mu_2$. 
 Fig. \ref{fig4} illustrates the displacements of the rightmost peak by 9, the next by 6 and the third by 4. The weight of this configuration, with respect to that of minimal weight, is $9+6+4$.
 
The generating factor for these displacements is simply  the number of
partitions with at most three parts (`at most' since some numbers in $(\mu_1,\mu_2,\mu_3)$  are allowed to be 0, corresponding to no displacement). This generating function is simply \cite{Andr} (see e.g., Theo. 1.1 or Theo. 3.2 with $N\rw\y$):
\begin{equation}
\frac{1}{(1-q)(1-q^2)(1-q^3)}\equiv \frac1{(q)_3}.
\end{equation}
The generalization to the case of $m_1$ particles of charge 1 is immediate: the  generating factor is
$(q)_{m_1}^{-1}$, where
\begin{equation}\label{defqm}
(q)_m= \prod_{i=1}^m (1-q^i).
\end{equation}



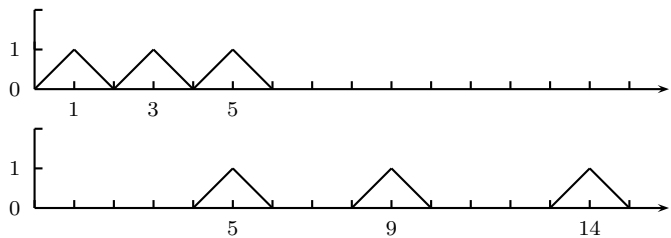
\begin{figure}[ht]
\caption{The minimal weight configuration for $a=0$ and $m_1=3$, and the displacement of three peaks (ordered from right to left) by 9, 6 and 4.}
\label{fig4}
\begin{center}
\vskip-1.9cm
\begin{pspicture}(0,-1.5)(11,3)

{\psset{yunit=15pt,xunit=15pt,linewidth=.8pt}

\psline(0,2)(0.2,2)
\psline(0,1)(0.2,1.)
\psline(0,1)(0,1.2)
\psline(1,0)(1,.2)     \psline(2,0)(2,.2)        \psline(3,0)(3,.2)
\psline(4,0)(4,.2)     \psline(5,0)(5,.2)        \psline(6,0)(6,.2)
\psline(7,0)(7,.2)     \psline(8,0)(8,.2)        \psline(9,0)(9,.2)
\psline(10,0)(10,.2)     \psline(11,0)(11,.2)        \psline(12,0)(12,.2)
\psline(13,0)(13,.2)     \psline(14,0)(14,.2)        \psline(15,0)(15,.2)

\psline{-}(0,0)(0,2)
\psline{->}(0,0)(16,0)


\rput(-.5,0){\scriptsize 0}

\rput(-.5,1){\scriptsize 1}
\rput(1,-.5){\scriptsize 1}
\rput(3,-.5){\scriptsize 3}
\rput(5,-.5){\scriptsize 5}

\psline(0,0)(1,1)
\psline(1,1)(2,0)
\psline(2,0)(3,1)
\psline(3,1)(4,0)
\psline(4,0)(5,1)
\psline(5,1)(6,0)
}


{\psset{yunit=15pt,xunit=15pt,linewidth=.8pt}

\psline(0,-1)(0.2,-1)
\psline(0,-2)(0.2,-2.)
\psline(0,-3)(0,-2.8)
\psline(1,-3)(1,-2.8)     \psline(2,-3)(2,-2.8)        \psline(3,-3)(3,-2.8)
\psline(4,-3)(4,-2.8)     \psline(5,-3)(5,-2.8)        \psline(6,-3)(6,-2.8)
\psline(7,-3)(7,-2.8)     \psline(8,-3)(8,-2.8)        \psline(9,-3)(9,-2.8)
\psline(10,-3)(10,-2.8)     \psline(11,-3)(11,-2.8)        \psline(12,-3)(12,-2.8)
\psline(13,-3)(13,-2.8)     \psline(14,-3)(14,-2.8)        \psline(15,-3)(15,-2.8)

\psline{-}(0,-3)(0,-1)
\psline{->}(0,-3)(16,-3)

%
\rput(-.5,-3){\scriptsize 0}
\rput(-.5,-2){\scriptsize 1}
\rput(9,-3.5){\scriptsize 9}
\rput(14,-3.5){\scriptsize 14}
\rput(5,-3.5){\scriptsize 5}

\psline(4,-3)(5,-2)
\psline(5,-2)(6,-3)
\psline(8,-3)(9,-2)
\psline(9,-2)(10,-3)
\psline(13,-3)(14,-2)
\psline(14,-2)(15,-3)
}
\end{pspicture}
\end{center}
\end{figure}


Let us now consider the displacements of the particles of a given type in presence of other particles with different charges. Consider first the simple case of particles of charge 1 and 3. The minimal-weight configuration is displayed in Fig. \ref{fig5}(a).
The subsequent figures display the displacements of the particle of charge 1, by unit steps, into the larger one, illustrating  the various stages of interpenetration. The basic constraint is that the individual charges must be preserved. In particular, no  peak of charge $>3$ should be generated. In the first two steps (pictured in (b) and (c)), the particle 1 climbs in the straight-up segment of the particle 3.
The pattern (c) is clearly allowed since this still unambiguously describes a configuration of charge 1 and 3. However, our convention for reading the charge ascribes charge 3 to the leftmost peak and not to the rightmost one. The rule is thus that after this second move, it is understood that the identity of the two peaks is swapped. 
This rule allows for further moves of the charge-1 particle through the larger one, moving now in its straight-down segment, as in (d), until it gets detached of it,  in (e), from which the successive  displacements are mere translation on the $x$-axis. We stress that the configuration (d) is the one obtained by displacing the peak 1, from the minimal-weight configuration (a), by  three units. The main points here are that: (1) this procedure allows for an unrestricted sequence of successive displacements; and (2) that each step modifies the weight of the  path by 1.

Note that in this example, the peak of charge 3, initially at position 5, sees its position shifted at 3 after the passage of the charge-1 particle. This is the closest it can be from the  origin for the boundary condition under consideration, here $a=0$. Hence, the displacement of the smaller particles through the larger ones gives to these larger peaks access to the positions smaller than those fixed by the minimal-weight configuration.

\begin{figure}[ht]
\caption{{The various configurations obtained by displaying the peak of charge 1 through a peak of charge 3, in unit steps. The weight increase is 1 at each step.}}
\label{fig5}
\vskip-1.4cm\begin{center}
\begin{pspicture}(0,-5.5)(11,2)

{\psset{yunit=15pt,xunit=15pt,linewidth=.8pt}

\psline(0,1)(0,1.2)
\psline(1,0)(1,.2)     \psline(2,0)(2,.2)        \psline(3,0)(3,.2)
\psline(4,0)(4,.2)     \psline(5,0)(5,.2)        \psline(6,0)(6,.2)
\psline(7,0)(7,.2)     \psline(8,0)(8,.2)        \psline(9,0)(9,.2)
\psline(10,0)(10,.2)     \psline(11,0)(11,.2)        \psline(12,0)(12,.2)
\psline(13,0)(13,.2)     \psline(14,0)(14,.2)        \psline(15,0)(15,.2)
\psline(16,0)(16,.2)     \psline(17,0)(17,.2)        \psline(18,0)(18,.2)

\psline{-}(0,0)(0,3.)
\psline{->}(0,0)(8.5,0)
\psline{->}(9,0)(19.,0)
\psline(0,2)(0.2,2)
\psline(0,1)(0.2,1.)
\psline(0,3)(0.2,3)


\rput(4,3.4){\scriptsize (a)}
\rput(13,3.5){\scriptsize (b)}
\rput(4,-1.5){\scriptsize (c)}
\rput(13,-1.5){\scriptsize (d)}
\rput(4,-6.5){\scriptsize (e)}
\rput(13,-6.5){\scriptsize (f)}
\rput(4,-.5){\scriptsize$w=6$}
\rput(13,-0.5){\scriptsize$w=7$}

\rput(-.5,0){\scriptsize 0}
\rput(-.5,1){\scriptsize 1}
\rput(-.5,2){\scriptsize 2}
\rput(-.5,3){\scriptsize 3}
\psline(0,0)(1,1)
\psline(1,1)(2,0)
\psline(2,0)(5,3)
\psline(5,3)(8,0)

\psline{-}(9,0)(9,3)
\psline(9,0)(11,2)
\psline(11,2)(12,1)
\psline(12,1)(14,3)
\psline(14,3)(17,0)

\psline(9,2)(9.2,2)
\psline(9,1)(9.2,1.)
\psline(9,3)(9.2,3)
}


{\psset{yunit=15pt,xunit=15pt,linewidth=.8pt}

\psline(1,-5)(1,-4.8)     \psline(2,-5)(2,-4.8)        \psline(3,-5)(3,-4.8)
\psline(4,-5)(4,-4.8)     \psline(5,-5)(5,-4.8)        \psline(6,-5)(6,-4.8)
\psline(7,-5)(7,-4.8)     \psline(8,-5)(8,-4.8)        \psline(9,-5)(9,-4.8)
\psline(10,-5)(10,-4.8)     \psline(11,-5)(11,-4.8)        \psline(12,-5)(12,-4.8)
\psline(13,-5)(13,-4.8)     \psline(14,-5)(14,-4.8)        \psline(15,-5)(15,-4.8)
\psline(16,-5)(16,-4.8)     \psline(17,-5)(17,-4.8)        \psline(18,-5)(18,-4.8)

\psline(0,-2)(0.2,-2)
\psline(0,-3)(0.2,-3)
\psline(0,-4)(0.2,-4)

\psline{-}(0,-5)(0,-2)
\psline{->}(0,-5)(8.5,-5)
\psline{->}(9,-5)(19.,-5)


\rput(-.5,-5){\scriptsize 0}
\rput(-.5,-4){\scriptsize 1}
\rput(-.5,-3){\scriptsize 2}
\rput(-.5,-2){\scriptsize 3}
\psline(0,-5)(3,-2)
\psline(3,-2)(4,-3)
\psline(4,-3)(5,-2)
\psline(5,-2)(8,-5)

\rput(4,-5.5){\scriptsize $w=8$ + identity flip}
\rput(13,-5.5){\scriptsize $w=9$}

\psline{-}(9,-5)(9,-2)
\psline(9,-5)(12,-2)
\psline(12,-2)(14,-4)
\psline(14,-4)(15,-3)
\psline(15,-3)(17,-5)
}



{\psset{yunit=15pt,xunit=15pt,linewidth=.8pt}

\psline(0,1)(0,1.2)
\psline(1,-10)(1,-9.8)     \psline(2,-10)(2,-9.8)        \psline(3,-10)(3,-9.8)
\psline(4,-10)(4,-9.8)     \psline(5,-10)(5,-9.8)        \psline(6,-10)(6,-9.8)
\psline(7,-10)(7,-9.8)     \psline(8,-10)(8,-9.8)        \psline(9,-10)(9,-9.8)
\psline(10,-10)(10,-9.8)     \psline(11,-10)(11,-9.8)        \psline(12,-10)(12,-9.8)
\psline(13,-10)(13,-9.8)     \psline(14,-10)(14,-9.8)        \psline(15,-10)(15,-9.8)
\psline(16,-10)(16,-9.8)     \psline(17,-10)(17,-9.8)        \psline(18,-10)(18,-9.8)

\psline(9,-7)(9.2,-7)
\psline(9,-8)(9.2,-8)
\psline(9,-9)(9.2,-9)

\psline(9,-4)(9.2,-4)
\psline(9,-3)(9.2,-3)
\psline(9,-2)(9.2,-2)

\psline(0,-7)(0.2,-7)
\psline(0,-8)(0.2,-8)
\psline(0,-9)(0.2,-9)

\psline{-}(0,-10)(0,-7)
\psline{->}(0,-10)(8.5,-10)
\psline{->}(9,-10)(19.,-10)


\rput(-.5,-10){\scriptsize 0}
\rput(-.5,-9){\scriptsize 1}
\rput(-.5,-8){\scriptsize 2}
\rput(-.5,-7){\scriptsize 3}
\psline(0,-10)(3,-7)
\psline(3,-7)(6,-10)
\psline(6,-10)(7,-9)
\psline(7,-9)(8,-10)

\rput(4,-10.5){\scriptsize$w=10$ }
\rput(13,-10.5){\scriptsize $w=11$}

\psline{-}(9,-10)(9,-7)
\psline(9,-10)(12,-7)
\psline(12,-7)(15,-10)
\psline(16,-10)(17,-9)
\psline(17,-9)(18,-10)
}

\end{pspicture}
\end{center}
\end{figure}
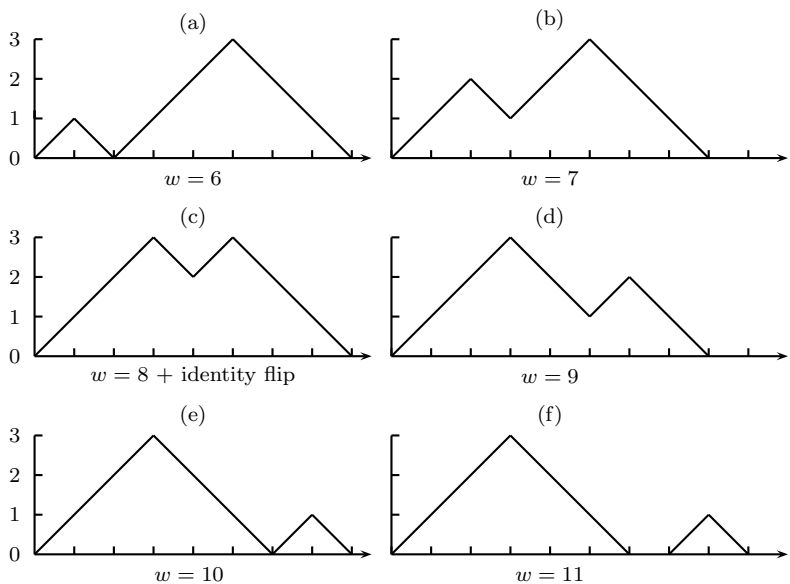

We have thus seen that every displacement of one unit
of the charge-1 particle increases the weight by 1 independently of the presence of higher charged particles. Therefore, the generating factor
$(q)_{m_1}^{-1}$, that captured the weight increase of  the displacements of particles 1 in absence of higher-charge particles, is actually generic.

Once the charge-1 particles have been displaced in all possible ways, we consider  the similar displacements of the particles of charge $2,\,3,\cdots, $ up to $k-1$. In each case, enumerating the displacements of the $m_j$ charge-$j$ particles is the same as enumerating the partitions with at most $m_j$ parts and this is given by   
$(q)_{m_j}^{-1}$. 

The generating functions for all paths with fixed particle content  is thus $q^{w_{\text{mwc}}(a)}$ times all the generating factors that enumerate all possible displacements of the peaks while keeping track of the weight relative to the minimal-weight configuration. It is thus given by
 \begin{equation}\label{gfa}
 G_a(\{m_j\};q)= \frac{q^{w_{\text{mwc}}(a)} } {(q)_{m_1} \dots (q)_{m_{k-1}}},
  \end{equation}
  with $w_{\text {mwc}}(a)$ defined in (\ref{mwca}).

The full generating function is obtained by summing over all values of the $m_j$:
 \begin{equation}
 G_a(q)=\sum_{m_1,\cdots,m_{k-1}=0}^\y G_a(\{m_j\};q).
   \end{equation}
This can be writen in the following form \cite{BreL}:
 \begin{equation}\label{gfBka}
 G_a(q)=
\sum_{m_1,\cdots,m_{k-1}=0}^\y \frac{
q^{N_1^2+\cdots+ N_{k-1}^2+N_{k-a}+\cdots +N_{k-1}} } { (q)_{m_1}\cdots (q)_{m_{k-1}}
},Ê  \end{equation}
with $N_j$ defined as
 \begin{equation}\label{defN}
 N_j= m_j+\cdots
+m_{k-1}.  \end{equation}


\section{Restricted partitions}\label{SBvsP}

In this section, we introduce partitions subject to a restriction that is akin to a generalized exclusion principle. These restricted partitions are defined in 
Section \ref{Sdefrp}. The Burge correspondence, relating restricted partitions and $\B$ paths, is reviewed in Section \ref{Sbur}.
The restricted partitions are precisely those partitions that capture the combinatorial interpretation of the sum side of the famous Rogers-Ramanujan identities and their generalization, as pointed out in Section \ref{SRR}. This is followed by a brief discussion explaining the difficulty of a direct approach to the enumeration of restricted partitions. As an aside, the many faces of restricted partitions are both summarized and amplified by the mention of their noteworthy occurrence as a vanishing condition for the Jack polynomials at special values of their `coupling constant'.



\subsection{Defining the restricted partitions}\label{Sdefrp}

Although the meaning of a partition is  assumed to be be known (and it has actually be used in the previous sections), we provide here a definition in order to introduce related terminology. A {\bf partition} $\la= (\la_1,\cdots , \la_m)$ of $n$ is a weakly decreasing sequence of integers  $\la_i\geq \la_{i+1}$, with $\la_m\geq 1$, whose sum add up to $n$. $n$ is called the {\bf weight} of the partition and the $\la_i$ are its {\bf parts}. A number of 0 can be added at the end of  partition, but the `parts' refer to the non-zero entries of a partition.
It is useful to  introduce a notion of {\bf distance} between the parts of a partition: the parts $\la_i$ and $\la_{i+j}$ are said to be separated by the distance $j$.

{\bf Restricted partitions} refer to partitions satisfying the
{\bf `difference-2 condition at distance $k-1$' }: 
\begin{equation}\label{difk}
\la_l\geq Ê\la_{l+k-1}+2. 
\end{equation}
(By the `difference-2 condition' we always understand that the difference is at least 2.)
Restricted partitions are conveniently subdivided into classes specified by
a condition on the tail of the partition, namely, an upper bound, $i-1$, on the possible occurrences of the part 1. This is ensured by enforcing 
\begin{equation}\label{bd}
 \la_{m-i+1}\geq 2 .
\end{equation}
Take for instance $(8,8,7,7,5,3,3,2,2,1,1)$: the difference-2 condition is not satisfied at distance $1,\,2$ or 3 but it is at distance 4: with $\la_{l,j}=\la_l-\la_j$, one has
\begin{equation}
\la_{1,5}=3,\quad \la_{2,6}=5,\quad \la_{3,7}=4,\quad \la_{4,8}=5,\quad \la_{5,9}=3,\quad \la_{6,10}=2,\quad \la_{7,11}=2,
\end{equation}
and indeed $\text{min}_l (\la_{l,l+4})\geq 2$. Here $i\geq 3$.

Notice the following equivalent characterization of
the restricted partitions in terms of the {\bf frequency conditions}:
\begin{equation}\label{fre}
f_j+f_{j+1}\leq k-1\qquad {\rm and}\qquad  f_1\leq i-1\;, 
\end{equation}
 where $f_j$ is the frequency of the part $j$ in the partition. The equivalence between the two formulations of the restrictions is obvious:  if a partition satisfying (\ref{difk}) contains the sequence $(j+1,\ldots,j+1,j,\ldots j)$, the multiplicity of $j+1$ plus that of $j$ cannot be larger than $k-1$ to respect the difference-2 condition at distance $k-1$. Moreover, the boundary condition $\la_{m-i+1}\geq 2$ is precisely designed to set up an upper bound, $i-1$, on the number of parts equal to 1, i.e., the frequency $f_1$ should be $\leq i-1$.
 For instance, the frequencies for  the partition $(5,4,3,3,3,2,1,1)$ are $f_2=f_4=f_5=1, f_1=2, f_3=3.$ The maximal value of $f_j+f_{j+1}$ is  4 (obtained either with $j=2$ or $j=3$), meaning that  this partition satisfies the condition (\ref{fre}) when $k\geq 5$.

Below, in describing the Burge correspondence, it will be convenient to add the frequency $f_0$, which is immaterial in specifying a partition since a number of 0's can always be added at the end of a partition.

\subsection{Paths and restricted partitions: the Burge correspondence}\label{Sbur}

The bijection between the paths and the restricted partitions is described by the Burge correspondence \cite{Bu}.
 A $\B$ path, viewed as a sequence of SE, H  and NE edges, is  a binary word in $\a$ and $\b$, which are the edges linking the points:
\begin{equation}  \label{defab}
\a:(i,j)\rw(i+1,{\rm max}\, (0,j-1))
\qquad {\rm 
or }\qquad
\b:(i,j)\rw(i+1,j+1).
\end{equation} 
The Burge correspondence is a weight-preserving bijection relating  a binary word to a partition.

The correspondence  relies on the characterization of a partition in terms of non-overlapping pairs of adjacent frequencies $(f_j,f_{j+1})$ with $f_{j+1}>0$, starting the pairing from the largest part. Take for instance the partition $(5,4,3,3,3,2,1,1)$ with $k=5$ and $i=4$. 
 We have the following frequency pairing:
\begin{equation}
(5,4,3,3,3,2,1,1) : \quad \begin{matrix}
j: & 0&\phantom{}1&2\phantom{}&\phantom{}3
&4\phantom{}&5\\
f_j:& (0&2)&(1&3)&(1&1)\end{matrix}.
\end{equation}
Let us now define a sequence of two operations, $\ah$ and $\bh$, on the set of paired frequencies. If $(f_0,f_1)$ is not a pair, we act with $\ah$ defined as follows:
\begin{equation}\ah: (f_j,f_{j+1})\rw (f_j+1,f_{j+1}-1) \qquad \forall\, j\geq 1 .
\end{equation}
If $(f_0,f_1)=(0,f_1)$ is a pair, we act with $\bh$ defined as follows:
\begin{equation}\bh: \left\{ \begin{matrix}
&(0,f_1)\rw (0,f_1-1)\phantom{\qquad\qquad\;}& \\  & (f_j,f_{j+1})\rw (f_j+1,f_{j+1}-1) & \forall\, j>1 . \end{matrix} \right.
\end{equation}
After each operation, the pairing is modified according to the new values of the frequencies. We then act successively with $\ah$ or $\bh$ on the partition until all frequencies become zero. The ordered sequence of $\ah$'s and $\bh$'s so obtained is then reinterpreted as a path, with $\ah$ and $\bh$ replaced respectively by $\a$ and $\b$, by considering  $\a$ to be a SE or H edges and $\b$  a NE edge, as given in (\ref{defab}) \cite{BreL}.  The path starts at a prescribed initial vertical position $a$, whose relation to $i$ is worked out below. Once this positioning is fixed, we add to the end of the sequence the number of $\a$'s needed to reach the horizontal axis.

Returning to our example: since $f_0$ is  paired, the first acting operator is  $\bh$. Writing the result of the successive operations followed by the repairing, one has
\begin{equation}\label{exB}
 \begin{matrix}
j: &0&1&2&3&4&5 \\
f_j:& (0&2)&(1&3)&(1&1)\\
\bh:& 0&(1&2)&(2&2)&\\
\ah^2:& (0&3)&(0&4)&&\\
\bh^3:& 0&0&(3&1)&&\\
\ah^5:& (0&4)&&&&\\
\bh^4:& 0&0&&&&.\\
\end{matrix}
\end{equation}
To be clear: once the  action of an operator is completed, the parentheses are retired and a new pairing is done before a further action is considered. For instance, after the first action of $\bh$, the frequencies are: $(0,1),(2,2),(2,0)$; the parentheses are retired and the last 0 is omitted; this yields: $0,1,2,2,2$. Then these are paired anew as: $0,(1,2),(2,2)$. This is the new pairing that is given on the third line.
The corresponding path is thus associated to the word $\b\a^2\b^3\a^5\b^4$, up to the missing final power of $\a$ required to reach the $x$-axis.

It remains to settle the relation between the initial position $a$ and the maximal  bound on the frequency $f_1$, namely $ i-1$.
If $f_1>0$, $f_0$ is paired so that the above construction starts with $\bh^{f_1}$. The corresponding path starts with $f_1$ NE edges. When the initial point is $a$, the sum $a+f_1$ should not exceed the maximal allowed height $ k-1$, that is,
\begin{equation}
 a+f_1\leq a+i-1\leq k-1.
\end{equation}
This and the bound $0\leq a\leq k-1$,
 entail the precise relation:
\begin{equation}
a= k-i.
\end{equation}
As a verification, let us check the correspondences between the extremal values. When $i=1$,  there is no partition of weight 1, and similarly, there is no path of weight 1 when $a=k-1$. When $i=k$, there is no additional constraint on the number of 1 (the maximal number is fixed by the difference-2 condition to be $k-1$); correspondingly, with $a=0$, there is no special constraint on the height of the first peak apart form the generic one ($y \leq k-1$). Note finally, that the relation between $a$ and $i$ also follows directly from the comparison of the generating function for paths and that for restricted partitions (cf. Section \ref{SRR}).

The path considered in our example starts thus at $a=1$ (recall that $k=5$ and $i=4$) and it must finish on the $x$ axis. The word $\b\a^2\b^3\a^5\b^4$ must thus be completed by $\a^4$. The resulting path is displayed in Fig. \ref{fig6}.
 In this example, $m_1=m_3=m_4=1$ and $m_2=0$, so that $m=\sum jm_j=8$.

 \begin{figure}[ht]
\caption{{\footnotesize The B$_{4,1}$ path associated to the partition $(5,4,3,3,3,2,1,1)$ via the Burge correspondence worked out in (\ref{exB}). It weight is 1+6+15=22, which is precisely the sum of the parts in the partition. The total charge 8 equals to the number of parts of the partition.}}
\label{fig6}\begin{center}
\begin{pspicture}(4,0)(7.5,2.5)
{\psset{yunit=40pt,xunit=40pt,linewidth=.8pt}
\psline{-}(0.3,0.3)(0.3,1.5) \psline{->}(0.3,0.3)(7.0,0.3)
\psset{linestyle=solid}
\psline{-}(0.3,0.3)(0.3,0.35) \psline{-}(0.6,0.3)(0.6,0.35)
\psline{-}(0.9,0.3)(0.9,0.35) \psline{-}(1.2,0.3)(1.2,0.35)
\psline{-}(1.5,0.3)(1.5,0.35) \psline{-}(1.8,0.3)(1.8,0.35)
\psline{-}(2.1,0.3)(2.1,0.35) \psline{-}(2.4,0.3)(2.4,0.35)
\psline{-}(2.7,0.3)(2.7,0.35) \psline{-}(3.0,0.3)(3.0,0.35)
\psline{-}(3.3,0.3)(3.3,0.35) \psline{-}(3.6,0.3)(3.6,0.35)
\psline{-}(3.9,0.3)(3.9,0.35) \psline{-}(4.2,0.3)(4.2,0.35)
\psline{-}(4.5,0.3)(4.5,0.35) \psline{-}(4.8,0.3)(4.8,0.35)
\psline{-}(5.1,0.3)(5.1,0.35) \psline{-}(5.4,0.3)(5.4,0.35)
\psline{-}(5.7,0.3)(5.7,0.35) \psline{-}(6.0,0.3)(6.0,0.35)
\psline{-}(6.3,0.3)(6.3,0.35) \psline{-}(6.6,0.3)(6.6,0.35)
\rput(0.6,-0.05){{\scriptsize $1$}}
\rput(2.1,-0.05){{\scriptsize $6$}} 
\rput(4.8,-0.05){{\scriptsize $15$}}


 \psline{-}(0.3,0.6)(0.35,0.6)
\psline{-}(0.3,0.9)(0.35,0.9) \psline{-}(0.3,1.2)(0.35,1.2)
\psline{-}(0.3,1.5)(0.35,1.5) 

\rput(0.05,0.9){{\scriptsize $2$}} \rput(0.05,1.5){{\scriptsize $4$}}
\rput(0.05,0.6){{\scriptsize $1$}} \rput(0.05,1.2){{\scriptsize $3$}}
\psline{-}(0.3,0.6)(0.6,0.9) \psline{-}(0.6,0.9)(1.2,0.3)
\psline{-}(1.2,0.3)(2.1,1.2) \psline{-}(2.1,1.2)(3,0.3)
\psline{-}(3.6,0.3)(4.8,1.5) \psline{-}(4.8,1.5)(6,0.3)

}
\end{pspicture}
\end{center}
\end{figure}
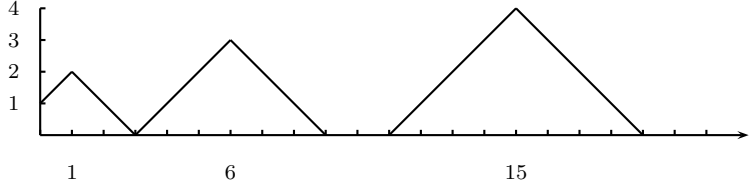


 This relation between a restricted partition and a $\B$ path is actually a bijection. To recover a partition from a path, one simply runs the above algorithm in the reverse direction \cite{Bu}.
 

As already mentioned, this correspondence is weight preserving: the weight of a path is equal to that of the corresponding partition. 
Moreover, the total charge of the path is equal to the number of parts $m$ in the restricted  partition. In this version of the correspondence, both features are not 
transparent, but this will be remedied later, when restricted partitions and paths will both be related bijectively to multiple partitions.

\subsection{Some remarks concerning restricted partitions}\label{Srem}

\subsubsection{Restricted partitions and generalized Rogers-Ramanujan identities}\label{SRR}

The partitions satisfying (\ref{difk}) for $k=2$ describe the combinatorics of the sum side of the Rogers-Ramanujan identities.
For $i=1,2$ these identities are respectively (see e.g., \cite{Andr} chap. 7):
\begin{align}\label{RR}
&\sum_{m=0}^\y \frac{{q^{m^2+m}}}{(q)_m} =\prod_{n=0}^\y \frac1{(1-q^{5n+2})(1-q^{5n+3})}\nonumber \\ 
&\sum_{m=0}^\y \frac{{q^{m^2}}}{(q)_m}= \prod_{n=0}^\y \frac1{(1-q^{5n+1})(1-q^{5n+4})}, 
\end{align}
using the notation introduced in (\ref{defqm}).
The coefficient of $q^n$ on the left-hand side enumerates the total number of partitions of $n$ satisfying $\la_j\geq \la_{j+1}+2$ with at most $i-1$ parts equal to 1. The coefficient of $q^n$ on the right-hand side similarly enumerates all the partitions of $n$ that have no part equal to 0, 1 or 4 mod 5 in the first case, e.g.,
\begin{equation}\prod_{n=0}^\y \frac{1}{(1-q^{5n+2})(1-q^{5n+3})}=  \prod_{\substack{n\not
\equiv 0,
\pm 1\\
\;{\rm mod}\, 5}}^\infty  \frac1{(1- q^n)}\; ,
\end{equation}
 and with no part equal to 0, 2 or 3 mod 5 in the second case. Take for instance $i=1$ and $n=8$: the partitions of 8 with parts $\geq 2 $ and differing at least by 2 are: $(8),\, (6,2),\,(5,3)$. Partitions of 8 with no part equal to either $1,\, 4,\,5,\, 6$ are: $(8),\,(3,3,2),\, (2,2,2,2)$. One gets an equal number of partitions of each type, which was to be illustrated.

The generalization of these identities to all $k$ is known as the  Andrews-Gordon identity \cite{An,Andr}:
\begin{equation}  \label{AG}
 \sum_{m_1,\cdots,m_{k-1}=0}^\y \frac{
q^{N_1^2+\cdots+ N_{k-1}^2+N_i+\cdots +N_{k-1} } } { (q)_{m_1}\cdots (q)_{m_{k-1}} }
=   \prod_{\substack{n\not
\equiv 0,
\pm i\\
\;{\rm mod}\, (2k +1)}}^\infty  \frac1{(1- q^n)}  ,
\end{equation}
with $
 N_j= m_j+\cdots
+m_{k-1}$.
 The right-hand side  enumerates partitions into parts that are not equal to 0 or $\pm i$ mod $2k+1$.
If $p_{k,i}(n)$ stands for the number of partitions of $n$ that satisfy (\ref{difk}) and (\ref{bd}), then
the left-hand side is the generating function $\sum p_{k,i}(n)q^n$.
It agrees with the generating function for $\Ba$ paths -- cf . eq. (\ref{gfBka}) --,
with $i=k-a$.

\subsubsection{Comments  on the direct enumeration of restricted partitions}\label{Senu}

 In view of constructing the generating function of restricted partitions, the bijection with $\B$ paths 
 is quite useful because a direct attack of the enumeration problem for restricted partitions is particularly hard. In fact the original proof of (\ref{AG}) is  basically a verification proof \cite{An} (see also \cite{Andr}, chapter 7).

 To explain the difficulty underlying a direct enumeration procedure, consider first the case $k=2$, for which such a direct derivation is possible, and identify the obstruction that prevents to argument to be generalized to larger values of $k$ \cite{JM.A}.
 
Let us thus take $k=i=2$. The problem is then to enumerate all partitions subject to the condition
$\la_j\geq \la_{j+1}+2 $. For this, it is convenient to first count such partitions with precisely  $m$  parts
 and then sum over $m$. For the first step, we use the {\bf staircase trick}: Every partition satisfying the  difference-2 condition at distance 1 and having exactly $m$ parts can be described as the sum of an unrestricted partition with at most $m$ parts plus the `staircase' $(2m-1,\cdots, 5,3,1)$. 
 We now use a basic result already quoted: the generating function for partitions with at most $m$ parts is $(q)_m^{-1}$.
 Since the weight of the staircase is $q^{m^2}$, we conclude that  $q^{m^2}/(q)_m$ is  the generating function of partitions with $m$ parts and satisfying $\la_j\geq \la_{j+1}+2$. Summing over $m$ leads to the left-hand side of the second equation in (\ref{RR}).
 For $i=1$, there can be no 1, so that the required staircase is rather $(2m,\cdots, 6,4,2)$. This  produces an extra term $q^m$ within the sum. We recover then the left-hand side of the first equation in (\ref{RR}). We have thus obtained, by a simple and direct method, the generating functions for restricted partitions when $k=2$.

Now this simple argument cannot be extended to  $k>2$. Let us consider $k=3$ to illustrate the breaking point and set $i=3$. The `ground state' that replaces the staircase of the previous example, is now $(\cdots , 7,5,5,3,3,1,1)$. Trying the same  strategy as for the $k=2$ case would amount to describe all partitions of length $m$ with
$\la_j\geq \la_{j+2}+2$
in terms of the  partitions with at most $m$ parts, to which we add the contribution the flatten staircase $(\cdots , 7,5,5,3,3,1,1)$. But such a description is simply not correct in the present context. This can be seen most directly from a counter-example.  There are three allowed partitions with 3 parts and weight 7  with difference 2 at distance 2: $(5,1,1)$, $(4,2,1)$ and $(3,3,1)$. Subtracting the ground-state $(3,1,1)$, we are left with $(2,0,0),\ (1,1,0)$ and $(0,2,0)$. But $(0,2,0)$ is not a genuine partition. 
What happens here is that by starting with the partition $(\la_1,\la_2,\la_3)$ satisfying $\la_1\geq Ê\la_2\geq \la_3\geq 1$ together with  $\la_1\geq \la_3+2$, the removal of the staircase $(3,1,1)$ leads to parts $\mu_i$ such that $\mu_1\geq \mu_3$ but where in addition $\mu_1\geq \mu_2-2$. The conditions at distance 1 become thus incorrect. 

For a general $k$, the subtraction of the staircase with steps of width $k-1$ leads to conditions at distance $j<k-1$ that are no longer those of a weakly decreasing sequence. For $k=2$, there is simply no lower-distance condition that could be spoiled.
This shows neatly that the argument used for $k=2$ cannot be extended to higher value of $k$. 

The bottom line of this discussion is that the staircase trick can be used only when the difference condition is enforced between adjacent parts, i.e., at distance 1. Actually, for the applicability of the staircase trick, the value of the difference (which is 2 here) is irrelevant. This explains the usefulness of the multiple partitions to be described in Section \ref{Smp} (see in particular Section \ref{Sgfmp}).

\subsubsection{Interlude: the many occurrences of restricted partitions}\label{Sint}

A certain number of combinatorial objects are related to restricted partitions and unravelling some of these relations constitute an essential part of this paper. In addition to the Bressoud paths introduced in the previous section, the Andrews-Baxter-Forrester paths to be defined in Section \ref{SABF},
and the multiple partitions to be introduced in the next section, 
we have also encountered
partitions with parts not equal to $0,\pm i $ mod $(2k+1)$ (enumerated by the product side of the Andrews-Gordon identities in eq.  (\ref{AG})), and partitions with prescribed ranks (cf. the introduction).\footnote{To make this list complete, let us add the physical links.  At first, the restricted partitions are related of the basis of states of the $\Z_k$ parafermionic models, as explained in Section \ref{SZk}. It turns out that they also characterize the basis of states for the $\M(2,2k+1)$ minimal models -- cf. Section \ref{Sm2p} --, which, at first sight,  is somewhat surprising given that  these are in no way related to the parafermionic models. }

We also want to point out here a completely different (at least superficially) and striking occurrence of restricted partitions, this time in the context of symmetric polynomials. A particularly interesting basis of symmetric polynomials is provided by the Jack polynomials $J_\la(z_1,\cdots,z_N;\beta)$ \cite{Stan,Mac}. Here $\la$ stands for a partition,  $N$ is the number of variables, and $\beta$ is a real parameter.
Their precise definition is not required here, but it can be noted that they arise in physics as the eigenfunctions of the trigonometric version \cite{CMS} of the Calogero-Moser-Sutherland models \cite{LV} - and $\beta$ is then the model's coupling constant.

Here we simply draw the attention to the unexpected discovery made in \cite{Fetal} which is that $J_\la(z_1,\cdots,z_N;\beta)=0$ if 
the following three conditions are satisifed:
\begin{equation}
  \; z_1=z_2=\cdots=z_k, \qquad  \;\beta=-\frac{(r-1)}{k}, \quad\text{and}\quad \; \la_j\geq \la_{j+k-1}+r,
 \end{equation} 
for integers $r,k \geq 2$. 
Moreover, a basis for symmetric polynomials vanishing at $k$ coincident points is furnished by such Jack polynomials with $r=2$.\footnote{These polynomials are interpreted as fractional quantum Hall  wave-functions in \cite{BH} (and the sequence of frequencies $[f_0,f_1,f_2,\cdots ]$ corresponds the occupation numbers there), which are themselves related to parafermionic correlation functions \cite{RR}. }

\section{Multiple partitions}\label{Smp}

In the previous section, we have reviewed the relation between $\B$ paths and restricted partitions. But $\B$ paths are most naturally related to a completely different type of partitions, the so-called multiple partitions, defined in Section \ref{Sdefmp}.
The relation between paths and multiple partitions is presented in Section \ref{SBmp}. It relies on an operation introduced in Section \ref{Sexr}, the so-called exchange relation, that will play a central role from now on.  
This correspondence and the one just established between $\B$ paths and restricted partitions entail a relation between restricted and multiple partitions. Section \ref{Srp+mp} presents a direct formulation of this bijection.
Quite interestingly, this direct connection shows how a restricted partition can be viewed as a collection of `charged particles'.
Finally, a very simple derivation of the generating function for multiple partitions is presented  in Section \ref{Sgfmp}; this simplicity  clearly demonstrates the immediate advantage of formulating a problem (e.g., a basis of states) in terms of multiple partitions.


\subsection{Defining the multiple partitions}\label{Sdefmp}

\n A {\bf multiple partition} $\N$ is an ordered sequence of at most $k-1$ partitions of the form \cite{JM.A,Mult}:
\begin{equation}\label{mul}
\N= (n^{(1)}, n^{(2)}, \cdots ,n^{(k-1)})\qquad {\rm with}\qquad 
n^{(j)}= (n^{(j)}_1, \cdots , n^{(j)}_{m_j})\;,
\end{equation} 
that is parametrised by an integer $a$ in the range $0\leq a\leq k-1$, and which satisfy the following conditions:

\begin{enumerate}

\item
The parts within a given partition $n^{(j)}$ are all distinct, subject to the condition
\begin{equation}\label{difone}n^{(j)}_l \geq n^{(j)}_{l+1} + 2j\;.  
\end{equation}

\item The last part of a partition satisfy the following boundary condition: 
\begin{equation}\label{bon}   n^{(j)}_{m_j} \geq j+ {\rm max}\, (j+a-k+1,0)+
2j (m_{j+1}+\cdots +  m_{k-1}) \; .
\end{equation}

\end{enumerate}

\n (It is said that $\N$ is composed of at most $k-1$ partitions since if $m_j=0$, $n^{(j)}$ is void.)

The condition (\ref{difone}) essentially means that the parts in the $j$-th partition satisfy a difference-$2j$ condition at distance 1.  That parts are subject to a condition at distance 1 is the key simplifying feature of multiple partitions, as it  will be discussed in Section \ref{Sgfmp}.

The boundary condition  (\ref{bon}) captures a sort of interaction between the parts of the different partitions. The condition on $n^{(j)}_{m_j}$ does not depend upon the particular values of the parts of the following partitions, that is, $n^{(j)}_{m_j}$ is not bounded by $n^{(j+1)}_{1}$.
However, the value of $n^{(j)}_{m_j}$ does depend
upon the number of parts in these subsequent partitions.
 This can thus be viewed as a charge-dependent repulsion condition.

\subsection{Paths and multiple partitions}\label{SBmp}

\subsubsection{A path as a sequence of clusters}\label{Sclus}

With its initial point fixed, a path is completely  determined by the specification of the peak coordinates. We take the convention that the sequence of peaks is to be  read from right to left.  
 The peak data are equivalently given in terms of $\{x_n^{(c_n)} \}$, where $x_n$ is the $x$-coordinate of the $n$-th peak from the right and $c_n$ is its charge. We will also refer to $x_n^{(c_n)}$ as a {\bf cluster} of charge $c_n$ and weight $x_n$.
  For example, the path of in Fig. \ref{fig1}
corresponds to the sequence: 
\begin{equation}  39^{(2)} \,32^{(4)}\,29^{(1)}\,24^{(2)}\,18^{(3)}\,14^{(1)} 10^{(2)}\,6^{(3)}\,2^{(1)}.
 \end{equation} 
 The cluster sequence for the simpler path of Fig. \ref{fig6} is: $15^{(4)}\, 6^{(3)}\, 1^{(1)}$.

\subsubsection{The exchange relation}\label{Sexr}

We now introduce a formal {\bf exchange relation} that describes the interchange of two adjacent clusters $ x^{(i)} {x'}^{(j)}$.  It is defined as follows:
\begin{equation} \label{com}
 x^{(i)} {x'}^{(j)} \quad \rw \quad ({x'}+r_{ij}) ^{(j)} (x-r_{ij})^{(i)} \;, 
 \end{equation}
 where $r_{ij}$ is defined by
 \begin{equation} \label{rij}
 r_{ij} = 2\, {\rm min} \; (i,j)\; .
 \end{equation}
This  operation preserves the individual values of the charge and also the sum of the weights. 

Take a path written in the form $\{ x_j^{(c_j)} \}$. 
After a sequence of interchanges, such that $\{ x_j^{(c_j)} \}\rw
 \{ {x'_j}^{(c_j)} \}$, the new values of $x_j'$ are no longer necessarily decreasing and they no longer  correspond to peak positions in a modified path.  For instance, consider the path $7^{(3)}\, 3^{(1)}$; applying the exchange relation yields $5^{(1)}\, 5^{(3)}$, and this sequence of clusters obviously does not correspond to a path (the two peaks would be at the same position).  This is the reason for introducing the more general cluster terminology.
 
  The exchange relation (\ref{com}) appears in a different form in \cite{BreL} (under the name of the shuffle operation and with special application conditions that we do no have here). In its present form, it has been introduced in \cite{JM.A} as an schematic version of the parafermonic commutation relations.

\subsubsection{From paths to multiple partitions}\label{Sptomp}

Using (\ref{com}), one can reorder the peaks of a path in a canonical way, in increasing value of the charge (from left to right) and, within each sequence of clusters with identical charge, with decreasing value of the weight. The weights of the clusters of  charge $j$ are collected together  and they form the parts of a partition $n^{(j)}$. The result is an ordered sequence of partitions $(n^{(1)}, n^{(2)}, \cdots ,n^{(k-1)})$, namely, a multiple partition \cite{Mult}.
 For instance, for the path of Fig. \ref{fig7}, we have
\begin{equation}\label{exPM}  19^{(3)}\, 15^{(1)}\, 12^{(1)} \, 10^{(2)} \, 6^{(1)} \, 3^{(2)} \rw 
 17^{(1)}\, 14^{(1)}\, 10^{(1)} \, 12^{(2)} \, 7^{(2)} \, 5^{(3)} \,.\end{equation} 
The multiple partition thus obtained is
\begin{equation}\label{exfig7}
      (17,14,10)^{(1)},\, (12,7)^{(2)},\, (5)^{(3)} .\end{equation} 

 \begin{figure}[ht]\caption{{\footnotesize The B$_{3,0}$ path associated to the sequence of clusters: $19^{(3)}\, 15^{(1)}\, 12^{(1)} \, 10^{(2)} \, 6^{(1)} \, 3^{(2)},$ of weight $w=65$ }}
\label{fig7}
\begin{center}
\begin{pspicture}(4,0)(8,2.)
{\psset{yunit=40pt,xunit=40pt,linewidth=.8pt}
\psline{-}(0.3,0.3)(0.3,1.2) \psline{->}(0.3,0.3)(7.3,0.3)
\psset{linestyle=solid}
\psline{-}(0.3,0.3)(0.3,0.35) \psline{-}(0.6,0.3)(0.6,0.35)
\psline{-}(0.9,0.3)(0.9,0.35) \psline{-}(1.2,0.3)(1.2,0.35)
\psline{-}(1.5,0.3)(1.5,0.35) \psline{-}(1.8,0.3)(1.8,0.35)
\psline{-}(2.1,0.3)(2.1,0.35) \psline{-}(2.4,0.3)(2.4,0.35)
\psline{-}(2.7,0.3)(2.7,0.35) \psline{-}(3.0,0.3)(3.0,0.35)
\psline{-}(3.3,0.3)(3.3,0.35) \psline{-}(3.6,0.3)(3.6,0.35)
\psline{-}(3.9,0.3)(3.9,0.35) \psline{-}(4.2,0.3)(4.2,0.35)
\psline{-}(4.5,0.3)(4.5,0.35) \psline{-}(4.8,0.3)(4.8,0.35)
\psline{-}(5.1,0.3)(5.1,0.35) \psline{-}(5.4,0.3)(5.4,0.35)
\psline{-}(5.7,0.3)(5.7,0.35) \psline{-}(6.0,0.3)(6.0,0.35)
\psline{-}(6.3,0.3)(6.3,0.35) \psline{-}(6.6,0.3)(6.6,0.35)
\rput(1.2,-0.05){{\scriptsize $3$}}
\rput(2.1,-0.05){{\scriptsize $6$}} 

\rput(3.3,-0.05){{\scriptsize $10$}}
\rput(3.9,-0.05){{\scriptsize $12$}}
\rput(4.8,-0.05){{\scriptsize $15$}}
\rput(6,-0.05){{\scriptsize $19$}}

 \psline{-}(0.3,0.6)(0.35,0.6)
\psline{-}(0.3,0.9)(0.35,0.9) \psline{-}(0.3,1.2)(0.35,1.2)

\rput(0.05,0.9){{\scriptsize $2$}} 
\rput(0.05,0.6){{\scriptsize $1$}} \rput(0.05,1.2){{\scriptsize $3$}}
\psline{-}(0.6,0.3)(1.2,0.9) \psline{-}(1.2,0.9)(1.8,0.3)
\psline{-}(1.8,0.3)(2.1,0.6) \psline{-}(2.1,0.6)(2.4,0.3)
\psline{-}(2.7,0.3)(3.3,0.9) \psline{-}(3.3,0.9)(3.6,0.6)
\psline{-}(3.6,0.6)(3.9,0.9) \psline{-}(3.9,0.9)(4.5,0.3)
\psline{-}(4.5,0.3)(4.8,0.6) \psline{-}(4.8,0.6)(5.1,0.3)
\psline{-}(5.1,0.3)(6.,1.2) \psline{-}(6.,1.2)(6.9,0.3)

}
\end{pspicture}
\end{center}
\end{figure}
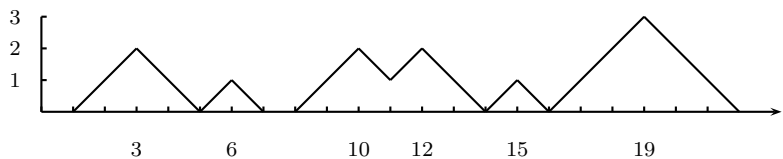

The construction of the multiple partition from a $\Ba$ path leads to conditions on their parts 
which are precisely (\ref{difone})  and (\ref{bon}).
Condition (\ref{difone}) reflects the minimal distance between two peaks of identical charge -- which is twice their individual charge (see for instance the peaks of Fig. \ref{fig2}) --, and the fact that clusters with the same charge are not reordered among themselves. The equation (\ref{bon}) captures the following boundary condition: if the path starts at its left with a peak of charge $j$, this peak position must be  $\geq j+ {\rm max}\, (j+a-k+1,0)$ -- cf. eq. (\ref{mxpos}). Consider thus its associated  multiple partition and use the exchange relation to move the cluster $n^{(j)}_{m_j}$ to the far right in order to reach its lowest allowed weight. It is thus displaced through all clusters of larger charge by means of the exchange relation (\ref{com}). This reduces its weight to $n^{(j)}_{m_j}-2j (m_{j+1}+\cdots +  m_{k-1})$. This number must satisfy the bound just indicated, which gives (\ref{bon}).


\subsubsection{From multiple partitions to paths}\label{mptop}

We have thus described a procedure for relating a path to an ordered sequence of $k-1$ partitions of respective lengths $m_1,\cdots , m_{k-1}$, where $m_j$ is the number of peaks of charge $j$.
We now show that this is a one-to-one correspondence  \cite{Mult}.

The inverse operation amounts to rewrite $(n^{(1)},\cdots,n^{(k-1)})$ as a sequence of clusters and then reorder them using (\ref{com}) until the sequence of clusters does form a path. But what conditions should a string of clusters satisfies in order to correspond to the peaks of a path? At first the clusters must be arranged in decreasing values of their weight in order to represent the sequence of peaks read from right to left:
\begin{equation}
x_1^{(c_1)}\cdots x_n^{(c_n)}\qquad \text{with}\qquad x_i> x_{i+1}.
\end{equation}
 But two adjacent  peaks in a path have a minimal distance in-between them. As already indicated, two peaks with the same charge $j$ must be separated by at least $2j$, while  two peaks of different charges, say $ x^{(i)}$ and $ {x'}^{(j)}$, must be separated by at least $r_{ij}$ -- defined in eq. (\ref{rij}) --
 if the highest charge peak is at the left and $r_{ij}+1$ otherwise. (The asymmetry is rooted in our rule of ascribing to the leftmost peak the highest charge whenever there are two peaks of the same height in a charge complex.)
 Stated more concisely, one requires the following condition on adjacent peaks $ x^{(i)}\, {x'}^{(j)}$:
 \begin{equation}\label{dist}
x- x' \geq r_{ij}+\begin{cases}
  1 \quad \text{if } \quad i>j \\ 0 \quad \text{otherwise.}\\
 \end{cases}
  \end{equation} 
For instance, in Fig. \ref{fig1}, those peaks that have the minimal separation are the pairs at positions 6 and 10, for which the distance is $2 \,\text{min}(2,3)=4$ since the largest peak is at the left,  and at positions 29 and 32, where now the distance is $2 \,\text{min}(1,4)+1=3$, the largest peak being at the  right.







 We thus start with a multiple partition and order their parts (which, we stress, have definite charge, meaning that these are actually clusters) in decreasing values of the weight (irrespectively of their charge). Clusters of the same charge do not need to be reordered among themselves. We then test the distance condition  (\ref{dist})  for adjacent pairs.
 We thus perform all the required  interchanges of the problematic adjacent pairs. Once all adjacent pairs are transformed in order to satisfy (\ref{dist}), due to the built-in constraints on the starting multiple partition, the resulting configuration is the corresponding path. 
 (This procedure is a somewhat simplified version of that in \cite{Mult}.)
 

Let us work out the example (\ref{exPM}) backward
and indicate in bold the pair of clusters that has been interchanged at each step:
\begin{equation}
\begin{matrix}
 17^{(1)} & 14^{(1)}& 10^{(1)} & 12^{(2)} & 7^{(2)} & 5^{(3)}\\
 17^{(1)} & 14^{(1)}&{\bf 14^{(2)} }&{\bf  8^{(1)}} & 7^{(2)} & 5^{(3)}\\
 17^{(1)} & {\bf 16^{(2)}}&{\bf  12^{(1)} }& 8^{(1)} & 7^{(2)} & 5^{(3)}\\
 {\bf 18^{(2)}} &{\bf  15^{(1)} }& 12^{(1)} & 8^{(1)} & 7^{(2)} & 5^{(3)}\\
18^{(2)} &  15^{(1)}& 12^{(1)} & 8^{(1)} & {\bf  9^{(3)} }&{\bf  3^{(2)}}\\
18^{(2)} &  15^{(1)}& 12^{(1)} &  {\bf 11^{(3)}} &  {\bf 6^{(1)} }& 3^{(2)}\\
18^{(2)} &  15^{(1)}& {\bf  13^{(3)}} &  {\bf 10^{(1)}} &  6^{(1)} & 3^{(2)}\\
18^{(2)} &  {\bf 15^{(3)}}& {\bf  13^{(1)}} &   10^{(1)} &  6^{(1)} & 3^{(2)}\\
{\bf 19^{(3)} }&  {\bf 14^{(2)}}& 13^{(1)} &  10^{(1)} &  6^{(1)} & 3^{(2)}\\
19^{(3)} &  {\bf 15^{(1)}}&{\bf 12^{(1)}} &  10^{(1)} &  6^{(1)} & 3^{(2)}\\
\end{matrix}
\end{equation}
The first two interchanges as well as the fifth one are done to  reorder the clusters in strictly decreasing values of the weight. The other ones are forced to  ensure the proper minimal weight difference ($r_{ij}$ or $r_{ij}+1$) between adjacent clusters in order for them to describe  genuine adjacent peaks.


Let us finally stress that  the bijection between $\B$ paths and multiple partitions is manifestly weight and charge preserving.

\subsection{Restricted and multiple partitions}\label{Srp+mp}

\subsubsection{From restricted partitions to multiple partitions}\label{Srptomp}

Let us first show how to associate a multiple partition to a partition $(\la_1,\cdots , \la_m)$ satisfying (\ref{difk}) \cite{Mult}.
This condition forbids subsequences of $k$ successive parts differing at most by 1, namely sequences  of the form 
\begin{equation}
\underbrace{(s,\cdots, s)}_k \qquad \text{or}\qquad \underbrace{(s+1,\cdots s)}_k.
\end{equation}
 However, shorter subsequences of parts differing at most by 1 are allowed. 
The first step amounts to identify (by scanning the partition from left to right) the sequences of $k-1$ adjacent parts that differ by at most 1.  
Each such sequence is then replaced by a cluster of charge $k-1$ whose weight is given by the sum of  its  parts, i.e.,
\begin{equation}( \la_j,\cdots, \la_{j+k-2} ) \rw 
\La^{(k-1)}, \qquad \text{with} \quad \La=\sum_{r=j}^{j+k-2} \la_r.
\end{equation}
(This operation is the rationale for the `cluster' terminology.)

Consider $(8,8,7,7,5,3,3,2,2,1,1)$ which satisfies (\ref{difk}) for  $k=5$;  there are two sequences of 4 parts that differ at most by 1, whose clustering yields:
\begin{align}
  (  8,8,7,7,5,3,3,2,2,1,1)
 \rw  ( {30^{(4)}}, 5,  {10^{(4)}}, 1,1) .
\end{align}
 
Once all clusters of charge $k-1$ are constructed, we move them (preserving their ordering) to the right of the sequence formed by the remaining parts. Each displacement is done using the exchange operation (\ref{com}),  by treating all parts which are crossed as clusters of charge 1. 
For our example, this reads
\begin{align}
  ( {30^{(4)}}, 5,  {10^{(4)}}, 1,1)  
  \rw  ( {30^{(4)}}, 5, 3,3) \,   {6^{(4)}}
 \rw ( 7, 5,5) \, {24^{(4)}\,6^{(4)}}.
\end{align}

Once this is completed, one is  left with  a smaller partition and a sequence of ordered clusters of charge $k-1$ at its right. For the resulting partition, one repeats the previous analysis but with $k-1$ replaced by $k-2$. Once all clusters of charge $k-2$ are identified, they are moved to the right of the partition.
This procedure is repeated for lower-charge clusters until all clusters of charge 2 are formed and moved outside of the decimated partition, at the left extremity of the sequence of ordered clusters of charge $3,\cdots, k-1$. The remaining parts of the partition are the clusters of charge 1. 

In our example, there is no cluster of charge 3 but one of  charge  2 and one of charge 1:
\begin{align}
 (7, { 5,5}) \,24^{(4)}\,6^{(4)} 
\rw (7)\,   {10^{(2)}} \,24^{(4)}\,6^{(4)}=  {(7)^{(1)}}\, (10)^{(2)}\, (24,6)^{(4)}
\end{align}

Notice that in the above example, the simper procedure which consists in clustering parts directly (starting from larger to smaller clusters) but without removing the clusters one by one,  yields the correct sequence of clusters once reordered:
\begin{align} (8,8,7,7,5,3,3,2,2,1,1) \rw   {30^{(4)}} 5^{(1)}  {10^{(4)}} 2^{(2)} \rw  7^{(1)}\, 10^{(2)} \,24^{(4)}\,6^{(4)}.\end{align}
This is most often true. However, the example $(4,3,3,2,2,1)$  shows the necessity of the above more systematic set of operations, which yields:
\begin{equation}(4,3,3,2,2,1)\rw (4,10^{(4)}, 1)\rw (4,3)\, 8^{(4)} \rw 7^{(2)}\, 8^{(4)}, 
\end{equation}
while a direct clustering would give the incorrect $4^{(1)}\, 10^{(4)}\, 1^{(1)}$.

\subsubsection{From  multiple partitions to  restricted partitions}\label{Smptorp}

We now turn to the formulation of the
inverse operation \cite{Mult}  (which is partly related to a construction in \cite{BreL}).
The starting point is the re-writing of the multiple partition  as the partition $(n_1^{(1)}\cdots  n_{m_1}^{(1)})$ followed by the sequence of clusters 
 $n_1^{(2)} \cdots n_\ell^{(j)} \cdots n_{m_{k-1}}^{(k-1)}$. Each cluster is then inserted successively  (starting with $n_1^{(2)}$ up to $n_{m_{k-1}}^{(k-1)}$) within the partition, using the interchange rule  (\ref{com}), treating again each part as a cluster of charge 1. Once inserted within the partition (at a position subject to criteria to be specified), a cluster is unfolded into  the number of parts given by its charge, with parts as equal as possible. This decomposition is unique.
 
 Consider first the issue of the uniqueness of breaking the cluster $\La^{(j)}$ into $j$ parts differing at most by 1 and whose sum is $\La$. Given $\La$ and $j$, there are unique non-negative integers $s$ and $r$ such that $\La= sj+r$ with $r<j$. In the decomposition of $\La$, there are then $r$ parts equal to $s+1$ and $j-r$ parts equal to $s$:
 \begin{equation}
\La^{(j)} \rw( \underbrace{s+1,\cdots , s+1}_{r}, \underbrace{s, \cdots, s}_{j-r})\;.
 \end{equation}
 Equivalently, we have
  \begin{equation}
\La^{(j)} \rw \big(\underbrace{\l\lc \frac{\La}{j} \r\rc,\cdots , \l\lc \frac{\La}{j} \r\rc, \l \lf \frac{\La}{j} \r\rf, \cdots, \l \lf \frac{\La}{j} \r\rf }_{j}\big) 
\end{equation}
and $r=0$ if $\lc {\La}/{j} \rc= \lf \La/j \rf$.\footnote{$\lfloor z \rfloor$ denotes the largest integer smaller than $z$ and $\lc z\rc $ stands for the smallest integer larger than $z$.}

 The position at which the cluster  of charge $j$ is moved within the partition is determined by requiring
the new sequence of numbers  that results after the unfolding  to be  a weakly decreasing sequence, which, in addition, should satisfy the difference-2 condition at distance $j$:
 \begin{equation}\label{disj}
 \la_l-\la_{l+j}\geq 2\;.
 \end{equation} 
Since $j\leq k-1$, (\ref{disj})  ensures the validity of the condition (\ref{difk}) at every intermediate stage of the  construction.

To illustrate this criterion,  let us rework backward the above example $7^{(1)}\, 10^{(2)} \,24^{(4)}\,6^{(4)}$. The first step amounts to insert $10^{(2)}$ into the partition (7) (from the left) and unfold it into two parts; there are two insertion points: either $(7,10^{(2)})\rw (7,5,5)$ or $(12^{(2)},5)\rw (6,6,5)$. The second choice contradicts the condition (\ref{disj}) for $j=2$: 
it generates a sequence of three parts that differs by 1. Taking thus the first choice (which does satisfy (\ref{disj})), we now have to insert, into the partition $(7,5,5)$, the charge-4 cluster $24^{(4)}$. This cannot be placed after the second 5 or in-between the two 5 and then unfolded into four parts since, in each case, this would generate a  non-weakly-decreasing sequence of integers. The insertion after the 7 leads to $(7,28^{(4)},3,3)\rw (7,7,7,7,7,3,3)$, which contradicts (\ref{disj}) for $j=4$. The last choice is $(30^{(4)}, 5,3,3)\rw (8,8,7,7,5,3,3)$, which is allowed. It only remains to insert $6^{(4)}$. This can be done after the second 3:  
\begin{equation}
(8,8,7,7,5,3,3, 6^{(4)})\rw(8,8,7,7,5,3,3,2,2,1,1),
\end{equation} 
 or in-between the 5 and 3: 
 \begin{equation}(8,8,7,7,5, 10^{(4)},1,1)\rw(8,8,7,7,5,3,3,2,2,1,1),
 \end{equation}
 both choices leading to the same result.
 
The insertion  position is not necessarily unique but when there are two allowed choices, the resulting restricted partitions are equal \cite{Mult}, a point illustrated by the last step in the above example.



The bijection between restricted and multiple partitions is thus simple and systematic. Moreover it manifestly preserves the weight, as well as the charge, if the charge of a restricted partition is  identified with the number of its parts.

\subsection{The generating function for multiple partitions}\label{Sgfmp}

As  explained in Section \ref{Senu}, the enumeration of partitions with difference ($=2$ there) condition at distance 1 can be handled very easily with the staircase trick. Quite fortunately, the parts of a multiple partition are precisely subject to a difference condition at distance 1: that this difference is generally larger than 2 does not prevent the applicability of the staircase trick. To demonstrate this, consider the partition $(\la_1,\cdots , \la_m)$ such that $\la_i\geq \la_{i+1}+p$. Subtracting from this the staircase partition $((m-1)p,\cdots, p,0)$ and defining $\mu_i=\la_i-(m-i)p$, ones gets $\mu_i\geq \mu_{i+1}$, the defining condition for an ordinary partition.

 Consider first the number of partitions $n^{(j)}$ with  $m_j$ parts satisfying (\ref{difone}). Allowing for the moment the value 0 for the last part, then each such partition is the sum of a partition with at most $m_j$ parts to which we add the staircase with steps of height  $2j$: $(2(m_j-1)j,\cdots,2j,0)$. Since the staircase has weight ${jm_j(m_j-1)}$, the sought-for generating function is
 \begin{equation}
\frac {q^{jm_j(m_j-1)}}{ (q)_{m_j}}.
\end{equation}
As already indicated, at this point we have not yet enforced the number of (non-zero) parts to be $m_j$ since the last part of the staircase is 0. This is cured by the next step where the boundary condition  $ n^{(j)}_{m_j} \geq\Delta_{j}(a)$ is taken into account, with $\Delta_{j}(a)$ defined as
\begin{equation}\label{bond}    \Delta_{j}(a)= j+ {\rm max}\, (j+a-k+1,0)+
2j (m_{j+1}+\cdots +  m_{k-1}).
\end{equation}
The condition $ n^{(j)}_{m_j} \geq\Delta_{j}(a)$ is implemented by  a shift of  each  part by $ \Delta_{j}(a)$ (ensuring thereby the strict positivity of the last part since $\Delta_{j}(a)>0$). This modifies the total weight 
 by $m_j\Delta_{j}(a) $. The generating function becomes thus 
\begin{equation}
\frac{q^{jm_j(m_j-1)+ m_j  \Delta_{j}(a) } }{ (q)_{m_j}} .
\end{equation}
By summing over all values of $m_j$ for $j=1,\ldots , k-1$, one obtains the full generating function in the form precisely given by  the multiple-sum expression in (\ref{gfBka}) -- or in (\ref{AG}), with $i=k-a$. This is, by any standard, a very simple derivation. 

\section{ABF paths}\label{SABF}


In this section we introduce another type of paths, which we call the $\F$ paths. These are defined as the countour of the configurations appearing in the corner-transfer-matrix solution of the restricted-solid-on-solid (RSOS) model of Andrews-Baxter-Forrester  in the regime II.
These new paths are defined in Section \ref{defABF}. They are subdivided into classes specified by their initial vertical point as well as their relative charge sector,
and each class is characterized by a ground state as shown in Section \ref{SgsABF}. This makes precise the notion of relative weight in definite sectors. The original weighting of the $\F$ paths is then shown to be expressible in terms  of the $\B$-type weight function in Section \ref{Snew}. This simple but key step is subsequently used to related $\F$ paths to multiple partitions in Section \ref{SABFmp}. Our  central result,  the bijective relation between the $\F$ and $\B$ paths, follows then rather directly, as detailed in Section \ref{SBvsAFB}.


\subsection{Defining the ABF paths}\label{defABF}

 {\bf ABF paths} in regime II of the RSOS model of \cite{ABF}, referred to as the ABF$_k$ paths for short, are defined as follows:\footnote{The parameter  $k$ refers to the model's basic parameter $r$ of \cite{ABF}, taken to be $k+2$ here.}

\n {\bf Shape}: A  ABF$_k$ path 
is a sequence of integral points $(x,y)$ within the strip $0\leq x \leq L$ and $0\leq y\leq k$, with adjacent points $(x,y)$ and $(x+1,y\pm1)$  linked by NE or SE edges.

\n {\bf Initial and end points:} For a fixed value of $k$, the class of paths is specified by the  initial point $y_0=\ell$, with $0\leq \ell\leq k$.  $\F$ paths are required to terminate on the axis:  the final point is thus $y_L=0$.

 \n
 A $\F$ path with initial point $\ell$ will be denoted $\Fl$. An example of a ABF$_{4,2}$ path is illustrated in Fig. \ref{fig8}.

\n Contrary to $\B$ paths, the $\F$ ones have height up to $k$ (and not $k-1$) and they have no H edges. Moreover, they have finite length.

\n {\bf Weight}: The {\it weight} $\w $  of a $\F$ path is 
\begin{equation}\label{weig}
\w = \sum_{x=1}^{L-1} \w (x)\qquad \text{where} \qquad \w (x)=\begin{cases}
\frac{x}2 &\text{if $x$ is an extremum of the path} \\
0& \text{otherwise}\;. 
\end{cases} \end{equation}
We  stress that not only the peaks but also the valley positions do contribute to the weight.
Also, as their explicit exclusion from the sum indicates, the initial and final points are not considered as local extrema. 
 In Fig. \ref{fig8}, the vertices contributing to the weight are indicated by dots.

\n {\bf Charge}: The charge of a peak is defined exactly as for $\B$ paths. Again, we will denote by $m_j$ the number of peaks of charge $j$. The total charge of a path is: 
\begin{equation}\label{cha}
\m= \sum_{j=1}^k j m_j .
\end{equation} 

\n The length of a path is twice its total charge plus the initial point:
\begin{equation} 
L=\ell+\sum_{j=1}^k 2j m_j= \ell+2\m .\end{equation} 


 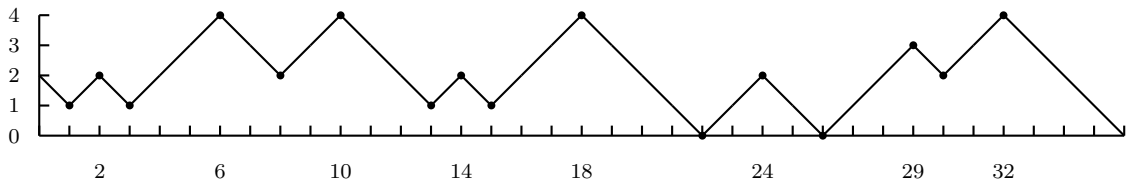
\begin{figure}[ht]\
\caption{{\footnotesize The pictured  ABF$_{4,2}$ path is the path of Fig. \ref{fig1} truncated at $L=36$, before the occurrence of the first H edge. The charge assignment of the peaks is given in Fig. \ref{fig1}. The weight is the half-sum of the horizontal positions of the dotted vertices: $\w=(1+2+\cdots +30+32)/2$. The total charge $\m$ is 17 and the length is $L=2\m+2$.}} 
\label{fig8}
\begin{center}
\begin{pspicture}(2.5,0)(10.5,2.1)
{\psset{yunit=38pt,xunit=38pt,linewidth=.8pt}
\psline{-}(0.3,0.3)(0.3,1.5) \psline{-}(0.3,0.3)(11.1,0.3)
\psset{linestyle=solid}
\psline{-}(0.3,0.3)(0.3,0.4) \psline{-}(0.6,0.3)(0.6,0.4)
\psline{-}(0.9,0.3)(0.9,0.4) \psline{-}(1.2,0.3)(1.2,0.4)
\psline{-}(1.5,0.3)(1.5,0.4) \psline{-}(1.8,0.3)(1.8,0.4)
\psline{-}(2.1,0.3)(2.1,0.4) \psline{-}(2.4,0.3)(2.4,0.4)
\psline{-}(2.7,0.3)(2.7,0.4) \psline{-}(3.0,0.3)(3.0,0.4)
\psline{-}(3.3,0.3)(3.3,0.4) \psline{-}(3.6,0.3)(3.6,0.4)
\psline{-}(3.9,0.3)(3.9,0.4) \psline{-}(4.2,0.3)(4.2,0.4)
\psline{-}(4.5,0.3)(4.5,0.4) \psline{-}(4.8,0.3)(4.8,0.4)
\psline{-}(5.1,0.3)(5.1,0.4) \psline{-}(5.4,0.3)(5.4,0.4)
\psline{-}(5.7,0.3)(5.7,0.4) \psline{-}(6.0,0.3)(6.0,0.4)
\psline{-}(6.3,0.3)(6.3,0.4) \psline{-}(6.6,0.3)(6.6,0.4)
\psline{-}(6.9,0.3)(6.9,0.4) \psline{-}(7.2,0.3)(7.2,0.4)
\psline{-}(7.5,0.3)(7.5,0.4) \psline{-}(7.8,0.3)(7.8,0.4)
\psline{-}(8.1,0.3)(8.1,0.4) \psline{-}(8.4,0.3)(8.4,0.4)
\psline{-}(8.7,0.3)(8.7,0.4)
\psline{-}(9.0,0.3)(9.0,0.4)
\psline{-}(9.3,0.3)(9.3,0.4) \psline{-}(9.6,0.3)(9.6,0.4)
\psline{-}(9.9,0.3)(9.9,0.4) \psline{-}(10.2,0.3)(10.2,0.4)
\psline{-}(10.5,0.3)(10.5,0.4) \psline{-}(10.8,0.3)(10.8,0.4)
\psline{-}(11.1,0.3)(11.1,0.4)

\psset{dotsize=3pt}
\psdots(0.6,0.6)(0.9,0.9)(1.2,0.6)(2.1,1.5)
(2.7,0.9)(3.3,1.5)(4.2,0.6)(4.5,0.9)
(4.8,0.6)(5.7,1.5) (6.9,0.3)(7.5,0.9)(8.1,0.3)
(9,1.2)(9.3,0.9)(9.9,1.5)

\rput(0.9,-0.05){{\scriptsize $2$}}
\rput(2.1,-0.05){{\scriptsize $6$}} \rput(3.3,-0.05){{\scriptsize $10$}}
\rput(4.5,-0.05){{\scriptsize $14$}}\rput(5.7,-0.05){{\scriptsize $18$}}
\rput(7.5,-0.05){{\scriptsize $24$}}
\rput(9,-0.05){{\scriptsize $29$}}\rput(9.9,-0.05){{\scriptsize $32$}}
 \psline{-}(0.3,0.6)(0.4,0.6)
\psline{-}(0.3,0.9)(0.4,0.9) \psline{-}(0.3,1.2)(0.4,1.2)
\psline{-}(0.3,1.5)(0.4,1.5) 

\rput(0.05,0.9){{\scriptsize $2$}} \rput(0.05,1.5){{\scriptsize $4$}}
\rput(0.05,0.3){{\scriptsize $0$}} 
\rput(0.05,0.6){{\scriptsize $1$}} \rput(0.05,1.2){{\scriptsize $3$}}
\psline{-}(0.3,0.9)(0.6,0.6) \psline{-}(0.6,0.6)(0.9,0.9)
\psline{-}(0.9,0.9)(1.2,0.6) \psline{-}(1.2,0.6)(1.5,0.9)
\psline{-}(1.5,0.9)(1.8,1.2) \psline{-}(1.8,1.2)(2.1,1.5)
\psline{-}(2.1,1.5)(2.4,1.2) \psline{-}(2.4,1.2)(2.7,0.9)
\psline{-}(2.7,0.9)(3.0,1.2) \psline{-}(3.0,1.2)(3.3,1.5)
\psline{-}(3.3,1.5)(3.6,1.2) \psline{-}(3.6,1.2)(3.9,0.9)
\psline{-}(3.9,0.9)(4.2,0.6) \psline{-}(4.2,0.6)(4.5,0.9)
\psline{-}(4.5,0.9)(4.8,0.6) \psline{-}(4.8,0.6)(5.1,0.9)
\psline{-}(5.1,0.9)(5.4,1.2) \psline{-}(5.4,1.2)(5.7,1.5)
\psline{-}(5.7,1.5)(6.0,1.2)
 \psline{-}(6.0,1.2)(6.3,0.9)\psline{-}(6.3,0.9)(6.6,0.6)
\psline{-}(6.6,0.6)(6.9,0.3) \psline{-}(6.9,0.3)(7.2,0.6)
\psline{-}(7.2,0.6)(7.5,0.9) \psline{-}(7.5,0.9)(7.8,0.6)
\psline{-}(7.8,0.6)(8.1,0.3) \psline{-}(8.1,0.3)(9,1.2)
\psline{-}(9.0,1.2)(9.3,0.9) \psline{-}(9.3,0.9)(9.6,1.2)
\psline{-}(9.6,1.2)(9.9,1.5) \psline{-}(9.9,1.5)(11.1,0.3)

}
\end{pspicture}
\end{center}
\end{figure}

\subsection{The ground-state configuration}\label{SgsABF}

The absence of H edges in $\F$ paths brings a radical difference with $\B$ paths: there is no configuration with zero weight $\w$. The ground-state for $\Ba$ paths is described by the path composed of $a$ SE edges followed by H edges and it has $w=0$. In contrast, 
 a $\F$ path of non-zero length necessarily  has a number of peaks and valleys and these all contribute to the weight $\w$. 
  There is actually a unique path, called the {\bf ground-state configuration}, which, for fixed values of $\ell$ and $L$, minimizes the weight.
  
Let us determine the shape of this ground-state path and thereby demonstrating its uniqueness. Given that all peak and valley positions do contribute to the weight, in order to minimize $\w$, one should minimize the number of contributing vertices. This means that for a fixed length $L$, one should  introduce as many peaks as possible having the maximal allowed charge, which is $k$. Recalling that $L=2\m+\ell$, we thus look for a path of total charge $\m$ with largest number of $k$-th components.
Setting
 \begin{equation}\label{sector}
\m= pk+r, \qquad 0\leq r\leq k-1,
\end{equation} 
this corresponds to the path with $p$ peaks of charge $k$, plus a remaining portion of charge $r$. It is easily checked that this portion should lie at the beginning of the path. Indeed,  given a fixed charge content, the weight is minimized by having the peaks ordered in increasing value of the charge from left to right. Compare for instance the weight of the configurations  of Fig. \ref{fig5} (a) and (e) viewed as $\F$ paths: they have respectively weight 4 and 8. (The situation in that regard is similar to that for $\B$ paths, despite the fact that the weight function is different here.)
 It only remains to describe the initial portion of charge $r$. Trying again to  minimizing the number of contributing vertices, it is clear that this must corresponds to a single peak of charge $r$. 
 The ground-state path is now fully characterized;  it is specified by $\ell$ and $r$ and it will be denoted as gs$(\ell;r)$ and its weight as $\w_{{\rm gs}(\ell;r)}$.

All the weights in a class of paths  must be defined relative to the ground-state configuration of that class.
The paths in the class ($\ell;r)$ 
are thus naturally weighted by 
\begin{equation}\Delta \w= \w- \w_{{\rm gs}(\ell;r)}.
\end{equation}


\subsection{A new weight function}\label{Snew}

As a key step toward establishing the relation between  $\F$ and $\B$ paths, 
we will now show that the relative weight $\Delta \w$ can be computed in an alternative way, using  the finitized version of the usual B-type weight function:
\begin{equation} \label{defwpatha} w= \sum_{x=1}^{L-1} w(x)\qquad {\rm where} \qquad w(x)=
\begin{cases}
x & \text{if $x$ is the position of a peak} ,\\
 0 &\text{otherwise}\;, \end{cases}
\end{equation}
to read
\begin{equation} \label{wdif} 
  \w -\w_{{\rm gs}(\ell;r)}= w-w_{{\rm gs}(\ell;r)}\;.
\end{equation} 

The proof of  (\ref{wdif}) proceeds as follows.
This equality is obviously true for the special path gs$(\ell;r)$ (but notice that $\w_{{\rm gs}(\ell;r)}\ne w_{{\rm gs}(\ell;r)}$).
Now, any configuration can be obtained from gs($\ell;r$) by combining the two length preserving elementary operations:

\begin{enumerate}

\item Change the charge content by breaking a peak of charge $s$ into two peaks of charge $j$ and $s-j$.

\item Keep the charge content fixed and displace  a peak of charge $j\leq k-1$ by one unit.

\end{enumerate}

\n We will verify that the modification in the weight, resulting from these two operations, is the same whether it is computed from  either side of (\ref{wdif}).

Let us first compare the weight of a peak with respect to that of two peaks with same total charge $s$ and located in-between the same initial and final positions, respectively $x$ and $x+2s$. Consider first a single peak of height $s$ centered at $x+s$ and ignore the contribution of the two minima at $x$ and $x+2s$ in $\w$ since they cancel in this computation. We have, using the cluster notation:
\begin{equation} 
(s+x)^{(s)} : \qquad  \w_1= \frac12(s+x) \,,\quad  w_1= s+x\;.
\end{equation} 
 Compare this with the weight  of the path described by a peak of height $j$ at $j+x$ followed by a peak of weight $s-j$  at $s+j+x$: 
\begin{equation} (s+j+x)^{(s-j)}(j+x)^{(j)}  :\qquad   \w_2= \frac12(s+4j+3x) \,,\quad  w_2=  s+2j+2x    \;.
\end{equation} 
The two differences are thus
\begin{equation}
\w_2-\w_1=  2j+x \quad  \text{and}\quad w_2-w_1=2j+x\qquad \Rw\qquad  \Delta \w=  \Delta w\;.
\end{equation} 
 Iterating this procedure, one can break a peak into more than two peaks  of  lower charges and  at each step we will preserve (\ref{wdif}).

  Consider now a configuration with given values $\{m_j\}$ obtained from gs$(\ell;r)$ by iterating the operation 1. Other configurations with the same charge content can be generated from this one
by a sequence of basic displacements of the peaks of charge $j\leq k-1$ in either direction. These displacements can be in the positive or negative directions since the starting configuration is not necessarily the minimal-weight configuration for that charge content. Consider then the displacement of a peak by one unit toward the right (the result is similar for a displacement toward the left). The weight $\w$ changes by 1 since the peak and the adjacent displaced minimum (either the one just at its right if the height of the peak increases or at its left otherwise) both increase by 1.  See for instance Fig. \ref{fig5} (a)-(e). Similarly, the weight $w$ is increased by 1 since the peak position is increased by 1. Therefore, a combination of such peak displacements always satisfies (\ref{wdif}).
This completes the proof of the relation (\ref{wdif}).

\subsection{ABF paths and multiple partitions}\label{SABFmp}

 An immediate consequence of the possible weighting of the $\F$ paths by means of the usual $\B$ weight function is that we can now associate a multiple partition to a $\F$ path \cite{JMpath}. The only difference with the multiple partition $\N$ obtained from a $\B$ path is that now there is an extra constituent partition due to the presence of peaks (clusters) of charge $k$. Let the resulting multiple partition be 
 denoted as $\Nk$, i.e.,
\begin{equation}\label{mulk}
\Nk= (\nt^{(1)}, \nt^{(2)}, \cdots ,\nt^{(k-1)},\nt^{(k)})\qquad {\rm with}\qquad 
\nt^{(j)}= (\nt^{(j)}_1, \cdots , \nt^{(j)}_{m_j})\;.
\end{equation}
Again these parts satisfy the condition 
\begin{equation}\label{difona}
\nt^{(j)}_l\geq \nt^{(j)}_{l+1} + 2j\;,  
\end{equation}
which is the same as (\ref{difone}) but now for $1\leq j\leq k$. The boundary condition reads
\begin{equation}\label{bona}   
\nt^{(j)}_{m_j} \geq j+ {\rm max}\, (j+\ell-k,0)+
2j (m_{j+1}+\cdots +  m_{k}) \; ,
\end{equation}
which is (\ref{bon}) but with $a\rw \ell$ and $k-1\rw k$ (the allowed height of a peak being now $k$ instead of $k-1$), and again $1\leq j\leq k$.  
But there is a novelty here which is caused by the finiteness of the length, or more precisely, the absence of H edges. This induces the following upper bound on the  parts:\footnote{This corrects an error in the second equation of (35) in \cite{JMpath}.}
\begin{equation} \label{upup}
\nt^{(j)}_1\leq -j +\ell+ 2\,[\, jm_j+(j+1)m_{j+1}+\cdots +km_k\,]\;.
\end{equation}

Let us then see how the bound (\ref{upup}) arises from the path connectivity, that is, the requirement that the charge complexes must be in contact.  All possible paths are obtained from the minimal-weight configuration by displacing the peaks of  charge $\leq k-1$ toward the right in all possible ways.
There is thus certainly a path whose rightmost peak has charge $j$ and which forms a charge complex by itself. Its $x$-position would then be $L$ minus half the particle diameter, which is $j$. This maximal $x$-value is clearly a consequence of the required path connectivity. Consider then the multiple partition associated to this specific path and determine the resulting constraints. The largest weight that a cluster of charge $j$ can have is obtained by moving the cluster $\nt^{(j)}_1$ in the leftmost position within the multiple partition viewed as a sequence of clusters, that is, by moving $\nt^{(j)}_1$ before all clusters of lower charges. This increases its weight by $ 2 [(j-1) m_{j-1}+\cdots +m_1]$. The resulting number, which corresponds to  the rightmost $x$-position of a peak of charge $j$ in the $\F$ path, must be $L-j$ (since by assumption, we are dealing with the precise multiple partition describing this path). 
In general, the path corresponding to a given multiple partition does not have its rightmost peak of charge $j$ at this maximal allowed position $L-j$. Therefore, $L-j$ must be viewed as an upper bound for $\nt^{(j)}_1+2\sum_{i=1}^{j-1} im_i$:
\begin{equation}\label{upua}
\nt^{(j)}_1+2\sum_{i=1}^{j-1} im_i \leq L-j=2\sum_{i=1}^k im_i + \ell-j,
\end{equation} 
and this indeed reduces to (\ref{upup}).

The combination of the inequalities (\ref{difona}), (\ref{bona}) and (\ref{upup}) entails a special consequence for the parts of the $k$-th partition in $\Nk$: these parts  are always of the form
\begin{equation}\label{nkpa}
\nt^{(k)}= ((2m_k-1)k+\ell,\cdots , 3k+\ell,k+\ell)\;.
\end{equation} 
This is seen as follows. Iterating (\ref{difona}) and using (\ref{bona}), both applied to $j=k$, one has
\begin{equation}\label{fff}
 \nt^{(k)}_1 \geq \nt^{(k)}_{m_k}+2k(m_k-1)  \geq k+\ell+ 2k(m_k-1),
\end{equation} 
while (\ref{upup}) implies
 \begin{equation}
 \nt^{(k)}_1 \leq -k+\ell +2km_k\;.
\end{equation}
Both together they enforce
 \begin{equation}
 \nt^{(k)}_1 = \ell +k(2m_k-1)\,,
\end{equation} 
as announced. 
Similarly, by combining this expression with the first inequality in (\ref{fff}),
\begin{equation}
k+\ell \leq  \nt^{(k)}_{m_k} \leq \nt^{(k)}_{1}-2k(m_k-1)  =-k+\ell+2km_k -2k(m_k-1)=k+\ell\;,
\end{equation} 
one gets $\nt^{(k)}_{m_k}=k+\ell$. With the difference between the first and the last part saturating the minimal allowed bound, the other parts are also completely determined by (\ref{difona}) to the values given in (\ref{nkpa}).

Let us illustrate this with the ABF$_{4,2}$ path of Fig. \ref{fig8}, whose description in terms of a sequence of clusters reads
\begin{equation}\label{exfig8}
 32^{(4)}\,29^{(1)}\,24^{(2)}\,18^{(3)}\,14^{(1)} 10^{(2)}\,6^{(3)}\,2^{(1)}\;.
\end{equation}
Displacing the charge-4 cluster to the far right using (\ref{com}) reduces its weight to 6, which is indeed $k+\ell=4+2$.


That clusters of charge $k$ are in a sense redundant for the description of the $\Nk$ multiple partition suggests that they could be eliminated. The systematization of this procedure is the strategy underlying the bijection between $\F$ and $\B$ paths.

\subsection{Bijective relation between ABF and Bressoud paths}\label{SBvsAFB}

\subsubsection{From $\F$ to $\B$ paths}\label{SAFBtoB}

The transformation of a  $\F$ path into a $\B$ one is very simple \cite{JMpath}: transform
 the $\F$ path into the multiple partition $\Nk$, reorder the clusters of charge $k$ to the far left (where they form the partition denoted by  ${\bar n}^{(k)}$) and drop them. This generates a multiple partition $\N$:
\begin{equation}\label{mula}
\Nk= (\nt^{(1)},  \cdots ,\nt^{(k-1)},\nt^{(k)})\rw ({\bar n}^{(k)},n^{(1)}, \cdots ,n^{(k-1)})\rw (n^{(1)},  \cdots ,n^{(k-1)})= \N,
\end{equation} 
from which a  $\B$ path is constructed. 


It only remains to clarify the issue of the initial point, that is, to determine the relation between $\ell$ and $a$.
By eliminating the peaks of charge $k$, all portions in the bulk of the path lying in the region $k-1\leq y\leq k$ disappear. Indeed, such sections necessarily result from peaks of charge $k$, possibly penetrated by peaks of lower charges. 
However, when $\ell>0$,  the $\F$ path can still reach the height $k$ in the first charge complex, delimitated at its left by the vertical axis, in spite of being composed of peaks of charge $< k$.
 The proper way to transform this portion of the path, without affecting its charge content, is simply to reduce the initial point by 1: the height of the whole initial complex is thus decreased by 1, rendering the height $k$ out of reach.  
 The final SE edge of the complex is transformed into a H edge in the $\B$ path. (This is illustrated in Fig. \ref{fig9} below.)
Therefore, the prescription is to replace $\ell$ by $a=\ell-1$, or more precisely, since $a$ must be non-negative, to let
\begin{equation}\label{avsl}
a=\text{max}\,(\ell-1,0).
\end{equation}   
For $\ell\geq 1$, this relation ensures the identity:
\begin{equation}\label{maxal}
    {\rm max}\, (j+\ell-k,0)={\rm max}\, (j+a-k+1,0) ,
   \end{equation}
that relates the boundary conditions   (\ref{bona}) and (\ref{bon})  for the multiple partitions $\Nk$ and $\N$ respectively, confirming the correctness of the reduction process (\ref{mula}). Note that for this relation to makes sense, the range of the charge-label $j$ must be the same in both sides, which means that the equality should be restricted to $j\leq k-1$, hence after the removal of the charge-$k$ clusters. Interpreted in this way, it also applies to the case $\ell=0$.

Note that $\ell=0$ and 1 are both related to $a=0$.
Therefore ABF$_{k,1}$ and  ABF$_{k,0}$ paths are both transformed into B$_{k-1,0}$ paths. Accordingly, we do not have a genuine injection between $\Fl$ and $\Ba$ paths since from $\ell=0,1$ to $a=0$, the relation is 2 to 1. To recover a injective relationship, one could simply prevent $\ell$ from being equal to either 0 or 1. 


In practice, to find the $\B$ path corresponding to a given $\F$ one, the construction of the  intermediate multiple  partitions
can be bypassed.
Instead, we can work directly with the sequence of clusters/peaks representing the $\F$ paths. To get $\F\rw\B$, we thus move all charge-$k$ clusters to the left of the $\F$ cluster sequence, drop them, and reorder the remaining clusters so that they represent a $\B$ path. Indeed, displacing the charge-$k$ clusters   at the left of the $\Nk$ multiple partition associated to the $\F$ path or at the left of the sequence of peaks describing the $\F$ path are clearly equivalent. Working directly with the sequence of peaks is likely to involve a smaller number of interchange operations in the reconstruction of the $\B$ path since we already start with a sequence  that satisfy the path characteristics. It should be stressed however that moving a cluster of charge $k$ within a sequence of lower-charge clusters representing peaks will generally destroy their peak  interpretation since, by means of the exchange operation, the weights are not all increased by the same amount. The increase depends upon the individual charge and consequently, the relative weights/distances between the clusters/peaks are modified. 
As a result, a number of interchanges is usually required to recover a $\B$ path.

Let us illustrate the transformation $\F\rw\B$ following the procedure just described. The ABF$_{4,2}$  path of Fig. \ref{fig8} presents a very simple example in that respect since there is only a single charge-4 cluster and it is already at the left in the sequence --
cf. eq. (\ref{exfig8}). Therefore, one simply has to drop it. Since this operation does not involve any interchange, the weight of the 
remaining clusters are not changed, and as a result no reordering  is required for these to represent a path. The outcoming B$_{3,1}$ path is presented in Fig. \ref{fig9}. It should be clear, however, that  the relative triviality of this example is somehow built in the short-cut used in the procedure, in which the intermediate construction of the multiple partitions $\Nk$ and $\N$ is avoided. Manifestly, the main role of the multiple partitions in the present context is to structure the argument underlying the path correspondence.

Finally, let us further qualify the injective nature of the correspondence, this times in regard to the length variable. Recall that the Bressoud paths have  infinite length while the ABF ones have finite length. 
To a given $\F$ path with finite length, we can associate an infinite number of equivalent paths by adding an arbitrary number of peaks of charge $k$ at the end of the path. All these paths have lengths differing by multiples of $2k$ (but they all have the same relative weight).
Manifestly,  $k$-peaks added to the path's tail play no role in the correspondence. This is made particularly clear  in the short-cut procedure, in which they are directly dropped. There are thus actually infinitely many $\F$ paths related to single $\B$ one. However, we can single out a representative among the class of equivalent $\F$ paths. A convenient choice is  the path with shortest length, meaning with no charge-$k$ peak at its right. Alternatively, 
 the $\F$ representative can be chosen to be the one of infinite length.

 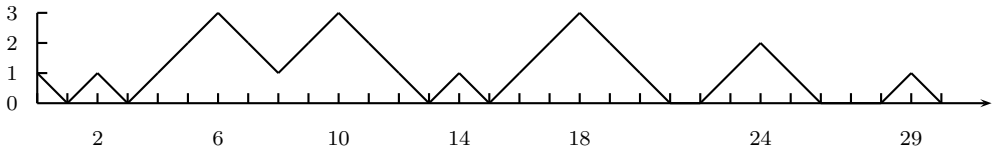
\begin{figure}[ht]\
\caption{{\footnotesize The B$_{3,1}$ path associated to the ABF$_{4,2}$ path of Fig. \ref{fig8}. The H edge between 21 and 22 results from a change of the vertical point, from 2 to 1. The end of this H edge marks the position of the right extremity of the first charge complex in the original ABF$_{4,2}$ path.  The two H edges between 26 and 28 are caused by the removal of the peak of charge 4. Clearly, the values of the $m_j$ for $1\leq j\leq 3$ are unaffected.}} 
\label{fig9}
\begin{center}
\begin{pspicture}(2.5,0)(10.5,2)
{\psset{yunit=38pt,xunit=38pt,linewidth=.8pt}
\psline{-}(0.3,0.3)(0.3,1.2) \psline{->}(0.3,0.3)(9.8,0.3)
\psset{linestyle=solid}
\psline{-}(0.3,0.3)(0.3,0.4) \psline{-}(0.6,0.3)(0.6,0.4)
\psline{-}(0.9,0.3)(0.9,0.4) \psline{-}(1.2,0.3)(1.2,0.4)
\psline{-}(1.5,0.3)(1.5,0.4) \psline{-}(1.8,0.3)(1.8,0.4)
\psline{-}(2.1,0.3)(2.1,0.4) \psline{-}(2.4,0.3)(2.4,0.4)
\psline{-}(2.7,0.3)(2.7,0.4) \psline{-}(3.0,0.3)(3.0,0.4)
\psline{-}(3.3,0.3)(3.3,0.4) \psline{-}(3.6,0.3)(3.6,0.4)
\psline{-}(3.9,0.3)(3.9,0.4) \psline{-}(4.2,0.3)(4.2,0.4)
\psline{-}(4.5,0.3)(4.5,0.4) \psline{-}(4.8,0.3)(4.8,0.4)
\psline{-}(5.1,0.3)(5.1,0.4) \psline{-}(5.4,0.3)(5.4,0.4)
\psline{-}(5.7,0.3)(5.7,0.4) \psline{-}(6.0,0.3)(6.0,0.4)
\psline{-}(6.3,0.3)(6.3,0.4) \psline{-}(6.6,0.3)(6.6,0.4)
\psline{-}(6.9,0.3)(6.9,0.4) \psline{-}(7.2,0.3)(7.2,0.4)
\psline{-}(7.5,0.3)(7.5,0.4) \psline{-}(7.8,0.3)(7.8,0.4)
\psline{-}(8.1,0.3)(8.1,0.4) \psline{-}(8.4,0.3)(8.4,0.4)
\psline{-}(8.7,0.3)(8.7,0.4)
\psline{-}(9.0,0.3)(9.0,0.4)
\psline{-}(9.3,0.3)(9.3,0.4)

\rput(0.9,-0.05){{\scriptsize $2$}}
\rput(2.1,-0.05){{\scriptsize $6$}} \rput(3.3,-0.05){{\scriptsize $10$}}
\rput(4.5,-0.05){{\scriptsize $14$}}\rput(5.7,-0.05){{\scriptsize $18$}}
\rput(7.5,-0.05){{\scriptsize $24$}}
\rput(9,-0.05){{\scriptsize $29$}}
 \psline{-}(0.3,0.6)(0.4,0.6)
\psline{-}(0.3,0.9)(0.4,0.9) \psline{-}(0.3,1.2)(0.4,1.2)

\rput(0.05,0.9){{\scriptsize $2$}} 
\rput(0.05,0.3){{\scriptsize $0$}} 
\rput(0.05,0.6){{\scriptsize $1$}} \rput(0.05,1.2){{\scriptsize $3$}}
\psline{-}(0.3,0.6)(0.6,0.3) \psline{-}(0.6,0.3)(0.9,0.6)
\psline{-}(0.9,0.6)(1.2,0.3) \psline{-}(1.2,0.3)(1.5,0.6)
\psline{-}(1.5,0.6)(1.8,.9) \psline{-}(1.8,.9)(2.1,1.2)
\psline{-}(2.1,1.2)(2.4,.9) \psline{-}(2.4,.9)(2.7,0.6)
\psline{-}(2.7,0.6)(3.0,.9) \psline{-}(3.0,.9)(3.3,1.2)
\psline{-}(3.3,1.2)(3.6,.9) \psline{-}(3.6,.9)(3.9,0.6)
\psline{-}(3.9,0.6)(4.2,0.3) \psline{-}(4.2,0.3)(4.5,0.6)
\psline{-}(4.5,0.6)(4.8,0.3) \psline{-}(4.8,0.3)(5.1,0.6)
\psline{-}(5.1,0.6)(5.4,.9) \psline{-}(5.4,.9)(5.7,1.2)
\psline{-}(5.7,1.2)(6.0,.9)
 \psline{-}(6.0,.9)(6.3,0.6)\psline{-}(6.3,0.6)(6.6,0.3)
\psline{-}(6.6,0.3)(6.9,0.3) \psline{-}(6.9,0.3)(7.5,0.9)
\psline{-}(7.5,0.9)(8.1,0.3) \psline{-}(8.1,0.3)(8.7,0.3)
\psline{-}(8.7,0.3)(9,0.6) \psline{-}(9,0.6)(9.3,0.3)

}
\end{pspicture}
\end{center}
\end{figure}

\subsubsection{From $\B$ to $\F$ paths}\label{SBtoAFB}

We have thus shown how to relate a $\Fl$ path to a $\Ba$ one. Let us now describe the reserve procedure \cite{JMpath}. This amounts to transform a multiple partition $\N$, that describes a path where H edges are allowed, into a $\Nk$ multiple partition, containing additional peaks of charge $k$ in sufficient number for the associated path to be connected. Recall however that if $a>0$, the first H edge, the one closest to the origin, will disappear by the mere lifting of the vertical point by 1.

The first step is thus to run (\ref{mula}) in reverse direction:
\begin{equation}\label{mulb}
\N= (n^{(1)},  \cdots ,n^{(k-1)}) \rw  ({\bar n}^{(k)},n^{(1)}, \cdots ,n^{(k-1)})\rw(\nt^{(1)},  \cdots ,\nt^{(k-1)},\nt^{(k)})=   \Nk .
\end{equation} 
The form of $\nt^{(k)}$ being fixed by (\ref{nkpa}), that of ${\bar n}^{(k)}$ is also fixed:
\begin{equation}\label{nkpaa}
{\bar n}^{(k)}= ((2m_k-1)k+2m+\ell,\cdots , 3k+2m+\ell,k+2m+\ell)\;,
\end{equation} 
where $m$ is the total charge of the original $\B$ path:
\begin{equation}
m=\sum_{j=1}^{k-1} \, jm_j.
\end{equation}  
Recall also that $\ell$ is fixed by $a$ unless $a=0$, in which case $\ell$ can either be 0 or 1. What is not fixed so far, however, is the value of $m_k$. This, we remind, is determined by requiring the corresponding $\F$ path, with initial point $\ell$, to be free of H edges. This is the same as enforcing the bound (\ref{upup}), using the fact that $\nt^{(j)}$ and $n^{(j)}$ are related by
\begin{equation}
\nt^{(j)}= n^{(j)}-2jm_k.
\end{equation} 
In terms of the original $\N$ data (but keeping $\ell$ since it is not completely fixed by $a$), the bound (\ref{upup}) reads:
\begin{equation} \label{upupa}
n^{(j)}_1-2jm_k\leq -j +\ell+ 2\,[\, jm_j+(j+1)m_{j+1}+\cdots +km_k\,]\;.
\end{equation} 
Isolating $m_k$ yields:
\begin{equation} 
m_k\geq  \l\lceil \frac{n^{(j)}_1+j-\ell-2\,[\, jm_j+\cdots +(k-1)m_{k-1} \,]}{2(k-j)} \r\rceil\;. 
\end{equation} 
Since this condition must hold for all values of $1\leq j\leq k-1$, this fixes $m_k$ to be 
\begin{equation} \label{delfmk}
m_k= {\rm max}\; \l(  \l\lceil \frac{n^{(j)}_1+j-\ell-2\,[\, jm_j+\cdots +(k-1)m_{k-1} \,]}{2(k-j)} \r\rceil  , 1  \leq j\leq k-1\r )\; .
\end{equation}
By construction, the length is fixed to be $L=2m+2km_k+\ell=2\m+\ell$, which thereby fixes the value of $r$ (to $\lf (L-\ell)/2k\rf$). This is needed to get gs$(\ell;r)$ and hence $w_{\text{gs}(r;\ell)}$, which enters in the determination of the relative weight.

Take the example of the B$_{3,0}$ path of Fig. \ref{fig7}, for which 
the multiple partition is given by (\ref{exfig7}). We look for the corresponding AFB$_{4,0}$ path, that is, we choose $\ell=0$. The value of $m_4$ is
\begin{equation} \label{exbc}
m_4= {\rm max}\; \l(  \l\lceil \frac{-2}6 \r\rceil  , \l\lceil \frac04 \r\rceil, \l\lceil \frac22 \r\rceil \r )= {\rm max}\; \l( 0,0,1\r) = 1\; .
\end{equation}
Consider then the addition of the cluster $24^{(4)}$ (this value of the weight is fixed by (\ref{nkpaa}) with $n+2m+\ell=4 +2(10)+0$) to the left of the sequence of clusters representing the path of Fig. \ref{fig7} (here we again bypass the construction of the multiple partitions).
The sequence of operations leading the to reconstruction of the AFB$_{4,0}$  path is described as follows (indicating again in bold the clusters that have been interchanged):
\begin{equation}
\begin{matrix}
 24^{(4)} & 19^{(3)}& 15^{(1)} & 12^{(1)} & 10^{(2)} & 6^{(1)}& 3^{(2)}\\
{\bf  25^{(3)}} & {\bf 17^{(1)} }& {\bf  14^{(1)} } & {\bf 14^{(2)} }& {\bf 10^{(4)}} & 6^{(1)}& 3^{(2)}\\
 25^{(3)} & 17^{(1)}&{\bf16^{(2)}} & {\bf12^{(1)}} & 10^{(4)} & 6^{(1)}& 3^{(2)}\\
 25^{(3)} & {\bf18^{(2)} }&{\bf 15^{(1)} }& 12^{(1)} & 10^{(4)} & 6^{(1)}& 3^{(2)}\\
    \end{matrix}
\end{equation}
In the first step, the cluster of charge 4 is placed in-between the peaks at position 10 and 6, a tentative positioning motivated by the location of the rightmost H edge in the B$_{3,0}$ path. The initial portion $10^{(4)}  6^{(1)} 3^{(2)}$ of the AFB$_{4,0}$  path under construction is then seen to be correct, that is, it involves no H edges (cf. Fig. \ref{fig10} for $0\leq x\leq 10$). In the next step we reorder 
$14^{(1)} 14^{(2)}\rw 16^{(2)} 12^{(1)} $ and finally $17^{(1)} 16^{(2)}\rw 18^{(2)} 15^{(1)} $. The resulting path is displayed in Fig. \ref{fig10}.

Still for the example of the B$_{3,0}$ path of Fig. \ref{fig7}, let us consider now the case $\ell=1$ and determine the corresponding AFB$_{4,1}$ path. The value of $m_4$ is again 1 as:
\begin{equation} \label{exba}
m_4= {\rm max}\; \l(  \l\lceil \frac{-3}6 \r\rceil  , \l\lceil \frac{-1}4 \r\rceil, \l\lceil \frac12 \r\rceil \r )= {\rm max}\; \l( 0,0,1\r) = 1\; .
\end{equation}
The added cluster is $25^{(4)}$ (the previous weight 24 being augmented by $\ell=1$ -- cf. (\ref{nkpaa})).
 The sequence of operations is now simply
\begin{equation}
\begin{matrix}
 25^{(4)} & 19^{(3)}& 15^{(1)} & 12^{(1)} & 10^{(2)} & 6^{(1)}& 3^{(2)}\\
{\bf 25^{(3)} }& {\bf19^{(4)}} & 15^{(1)} & 12^{(1)} & 10^{(2)} & 6^{(1)}& 3^{(2)},
    \end{matrix}
\end{equation}
that is, the cluster of charge 4 is moved to the second position. The resulting path is displayed in Fig. \ref{fig11}. Note that by reversing the procedure, the two paths of Fig. \ref{fig10} and \ref{fig11} are both related to the same B$_{3,0}$ path of Fig. \ref{fig7}.

 \begin{figure}[ht]\caption{{\footnotesize The ABF$_{4,0}$ path associated to the B$_{3,0}$ path of Fig. \ref{fig7}. }}
\label{fig10}
\begin{center}
\begin{pspicture}(4,0)(8,2.5)
{\psset{yunit=40pt,xunit=40pt,linewidth=.8pt}
\psline{-}(0.3,0.3)(0.3,1.5) \psline{-}(0.3,0.3)(8.7,0.3)
\psset{linestyle=solid}
\psline{-}(0.3,0.3)(0.3,0.35) \psline{-}(0.6,0.3)(0.6,0.35)
\psline{-}(0.9,0.3)(0.9,0.35) \psline{-}(1.2,0.3)(1.2,0.35)
\psline{-}(1.5,0.3)(1.5,0.35) \psline{-}(1.8,0.3)(1.8,0.35)
\psline{-}(2.1,0.3)(2.1,0.35) \psline{-}(2.4,0.3)(2.4,0.35)
\psline{-}(2.7,0.3)(2.7,0.35) \psline{-}(3.0,0.3)(3.0,0.35)
\psline{-}(3.3,0.3)(3.3,0.35) \psline{-}(3.6,0.3)(3.6,0.35)
\psline{-}(3.9,0.3)(3.9,0.35) \psline{-}(4.2,0.3)(4.2,0.35)
\psline{-}(4.5,0.3)(4.5,0.35) \psline{-}(4.8,0.3)(4.8,0.35)
\psline{-}(5.1,0.3)(5.1,0.35) \psline{-}(5.4,0.3)(5.4,0.35)
\psline{-}(5.7,0.3)(5.7,0.35) \psline{-}(6.0,0.3)(6.0,0.35)
\psline{-}(6.3,0.3)(6.3,0.35) \psline{-}(6.6,0.3)(6.6,0.35)
\psline{-}(6.9,0.3)(6.9,0.35) \psline{-}(7.2,0.3)(7.2,0.35)

\psline{-}(7.5,0.3)(7.5,0.35) \psline{-}(7.8,0.3)(7.8,0.35)
\psline{-}(8.1,0.3)(8.1,0.35)
\psline{-}(8.4,0.3)(8.4,0.35)
\psline{-}(8.7,0.3)(8.7,0.35)

\rput(1.2,-0.05){{\scriptsize $3$}}
\rput(2.1,-0.05){{\scriptsize $6$}} 

\rput(3.3,-0.05){{\scriptsize $10$}}
\rput(3.9,-0.05){{\scriptsize $12$}}
\rput(4.8,-0.05){{\scriptsize $15$}}
\rput(5.7,-0.05){{\scriptsize $18$}}
\rput(7.8,-0.05){{\scriptsize $25$}}

 \psline{-}(0.3,0.6)(0.35,0.6)
\psline{-}(0.3,0.9)(0.35,0.9) \psline{-}(0.3,1.2)(0.35,1.2)
\psline{-}(0.3,1.5)(0.35,1.5) 
 

\rput(0.05,0.9){{\scriptsize $2$}} \rput(0.05,1.5){{\scriptsize $4$}}
\rput(0.05,0.6){{\scriptsize $1$}} \rput(0.05,1.2){{\scriptsize $3$}}
\psline{-}(0.3,0.3)(1.2,1.2) \psline{-}(1.2,1.2)(1.8,0.6)
\psline{-}(1.8,0.6)(2.1,0.9) \psline{-}(2.1,0.9)(2.4,0.6)
\psline{-}(2.4,0.6)(3.3,1.5) \psline{-}(3.3,1.5)(3.6,1.2)
\psline{-}(3.6,1.2)(3.9,1.5)
 \psline{-}(3.9,1.5)(4.5,0.9)
\psline{-}(4.5,0.9)(4.8,1.2) \psline{-}(4.8,1.2)(5.1,0.9)
 \psline{-}(5.1,0.9)(5.7,1.5) \psline{-}(5.7,1.5)(6.9,0.3)
 \psline{-}(6.9,0.3)(7.8,1.2) \psline{-}(7.8,1.2)(8.7,0.3)

}
\end{pspicture}
\end{center}
\end{figure}
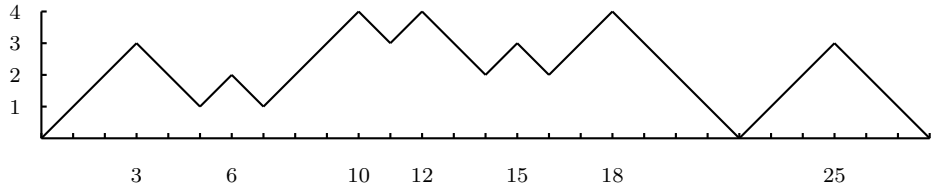

 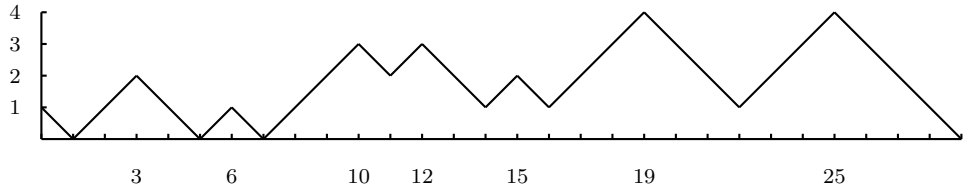
\begin{figure}[ht]\caption{{\footnotesize The ABF$_{4,1}$ path associated to the B$_{3,0}$ path of Fig. \ref{fig7}. The length of this path, compared with that of Fig. \ref{fig10}, is increased by 1 since $\ell=1$. We also observe that the position of the peak of charge 4 is different in the two pictures (it is the third peak from the left in Fig. \ref{fig10} and the sixth one here).}}
\label{fig11}
\begin{center}
\begin{pspicture}(4,0)(8,2.5)
{\psset{yunit=40pt,xunit=40pt,linewidth=.8pt}
\psline{-}(0.3,0.3)(0.3,1.5) \psline{-}(0.3,0.3)(9,0.3)
\psset{linestyle=solid}
\psline{-}(0.3,0.3)(0.3,0.35) \psline{-}(0.6,0.3)(0.6,0.35)
\psline{-}(0.9,0.3)(0.9,0.35) \psline{-}(1.2,0.3)(1.2,0.35)
\psline{-}(1.5,0.3)(1.5,0.35) \psline{-}(1.8,0.3)(1.8,0.35)
\psline{-}(2.1,0.3)(2.1,0.35) \psline{-}(2.4,0.3)(2.4,0.35)
\psline{-}(2.7,0.3)(2.7,0.35) \psline{-}(3.0,0.3)(3.0,0.35)
\psline{-}(3.3,0.3)(3.3,0.35) \psline{-}(3.6,0.3)(3.6,0.35)
\psline{-}(3.9,0.3)(3.9,0.35) \psline{-}(4.2,0.3)(4.2,0.35)
\psline{-}(4.5,0.3)(4.5,0.35) \psline{-}(4.8,0.3)(4.8,0.35)
\psline{-}(5.1,0.3)(5.1,0.35) \psline{-}(5.4,0.3)(5.4,0.35)
\psline{-}(5.7,0.3)(5.7,0.35) \psline{-}(6.0,0.3)(6.0,0.35)
\psline{-}(6.3,0.3)(6.3,0.35) \psline{-}(6.6,0.3)(6.6,0.35)
\psline{-}(6.9,0.3)(6.9,0.35) \psline{-}(7.2,0.3)(7.2,0.35)

\psline{-}(7.5,0.3)(7.5,0.35) \psline{-}(7.8,0.3)(7.8,0.35)
\psline{-}(8.1,0.3)(8.1,0.35)
\psline{-}(8.4,0.3)(8.4,0.35)
\psline{-}(8.7,0.3)(8.7,0.35)\psline{-}(9,0.3)(9,0.35)

\rput(1.2,-0.05){{\scriptsize $3$}}
\rput(2.1,-0.05){{\scriptsize $6$}} 

\rput(3.3,-0.05){{\scriptsize $10$}}
\rput(3.9,-0.05){{\scriptsize $12$}}
\rput(4.8,-0.05){{\scriptsize $15$}}
\rput(6,-0.05){{\scriptsize $19$}}
\rput(7.8,-0.05){{\scriptsize $25$}}

 \psline{-}(0.3,0.6)(0.35,0.6)
\psline{-}(0.3,0.9)(0.35,0.9) \psline{-}(0.3,1.2)(0.35,1.2)
\psline{-}(0.3,1.5)(0.35,1.5) 
 

\rput(0.05,0.9){{\scriptsize $2$}} \rput(0.05,1.5){{\scriptsize $4$}}
\rput(0.05,0.6){{\scriptsize $1$}} \rput(0.05,1.2){{\scriptsize $3$}}
\psline{-}(0.3,0.6)(0.6,0.3) \psline{-}(0.6,.3)(1.2,.9)
\psline{-}(1.2,0.9)(1.8,0.3) \psline{-}(1.8,.3)(2.1,0.6)
\psline{-}(2.1,0.6)(2.4,0.3) 
\psline{-}(2.4,0.3)(3.3,1.2) \psline{-}(3.3,1.2)(3.6,.9)
\psline{-}(3.6,.9)(3.9,1.2)
 \psline{-}(3.9,1.2)(4.5,0.6)
\psline{-}(4.5,0.6)(4.8,0.9) \psline{-}(4.8,0.9)(5.1,0.6)
 \psline{-}(5.1,0.6)(6,1.5) \psline{-}(6,1.5)(6.9,0.6)
 \psline{-}(6.9,0.6)(7.8,1.5) \psline{-}(7.8,1.5)(9,0.3)

}
\end{pspicture}
\end{center}
\end{figure}

The $\F$ path resulting from this construction is the one with the shortest length among the class of equivalent $\F$ paths, differing by the length of their tail composed of peaks of charge $k$ (cf. the remark ending the previous subsection).
If the number of added charge-$k$ clusters is larger than the minimum value given in (\ref{delfmk}) 
--  for instance, these extra peaks could be required to match a predetermined length --, then the supernumerary ones remain at the left of the sequence, meaning that they lie at the right of the path.


\section{Paths, partitions and  the $\Z_k$ parafermionic models}\label{SpvsZ}

In this section, we show how our paths and partitions do provide a combinatorial description of the states  in irreducible modules of the $\Z_k$ parafermionic models. 
 In Section \ref{SZk}
  we present some elements of the parafermionic theories \cite{ZFa}  that suffice to make comprehensible the specification of the two  bases of states that are introduced in Sections \ref{Sbasrp} and \ref{Sbasmp}.


\subsection{The $\Z_k$ parafermionic models}\label{SZk}

\subsubsection{General characteristics of the parafermionic algebra}\label{SZpa}

The parafermionic conformal algebra is spanned by $k-1$ parafermionic fields $\psi_\rho=\psi^\dagger_{k-\rho}$, $\rho=0,1,\cdots, k-1$, with dimension
 \begin{equation}
 h_{\psi_\rho}= \rho\left(1-\frac{\rho}{ k}\right)\;. 
 \end{equation}
Note that $\psi_0=I$, the identity field.
The defining OPEs are
\cite{ZFa}:
 \begin{align}\label{zkope}
\psi_\rho (z) \,\psi_{\s} (w) &\sim \frac{c_{\rho,\s}} {  (z-w)^{2\rho \s/k}}\;\left[
 \psi_{\rho+\s} (w) + {\frac{\rho}{\rho+\s}  } (z-w) \d \psi_{\rho+\s}(w)+\cdots\right] \qquad (\rho+\s\leq k) \;,\\
 \psi_\rho (z) \,\psi^\dagger_\rho (w) &\sim \frac1{(z-w)^{2\rho(k-\rho)/k}}
\left[I+(z-w)^2 \frac{2 h_{{\psi_\rho}}} {c}\, T(w)+\cdots\right].
 \end{align}
The structure constants $c_{\rho,\s}$ are fixed by associativity \cite{ZFa} but their values will not be needed.
  The  central charge is $c={2(k-1)/ (k+2)}$. 
The first OPE exhibits the cyclic $\Z_k$ symmetry and 
 the second one indicates that the Virasoro algebra lies in the enveloping parafermionic algebra. 
 
 By the parafermionic algebra we refer to the algebra generated by the parafermionic modes.
However, its formulation is somewhat tricky. At first, the decomposition of the parafermionic fields in modes depends upon the charge of the field on which they act.
Normalizing the charge  by setting that of $\psi_\rho$ to be $2\rho$,
the mode decomposition of $\psi_\rho$ and $\psi_\rho^\dagger$ acting on an arbitrary field $\phi_q$ of charge $q$ reads \cite{ZFa}:
 \begin{align}\label{modep}
  \psi_\rho(z)\phi_{q}(0)  =& \sum_{m=-\y}^\y
z^{-\rho q/k-m-\rho}A^{(\rho)}_{m+\frac{\rho(\rho+q)}{k}}\, \phi_{q}(0)  \\ 
  \psi_\rho^\dagger (z)\phi_{q}(0)  = &\sum_{m=-\y}^\y
 z^{\rho q/k-m-\rho} A_{m+\frac{\rho (\rho-q)}{k}}^{(\rho)^\dagger}\,\phi_{q}(0).
 \end{align}
 A second difficulty is that the commutation relations resulting from the OPE are actually generalized commutation relations, expressed in the form of infinite series \cite{LW,ZFa}. 
The explicit form of these commutation relations will not be needed here (but they are obviously required for the derivation of the quoted results) and can be found in \cite{ZFa,JMa,JM.A}. 

 
 To lighten the notation, it is convenient to omit
 the fractional part of the modes. It is easily recovered, being  fixed unambiguously by the charge of the state on which the corresponding operator acts. In order to keep in mind the omission their fractional modes, the operators will be
indicated by  calligraphic symbols, i.e., \cite{JMa,JMb}:
 \begin{equation}
 \A^{(\rho)}_n |\phi_{q}\R\equiv  A^{(\rho)}_{n+\frac{\rho(\rho+q)}{k}}|\phi_{q}\R\; . \end{equation}

\subsubsection{Parafermionic highest-weight modules}\label{Shwm}

The spectrum of the parafermionic theory is specified by its primary fields. 
The parafermionic primary fields $\varphi_\ell$ are labelled by their charge $\ell$,  with $ \ell= 0,\cdots ,k-1$, and they have conformal dimension \cite{ZFa}:
 \begin{equation}
 h_{\ell} =\frac {\ell(k-\ell) }{2k(k+2)} .
   \end{equation}
  To each
primary field, there corresponds a highest-weight state
$|\varphi_\ell\R$. In particular, $|0\R=|\varphi_0\R$.
The parafermionic highest-weight  
conditions read
 \begin{equation}
 \A_{n} | \varphi_\ell 
\rangle = \A^\dagger_{n+1} | \varphi_\ell \rangle 
=0  \qquad {\rm for}\quad n\geq 0,
\label{hiwe}
   \end{equation}
where $\A\equiv \A^{(1)}$.

The module over $| \varphi_\ell \rangle $ can be decomposed into a finite sum of modules with specific relative charge $2r$, defined modulo $2k$, in which all states
have conformal dimensions differing by integers. The highest-weight state $| \varphi_\ell^{(r)}\R$ in the module of relative charge $2r$ (and due to the field identifications, it suffices to consider the case where $0\leq r\leq k-\ell-1$ \cite{JMa}) has
 dimension: 
  \begin{equation}
h^{(r)}_{\ell}= h_{\ell}+\frac{r(k-\ell-r)}{ k}.
\label{bbpp}
   \end{equation}
$| \varphi_\ell^{(r)}\R$ actually  corresponds to the state $\A_{-1}^{r}| \varphi_\ell\R$ -- which, we stress, 
is not a parafermionic  highest-weight state when $r\not =0$.
The states in  the submodule of relative charge $2r$  are  described by sequences of $p+r$ $\A$ and $p$ $\A^\dagger$  modes, for all values of $p$.\footnote{Note that because the Virasoro algebra lives in the parafermionic enveloping algebra, Virasoro modes are not required; this is akin to the description of affine Lie algebra modules -- see for instance \cite{CFT}.} The two types of modes can be ordered separately and  their negated modes correspond to weakly decreasing sequences of positive or non-negative integers respectively. Without further constraints, this defines the Verma module \cite{JMa}. However, the parafermionic module with  highest-weight state $|\varphi_\ell\R$ has an infinite number of singular vectors,
 the  two primitive ones being
 \begin{equation}
  \A_{-1}^{k-\ell+1}|\varphi_\ell\R \, \qquad \text{and}\qquad 
 (\A^\dagger_{0})^{\ell+1}|\varphi_\ell\R .\label{sig}
    \end{equation}
 The irreducible modules can be described by an inclusion-exclusion procedure, where the successive exclusions and inclusions take care of the removal of the singular vectors, 
 and the subsequent corrections resulting from the overlap of their submodules.  This procedure is worked out in \cite{JMa}.
 %
The resulting character is an infinite alternating sum of Verma characters, which can be written in the concise form
(cf. eqs (1.9) and (1.10) of \cite{FMO} but renormalized here so that ${\chi}_{\ell,r}^{(k)}(0)=1$):
\begin{equation}\label{cara}
{\chi}_{\ell,r}^{(k)}(q)
=q^{-\frac{(\ell+1)^2}{4(k+2)}-r}\sum_{j=-\infty}^{\infty}
q^{(k+2)\l(j+\frac{(\ell+1)}{2(k+2)} \r)^2}
\Bigl\{V_{r-j(k+2)}(q)-V_{r+\ell+1+(k+2)j}(q)\Bigr\}
\end{equation}
where $V_t$ is the Verma character, 
given by
\begin{equation}\label{vermu}
V_t(q)=\sum_{j=0}^{\infty}
\frac{q^{j+t}}{(q;q)_j(q;q)_{j+t}}.
\end{equation}
 Such an alternating-sign character formula, deduced by the chiral algebra representation theory, is dubbed bosonic in \cite{KKMMa}.

\subsection{Parafermionic quasi-particle basis and restricted partitions}\label{Sbasrp}

\subsubsection{Formulation of a first quasi-particle basis}\label{Sbas1}

Given that $\psi_1^\dagger\sim \psi_1^{k-1}$, it is reasonable to expect that the parafermionic states, naturally expressed in terms of the modes of $\A$ and $\A^\dagger$, could all be described solely in terms of the $\A$ operators (which, we recall, are the modes of $\psi_1$).  The quasi-particle basis of the $\Z_k$ parafermionic models -- by which we refer to  a description of the states that is free of singular vectors -- is precisely of this type \cite{LP,JMb}. States in the highest-weight module $|\varphi_\ell\R$  are given by strings of $\A$ operators ordered in weakly decreasing values of minus the mode indices:
 \begin{equation}
 \A_{-n_1}\A_{-n_2}\cdots \A_{-n_m}|\varphi_\ell\R\; , \quad \quad  
n_i\geq n_{i+1}\geq 1\, .\label{ferbas}
   \end{equation}  
Again, all these states can be organized into modules with specific relative charge, those with relative charge $2r$ being selected by the condition:
 \begin{equation}   m=r \; {\rm mod}\; k .
   \end{equation}     

The  $\Z_k$ invariance  puts constraints on the
modes of $k$ consecutive operators. This  invariance translates into the condition $(\psi_1)^k\sim
 I $, which entails a linear relation for  sequences of
$k$  adjacent
$\A$ modes. 
To recover linear independence,  one state must be removed at every level (meaning minus the sum of the $k$ modes).  For this procedure to be consistent,
 all ordered sequences of $\A$ operators that contain any one of
the following $k$-string \cite{LP,JMb}
 \begin{equation}
 (\A_{-(n+1)})^{k-i}(\A_{-n})^{i}\qquad (i=0,\cdots , k-1, \quad n>0), \label{parexcl}
    \end{equation}
    must be forbidden
These are those $k$-strings which have mode indices `as equal as possible', namely, which they differ by at most 1. 
To be explicit: for $k=3$, this amounts to forbid
any string of states of the form (\ref{ferbas}) that contains one of  the following 3-strings:
 \begin{equation}
 \A_{-n}\A_{-n}\A_{-n}\;\;,\qquad \A_{-(n+1)}\A_{-n}\A_{-n} \;,\qquad
\A_{-(n+1)}\A_{-(n+1)}\A_{-n} . \label{extri}
   \end{equation}
This restriction rule is equivalent to the difference-2 condition
at distance $k-1$ for the mode indices negated \cite{LP,JMb}:
 \begin{equation}
    n_j- n_{j+k-1}\geq 2.
    \label{restru}
       \end{equation}
We stress that this exclusion rule is of a combinatorial nature: the excluded states do not identically vanish.

There is an additional constraint that 
distinguishes the different highest-weight modules. This comes from the singular vector
$\A_{-1}^{k-\ell+1}|\varphi_\ell\R$ (cf. (\ref{sig})), the unique singular vector expressed solely in terms of
the $\A$ modes.  In a quasi-particle basis, such as a state should be removed by an exclusion condition 
rather than a mere subtraction. The proper way to implement this condition is to enforce a constraint on the maximal number of
$\A_{-1}$ factors at the right end of the string, namely, that there can be at most  $(k-\ell$) $\A_{-1}$ operators.  
This constraint is fully captured by the condition:
 \begin{equation}
 n_{m-k+\ell}\geq 2,
 \label{ferbass}
    \end{equation}
(recall that $m$ stands for the total number of $\A$ operators in the string -- cf. (\ref{ferbas})). 
Note that
when $\ell=0$ or $1$, the condition (\ref{ferbass}) is superfluous, being  taken into account by the
restrictions  (\ref{restru}). 

The basis of states of the highest-weight module $|\varphi_\ell \R$ with relative charge $2r$ is thus obtained from all states of the form (\ref{ferbas}), for all $m=r$ mod $k$, and subject to the restrictions (\ref{restru}) and (\ref{ferbass}).

The defining conditions on this basis are precisely those for  the restricted partitions, namely (\ref{difk}) and (\ref{bd}), with $i$, in the latter  expression, related to $\ell$ by
 \begin{equation}\label{ivsl}
 i=\text{min}\; (k,k-\ell+1),
   \end{equation}
   where the minimum takes into account the upper bound $i\leq k$.

In a path description, the full irreducible module $|\varphi_\ell\R$ is generated by all $\Ba$ paths starting at the vertical value $a=k-i$, i.e., $a=\text{max}(\ell-1,0)$ (cf. \ref{avsl}).
   The paths describing the sector of relative (parafermionic) charge $2r$ are those whose total (path) charge $\sum_j jm_j=m$ is equal to $r$ mod $k$.  (Note that with the standard normalizations used here, the path charge is half the parafermionic charge.)
   
   At this point, the states in the modules $\ell=0$ and $\ell=1$ are described by the same B$_{k-1,0}$ paths or the same restricted partitions (with $i=k$). The degeneracy is lifted by the following correction.

   \subsubsection{The conformal dimension of a parafermionic state}

So far we have associated a string of parafermionic modes with a partition. This relation neglects the fractional part of the parafermionic modes. But this piece, as we already indicated, is completely fixed by the number of acting operators and the charge $\ell$ of the module under consideration. Therefore, this omission cannot spoil the 1-1 correspondence  between states and partitions. However, it matters for the determination of the conformal dimension of the state. Once this correction is made, the two modules $\ell=0$ and $\ell=1$ (which, as just noticed,  are described by the same B$_{k-1,0}$ paths and the same restricted partitions) are distinguished.

Reinserting the charge-dependent fractional part of each  mode in (\ref{ferbas}), one has
 \begin{equation}
 A_{-n_1+\frac{(1+2(m-1)+\ell)}{k}}\cdots A_{-n_{m-1}+\frac{(3+\ell)}{k}}\, A_{-n_m+\frac{(1+\ell)}{k}} \, | \varphi_\ell\R.
  \end{equation}
The fractional parts add up to 
 \begin{equation}\label{fra}
 F=\frac{1}{k}\sum_{j=0}^{m-1} (1+2j+\ell)= \frac{m}{k}(m+\ell).
  \end{equation}
The dimension of the string (\ref{ferbas}) is thus 
 \begin{equation}
  h(m)= \sum_{j=1}^m n_j-\frac{m}{k}(m+\ell).
  \label{defdes}
   \end{equation}  
$F$ is thus the correcting factor that needs to be subtracted to each $\B$ path to match the relative conformal dimension of the corresponding parafermionic  state (\ref{ferbas}). Note that it only depends upon the total parafermionic charge $2m$. With the above expression of $F$, we check  that all sates with $m=r$ mod $k$ have conformal dimension differing by integers, as it should.

 Finally, in order  for the dimension of the descendants in a module with specified value of $r$ to correspond to the conformal dimension relative to its top state $|\varphi_\ell^{(r)}\R$, we need to 
 subtract the contribution of the string $\A_{-1}^r$ which maps the highest-weight state $|\varphi_\ell\R$ to  $|\varphi_\ell^{(r)}\R$, that is,
  \begin{equation}\label{deltah}
\Delta h=  h(m)-(h_\ell^{(r)}-h_\ell)= \sum_{j=1}^m n_j-\frac{m}{k}(m+\ell)-\l( r-\frac{r}{k}(r+\ell)\r)   .\end{equation}
 
 
 \subsubsection{A detailed example: states in the $\Z_3$ vacuum module}\label{Sex1}
 
 Let us exemplify  the description of states in terms of restricted partitions for the $\Z_3$ model.
 Let us designate a state of the form (\ref{ferbas}) by the partition 
  \begin{equation}\label{paga}
  (n_1,\cdots,n_m)_{\gamma_{m,r,\ell}},
   \end{equation}
where the index $\gamma_{m,r,\ell}$, defined as 
  \begin{equation}\label{gade}
   \gamma_{m,r,\ell}=  \frac1{k}{(m-r)(m+r+\ell)}+r,
    \end{equation}
indicates the number that needs to be {\it subtracted} from  the partition weight (= the sum of its parts) to match the conformal dimension -- cf. (\ref{deltah}).
  
In Table \ref{tab1},  we list the states at the first few grades in the $\Z_3$ vacuum module, for the three relative charges.
  
\begin{table}[ht]
\caption{List of the states  in the $\Z_3$ vacuum module for grades up to 6. The  states are separated according to their relative charge $2r$ (so that the number of parts is equal to $r$ mod 3), with $r=0,1,2$. The states are written in the form of a restricted partition (without commas between the parts) as in (\ref{paga}), with an index (defined in (\ref{gade})) that gives the number that must be subtracted from the sum of the parts  to reproduce the relative conformal dimension of the first column.}
\label{tab1}
\begin{center}
\begin{tabular}{|c|c|c|c|} 

\hline
$\Delta h$ &$ r= 0$& $r=1$ & $r=2$\\
\hline
$0$ & $(\, )$ & $(1)_{1}$ & $(11)_{2}$\\
$1$ & $ -$ & $(2)_{1}$ & $(21)_{2}$\\
$2$ & $(311)_{3}$ & $(3)_{1}, (3311)_{6}$ & $(31)_{2},(22)_{2}$\\ 
$3$ & $(411)_{3},  (321)_{3}$& $(4)_{1}, (4311)_{6}$ & $(41)_{2},(32)_{2}$\\
$4$ & $(511)_{3} , (421)_{3} , (331)_{3}$& $(5)_{1},  (5311)_{6}, (4411)_{6}, (4321)_{6}$ & $(51)_{2},(42)_{2},(33)_{2}, (53311)_{9}$\\ 
$5$ & $(611)_{3} , (521)_{3} , (431)_{3},$& $(6)_{1},  (6311)_{6}, (5411)_{6}$& $(61)_{2},(52)_{2},(43)_{2}$\\ 

& $(422)_{3}$& $ (5321)_{6}, (4421)_{6}$  & $ (63311)_{9},  (54311)_{9}$\\
$6$ & $(711)_{3} , (621)_{3} , (531)_{3} , (441)_{3}$& $(7)_{1},  (7311)_{6}, (6411)_{6}, (5511)_{6}$& $(71)_{2},(62)_{2},(53)_{2},(44)_{2},  (73311)_{9}$\\ 

 & $(522)_{3}, (432)_{3}, (553311)_{12}$& $ (6321)_{6}, (5421)_{6}, (5331)_{6}, (4422)_{6}$  & $  (64311)_{9}, (55311)_{9}, 
 (54321)_{9}$\\

\hline

\end{tabular}
\end{center}
\end{table}

These states can also be described by paths. For instance, the B$_{2,0}$ paths associated to the 8 states at level 6 in the $r=1$ sector of the vacuum module (these all start at $a=0$ since $\ell=0$) correspond respectively to the cluster sequences: 
  \begin{equation}7^{(1)}, \; 7^{(1)} 4^{(2)} 1^{(1)},\; 6^{(1)}4^{(1)}2^{(2)},\; 10^{(2)}2^{(2)},\; 7^{(2)}4^{(1)}1^{(1)},\; 9^{(2)}3^{(2)},\; 8^{(2)}3^{(1)}1^{(1)},\; 8^{(2)}4^{(2)}.
   \end{equation}
For instance, the penultimate path is associated with the sequence of modes $\A_{-8}^{(2)}\, \A_{-3}^{(1)}\, \A_{-1}^{(1)}$ acting on the state $|0\R$. 
 That these paths pertain to the sector with $r=1$ can be seen from the value of their total charge, which is equal to  1 mod 3.
  Of course, the path weight (the sum of the cluster weights) must be adjusted by the subtraction of the appropriate ${\gamma_{m,1,0}}$ factor (${\gamma_{1,1,0}}=1 $ and ${\gamma_{4,1,0}}=6$).

 \subsubsection{The fermionic $\Z_k$ character}
 
 Consider the character $\chi_{\ell,r}^{(k)}(q)$ in the module $|\varphi_\ell^{(r)}\R$,  with the states weighted by ${\Delta h}$.
 The relation between the parafermionic states and restricted partitions unravelled above implies that the character $\chi_{\ell,r}^{(k)}(q)$ is:
 
\begin{itemize}

 \item
 the generating function  for restricted partitions into exactly $m$ parts,  with at most $k-\ell$ parts equal to 1, 
 \item modified by the correcting weight factor $q^{-\frac1{k}{(m-r)(m+r+\ell)}-r}$ (cf. (\ref{deltah})), and 
 \item summed over all values of $m=\sum_j jm_j =r\; \text{mod}\, k$. 
\end{itemize}

\n This gives (with the normalization $\chi_{\ell,r}^{(k)}(0)=1 $)
   \begin{equation} \chi_{\ell,r}^{(k)}(q)= 
  \sum_{\substack{m_1,\cdots,m_{k-1}=0 \\
m=r\;{\rm mod}\;k}}^\y G_{\text{max}(0,\ell-1)} (\{m_j\};q)\, q^{-\frac1{k}{(m-r)(m+r+\ell)}-r}
, \end{equation}
and, using (\ref{gfa}),
   \begin{equation}
 \chi_{\ell,r}^{(k)}(q)= 
\sum_{\substack{m_1,\cdots,m_{k-1}=0 \\
m=r\;{\rm mod}\;k}}^\y\frac{ q^{N_1^2+\cdots+ N_{k-1}^2- \frac1{k}{(m-r)(m+r+\ell)}-r+N_{k-\ell+1}+\cdots N_{k-1}}  }{
(q)_{m_1}\cdots (q)_{m_{k-1}} },
 \label{lepooo}
 \end{equation}
 with $N_j$ defined in (\ref{defN}).  This is the standard expression of the fermionic form of the $\Z_k$ character  \cite{LP,JMb}.


\subsection{Parafermionic quasi-particle basis and multiple partitions}\label{Sbasmp}

\subsubsection{A second  quasi-particle basis}\label{Sbas2}

The quasi-particle basis just presented is very economical in that it involves a single type of modes. This basis is directly related to the restricted partitions.  A more complicated-looking basis can be considered, that involves the modes of all $k-1$ parafermionic fields \cite{JM.A,Geo}. This is the basis which is related to the multiple partitions (and this is precisely in this context that multiple partitions have first appeared).

We thus consider a basis of states constructed out of ordered sequences of the $k-1$ parafermionic modes:
  \begin{equation}\label{secba}
  \A^{(1)}_{-n^{(1)}_1}\cdots \A^{(1)}_{-n^{(1)}_{m_1}} \A^{(2)}_{-n^{(2)}_1} \cdots\A^{(2)}_{-n^{(2)}_{m_2}}\cdots \A^{(k-1)}_{-n^{(k-1)}_1} \cdots\A^{(k-1)}_{-n^{(k-1)}_{m_{k-1}}} \; |\varphi_\ell\R\;. 
   \end{equation}
Such a state is conveniently represented by a multiple partition of the form (\ref{mul}).

In \cite{JM.A}, the  conditions on the  indices $n^{(j)}_l$ that ensure the linear independence of the states are determined. These are precisely the conditions (\ref{difone}) and (\ref{bon}) with $a$ related to $\ell$ via (\ref{avsl}) (so that
(\ref{maxal}) is satisfied), but here with $\ell$ labeling the highest-weight state.  The condition (\ref{bon})  incorporates  different features: 

\begin{itemize}
\item The repulsion term between the different types of modes: $  n^{(j)}_{m_j} \geq \cdots +
2j (m_{j+1}+\cdots +  m_{k-1})$ is rooted in the associativity requirement.  

\item The piece $  n^{(j)}_{m_j} \geq j+\cdots$ takes into account the highest-weight  
conditions 
${\cal A}_{-n}^{(j)} | \varphi_\ell 
\rangle 
=0 $ for $n<j$.
\item Finally, the part $  n^{(j)}_{m_j} \geq \cdots +{\rm max}\, (j+\ell-k,0)+\cdots$,
 the module selecting condition, comes from
the singular vector $({\cal A}_{-1})^{k-\ell+1} |  \varphi_\ell 
\rangle 
=0 $, using  $ \A_{-j}^{(j)}\propto  (\A_{-1})^j$.
\end{itemize}

The weight of a multiple partition is not the conformal dimension of  a string of operator of the form (\ref{secba}): the fractional part $F$ of the modes needs to be subtracted from the former. But the expression $F$ is exactly the one obtained previously (cf. (\ref{fra})), depending only upon the total charge $m$. The analysis at this point is the same as before, except that in (\ref{deltah})  $\sum_j n_j$ is replaced by $\sum_{l,j}n_l^{(j)}$.

The states in the three sectors of the vacuum module, up to grade 6, are presented in Table \ref{tab2} in the multiple-partition basis.

\begin{table}[ht]
\caption{The states of Table \ref{tab1}, rewritten  (in the same order) in terms of multiple partitions. These are given as two ordered partitions separated by a semi-column as $(n_1^{(1)},\cdots, n_{m_1}^{(1)};n^{(2)}_1\cdots , n_{m_2}^{(2)})$ (and generally without commas between the parts). The  number $\gamma_{m,r,0}$ is again given as a subindex. }
\label{tab2}
\begin{center}
\begin{tabular}{|c|c|c|c|} 

\hline
$\Delta h$ &$ r= 0$& $r=1$ & $r=2$\\
\hline
$0$ & $(\, )$ & $(1;)_{1}$ & $(;2)_{2}$\\
$1$ & $ -$ & $(2;)_{1}$ & $(;3)_{2}$\\
$2$ & $(3;2)_{3}$ & $(3)_{1}, (;62)_{6}$ & $(31;)_{2},(;4)_{2}$\\ 
$3$ & $(4;2)_{3},  (3;3)_{3}$& $(4;)_{1}, (;72)_{6}$ & $(41;)_{2},(;5)_{2}$\\
$4$ & $(5;2)_{3} , (4;3)_{3} , (3;4)_{3}$& $(5;)_{1},  (53;2)_{6}, (;82)_{6}, (;73)_{6}$ & $(51;)_{2},(42;)_{2},(;6)_{2}, (5;62)_{9}$\\ 
$5$ & $(6;2)_{3} , (5;3)_{3} , (3;5)_{3},$& $(6;)_{1},  (63;2)_{6}, (;92)_{6}$& $(61;)_{2},(52;)_{2},(;7)_{2}$\\ 

& $(4;4)_{3}$& $ (53;3)_{6}, (;83)_{6}$  & $ (6;62)_{9},  (5;72)_{9}$\\
$6$ & $(7;2)_{3} , (6;3)_{3} , (531;)_{3} , (3;6)_{3}$& $(7;)_{1},  (73;2)_{6}, (64;2)_{6}, (;10,2)_{6}$& $(71;)_{2},(62;)_{2},(53;)_{2},(;8)_{2},  (7;62)_{9}$\\ 

 & $(5;4)_{3}, (4;5)_{3}, (;10,6,2)_{12}$& $ (63;3)_{6}, (;93)_{6}, (53;4)_{6}, (;84)_{6}$  & $  (6;72)_{9}, (7;62)_{9}, 
 (5;73)_{9}$\\

\hline

\end{tabular}
\end{center}
\end{table}
\subsubsection{A modified form of the second  quasi-particle basis}\label{Sbas2m}


It is still possible to bring a little extra complication in the previous basis by incorporating the redundant $\A^{(k)}$ modes. This amounts to modify the
state (\ref{secba}) into
  \begin{equation}\label{secbb}
  \A^{(1)}_{-\nt^{(1)}_1}\cdots \A^{(1)}_{-\nt^{(1)}_{m_1}}\cdots \A^{(k-1)}_{-\nt^{(k-1)}_1} \cdots\A^{(k-1)}_{-\nt^{(k-1)}_{m_{k-1}}} \A^{(k)}_{-\nt^{(k)}_1} \cdots\A^{(k)}_{-\nt^{(k)}_{m_k}} \; |\varphi_\ell\R\;. 
   \end{equation}
   where the indices $\nt^{(j)}_l$ satisfy the conditions (\ref{difona}) and (\ref{bona}).
This   is clearly associated to the multiple partition $\Nk$ in (\ref{mulk}).

It is via this modified version that the second quasi-particle basis is connected to the $\Fl$ paths, with the module label $\ell $ corresponding to the path initial point. With $\m$ defined in (\ref{cha}), the fractional dimension of the above state becomes
  \begin{equation}\label{ftil}
\tilde{F}= \frac{\m (\m+\ell)}{k}.
   \end{equation}
   Quite remarkably, the relative conformal dimension of a state in the module $|\varphi_\ell\R$ (in any sector) described by a $\Fl$ path is given simply by
  \begin{equation}\label{wf}
h-h_\ell=w-\tilde{F}.
   \end{equation}
Given that  the relative conformal dimension  in the $r$-th sector, $h-h_\ell^{(r)}$, is equal to the relative weight $w-w_{\text{gs}(\ell;r)}$, the above relation
 can be rewritten as
\begin{equation}
h-h_\ell= (h-h_\ell^{(r)}) + (h_\ell^{(r)}- h_\ell) = (w-w_{\text{gs}(\ell;r)})+w_{\text{gs}(\ell;r)}-\tilde{F}.
   \end{equation}
Hence, to establish (\ref{wf}), one has to show that, say with $r+\ell\leq k-1$ (the condition selecting a set of independent modules),
  \begin{equation}\label{hlra}
w_{\text{gs}(\ell;r)}-\tilde{F}=h_\ell^{(r)}-h_\ell=\frac{r(k-r-\ell)}{k},
   \end{equation}
    (cf. eq. (\ref{bbpp})).
 With this restriction on $r+\ell$, the path describing the ground state gs$(\ell;r)$ starts with a peak of charge $r$ at position $r$,  then reaches the $x$ axis and is completed by $p$ peaks of charge $k$. The weight $w_{\text{gs}(\ell;r)}$ is simply (cf. eq. (\ref{aav}) with $x_0=2r+\ell$ and $j\rw k$):
  \begin{equation}
  w_{\text{gs}(\ell;r)}=r+(2r+\ell)p +kp^2.
   \end{equation}
With $\tilde{F}$ given by (\ref{ftil}), and taking $\m=r+pk$, we see that
$w_{\text{gs}(\ell;r)}-\tilde{F}$ is indeed given by (\ref{hlra}). 

For instance, the  ABF$_{4,0}$ path of Fig. \ref{fig10} describing a state in the vacuum module of the $\Z_4$ model, has $w=89$ and $\m=14$ so that the conformal dimension of the corresponding state is $h=89-(14)^2/4=40$.

The first few ABF$_{3,0}$ paths in the sector $r=0$ are displayed  in \cite{FP}.

\section{Other path-state correspondences}\label{Sother}

In order to place this work in a somewhat broader perspective, we review briefly some other path-state correspondences, restricting ourself to the related graded parafermionic model and the minimal models. Minimal models are not regarded as close relatives of the parafermionic theories (although two unitary minimal models are equivalent to parafermionic ones) but their path description is surely similar. In this way, we observe in Section \ref{Sm2p} the reappearance of the  Bressoud paths, this time in the context of the non-unitary minimal models $\M(2,2k+1)$. Similarly, the $\F$ paths come out again (albeit in a dual version), in relation to the unitary minimal models $\M(k+1,k+2)$ -- cf. Section \ref{Summ}. Through their $\F$ path formulation, the unitary minimal models and the $\Z_k$ parafermionic ones are thus found to be in a dual relationship \cite{FWa, OleJS,BMlmp}. Quite interestingly, a similar duality is observed between the graded version of the $\Z_k$ parafermions and the $\M(k+1,2k+3)$ minimal models \cite{JSTAT}, a duality for which there is still no clear rationale. The paths at work in this context are some sort of `graded' version (i.e., formulated on a half-integer lattice) of the $\F$ paths. This is briefly addressed in Section \ref{Sgp}.
Finally, the full generalization of the dual $\F$ paths, the so-called  Forrester-Baxter RSOS paths, which provide the description of the states in all minimal models $\M(p',p)$, are presented in Section \ref{SFB}. When $p\geq 2p'-1$, these can be transformed into new paths that represent natural generalizations of the Bressoud paths, as sketched in Section \ref{Sgb}. For $p'=2$ and $p=2k+1$, these  are genuine $\B$ paths.

\subsection{$\B$ paths and the $\M(2,2k+1)$ minimal models}\label{Sm2p}

The basis of states for the $\M(2,2k+1)$ minimal models\footnote{
The minimal models are characterized by two coprime integers $p'
,p$ both $\geq 2$, and we will choose $p>p'$. 
Their  central charge is
$
c=1- \frac{6(p-p')^2}{pp'}.
$
  For each model, there is a finite number of primary fields $\phi_{r,s}= \phi_{p'-r,p-s}$
with conformal dimension:
$h_{r,s}= \frac{(pr-p's)^2-(p-p')^2}{4 p p'}$ with $1\leq r\leq p'-1$ and $1\leq s\leq p-1$.
The bosonic (or Rocha-Caridi) form of the  irreducible character is \cite{Roc} (see e.g. \cite{CFT}):
\begin{equation*}
\chi^{(p',p)}_{r,s}(q)= \frac{1}{(q)_\y} \sum_{j=-\y}^{\y}\l (q^{j(p'p j+pr-p's)}-q^{(p'j+r)(pj+s)}\r).
 \end{equation*}
} is \cite{FNO}
\begin{equation}
L_{-n_1} \cdots L_{-n_m} | \phi_{1,s}\R \qquad\text{with}\qquad  {n_i\geq n_{i+k-1}+2}\qquad \text{and} \qquad n_{m-s+1} \geq 2.
 \end{equation}
The first condition  is  generic to all modules and results from the vacuum null field. The different modules $(1,s)$, with $1\leq s\leq k$ (which is not restrictive since $| \phi_{1,s}\R=| \phi_{1,2k-s}\R$), are distinguished by the  second condition which takes into account the singular vector at level $s$.
These conditions on the mode indices of the Virasoro operators are simply the defining conditions (\ref{difk})-(\ref{bd}) for restricted partitions, with $s=i$. Equivalently, every state in the $| \phi_{1,s}\R$ module is in correspondence with a $\Ba$ path with 
\begin{equation}
a=k-s.
\end{equation}
 Here,  the relative conformal dimension is given directly by the weight of the path: 
 \begin{equation}
w=h-h_{1,s}.
 \end{equation}
  The fermionic character $\chi_{1,s}(q)$ of the irreducible module $| \phi_{1,s}\R$  of the $\M(2,2k+1)$ minimal model is thus $G_{k-s}(q)$ given in (\ref{gfBka}).

\subsection{Dual $\F$ paths and the $\M(k+1,k+2)$ minimal models}\label{Summ}

The states in the unitary minimal models $\M(k+1,k+2)$ are related to the $\F$ paths of the RSOS model of \cite{ABF} in regime III \cite{Kyoto,Mel}. These are actually {\it dual} to  the previously introduced $\F$ paths. The duality is defined in terms of the weight function:
\begin{equation}\label{weigd}
\w^* = \sum_{x=1}^{L-1} \w^* (x)\qquad \text{where} \qquad \w^* (x)=\begin{cases}
\frac{x}2 &\text{$x$ is {\it not} an extremum of the path} \\
0& \text{otherwise}\;,
\end{cases} \end{equation}
(compare with (\ref{weig})). Paths are characterized by the specification of the values of $y_0,\, y_{L-1}$ and $y_L$. Choosing $y_{L-1}=y_L+1$ amounts to enforce the last edge of the path to be a  SE one. Then $0\leq y_0\leq k$ and $0\leq y_L\leq k-1$. These end points are related to the module parameters $(r,s)$ by
\begin{equation}
 s=y_0+1\quad \text{and}\quad r=y_L+1.
 \end{equation}
The fermi-gas analysis of these paths is presented in \cite{OleJS,OleJSb}, reproducing fermionic expressions conjectured in \cite{Mel} (and initially proved in \cite{Ber} for $s=1$ by a different method). In addition, the bosonic character formula (cf. the previous footnote) is given a nice path interpretation \cite{OleJSb}, where, in particular, the height constraints $y\geq 0$ and $y\leq k$ are shown to capture the factorization of the two primary singular vectors (see also related remarks in \cite{Rig}).

\subsection{Graded paths,  graded parafermions and the $\M(k+1,2k+3)$ minimal models}\label{Sgp}

The graded parafermions refer to the coset conformal field theory $\widehat{ops}(1,2)_k/\widehat {u}(1)$ \cite{CRS,JM} (while the ordinary parafermions are described by the coset $\widehat{su}(2)_k/\widehat {u}(1)$). Their basis of states is formulated in terms of the so-called jagged partitions \cite{JM,BFJM,FJM.R,FJM.E}. The corresponding paths are presented in \cite{JMan}. These are equivalent to B$_{k-1/2}$ paths (the maximal height being $k-1/2$) but defined on a half-integer lattice \cite{JSTAT}. There is a further constraint which is that the peak $x$-position and the initial point $y_0$ must be integer.
The weight is the sum of the $x$-position of the peaks. These paths could be called graded $\B$ paths. They can be related, via a suitable definition of the multiple partitions,  to a graded version of the $\F$ paths, as shown in \cite{JSTAT}. 

Quite remarkably, the dual version of these graded $\F$ paths, defined in terms of the dual weight function (\ref{weigd}), describes the non-unitary minimal models $\M(k+1,2k+3)$. These paths are still defined on a half-integer lattice, with peaks at integer $(x,y)$ positions, and with maximal height equal to $k$. They can be chosen to terminate with a SE edge; the end points $y_0$ (integer) and $y_L$ (half-integer) are related to the module labels $(r,s)$ by 
\begin{equation}
 s=2y_0+1 \quad \text{and}\quad  r=y_L+\frac12 .
 \end{equation}
The detailed analysis of these paths, including the fermi-gas-type construction of their generating function (which leads to a novel expression for their fermionic characters), is presented in \cite{JSTAT}.


\subsection{Forrester-Baxter RSOS paths and the  $\M(p',p)$ minimal  models}\label{SFB}


A generalization of the RSOS model of \cite{ABF} has been solved in \cite{FB}.  In the infinite length limit, the configurations in regime III provide the description of the states in all irreducible modules of the minimal models $\M(p',p)$ \cite{Huse,Kyoto,Mel,Rig,Nak}. The corresponding paths, called RSOS$(p',p)$ path, are sequence of  edges linking  adjacent vertices $(i,\ell_i)$ and $(i+1,\ell_{i+1})$ such that $|\ell_i-\ell_{i+1}|=1$. The height variable $\ell_i$ ranges over $1\leq \ell_i\leq p-1$, while 
its index  is bounded by  $0\leq i\leq L$.
Each path is specified by particular boundary conditions: the values of $\ell_0$ and those of $\ell_{L-1}$ and $\ell_L$ (the last two specify the RSOS ground state on which the configuration is built). 
As for the dual $\F$ paths, we will choose $\ell_{L-1}=\ell_L+1$ so that  the paths all terminate with a SE edge.
The relation between $(\ell_0,\ell_L)$ and the indices $(r,s)$ labeling the  irreducible modules is:
\begin{equation}\label{kacla}
\ell_0=s  \qquad \text{and} \qquad \ell_L= \l\lf \frac{rp}{p'}\r \rf.
\end{equation}

The weight of a path is \cite{FB}
\begin{equation}\label{wei}
\wf=\sum_{i=1}^{L-1} \wf_i,
\end{equation}
 where, for the vertex specified by the triplet $(\ell_{i-1},\ell_i, \ell_{i+1})$,
$\wf_i$ reads:
\begin{align}\label{wFB}
(d\mp1,d,d\pm1): \quad \wf_i = \frac{i}2,\qquad \quad
(d,d\mp1,d): \quad \wf_i =\pm i  \l\lf \frac{d(p-p')}{p} \r\rf.
\end{align} 
For $p'=p-1$, we recover the weight function (\ref{weigd}) (all vertices  contribute for $i/2$ except for the local extrema which do not contribute since $\l\lf d/{p}\r \rf=0$ for $d\leq p-1$). Indeed, the RSOS$(p-1,p)$ paths are precisely the dual $\F$ paths introduced in Section \ref{Summ}, with $y=\ell-1$ and $p=k+2$.

As for the dual $\F$ paths, 
with the  boundary conditions specified, there is a unique ground state configuration gs$(r,s)$ defined to be the path with lowest weight, say $\wf_{{\rm gs}(r,s)}$.  The relative weight of a path, $ \wf-\wf_{{\rm gs}(r,s)}$, is the relative conformal dimension:
\begin{equation}
\wf-\wf_{{\rm gs}(r,s)}= h-h_{r,s}.
\end{equation}

Following \cite{FLPW}, it is convenient to color in gray the $p'-1$ strips between the heights $d_{r'}$ and $d_{r'}+1$ with 
\begin{equation}\label{hr}
d_{r'}=  \l \lf \frac{r'p}{p'}\r\rf \qquad \text{for} \qquad 1\leq r'\leq p'-1 .
\end{equation}
The band structure is symmetric with respect to the up-down
reversal.
For unitary models, $p=p'+1$, all the bands are gray, while there is a single gray band for the $\M(2,p)$ models.
The band structure is illustrated in Fig. \ref{fig21} for a RSOS$(4,7)$ path and in Fig. \ref{fig22} for a RSOS$(2,9)$ path.


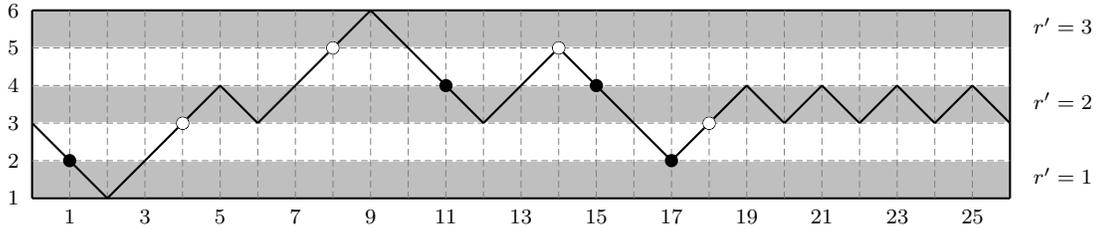
\begin{figure}[ht]
\caption{{\footnotesize A RSOS$(4,7)$ path   representing a state in the  module $r=2, s=3$.
The dots indicate the scoring vertices: the white dots have weights $u_i$ and the black ones have weights $v_i$, cf. eq. (\ref{defuv}).
The scoring vertices have weight $0,2,3,6,6,8,8,9$, so that $w^{\text{sc}}= 42$.}}
\vskip-1.4cm
\label{fig21}
\begin{center}
\begin{pspicture}(1,0)(13.0,4.5)




{
\psset{linestyle=none}
\psset{fillstyle=solid}
\psset{fillcolor=lightgray}
\psframe(0.5,0.5)(13.5, 1.0)
\psframe(0.5,1.5)(13.5, 2.0)
\psframe(0.5,2.5)(13.5, 3.0)
}

\psset{linestyle=solid}
\psline{-}(0.5,0.5)(13.5,0.5)
\psline{-}(0.5,3.0)(13.5,3.0)

\psline{-}(0.5,0.5)(0.5,3.0)
\psline{-}(13.5,0.5)(13.5,3.0)

{
\psset{linewidth=0.25pt,linestyle=dashed, dash=2.5pt 1.5pt,linecolor=gray}

\psline{-}(0.5,1.0)(13.5,1.0)
\psline{-}(0.5,1.5)(13.5,1.5)
\psline{-}(0.5,2.0)(13.5,2.0)
\psline{-}(0.5,2.5)(13.5,2.5)

\psline{-}(1.0,0.5)(1.0,3.0) \psline{-}(1.5,0.5)(1.5,3.0) 
\psline{-}(2.0,0.5)(2.0,3.0) \psline{-}(2.5,0.5)(2.5,3.0) \psline{-}(3.0,0.5)(3.0,3.0) 
\psline{-}(3.5,0.5)(3.5,3.0) \psline{-}(4.0,0.5)(4.0,3.0) \psline{-}(4.5,0.5)(4.5,3.0) 
\psline{-}(5.0,0.5)(5.0,3.0) \psline{-}(5.5,0.5)(5.5,3.0) \psline{-}(6.0,0.5)(6.0,3.0) 
\psline{-}(6.5,0.5)(6.5,3.0) \psline{-}(7.0,0.5)(7.0,3.0) \psline{-}(7.5,0.5)(7.5,3.0) 
\psline{-}(8.0,0.5)(8.0,3.0) \psline{-}(8.5,0.5)(8.5,3.0) \psline{-}(9.0,0.5)(9.0,3.0) 
\psline{-}(9.5,0.5)(9.5,3.0) \psline{-}(10.0,0.5)(10.0,3.0) \psline{-}(10.5,0.5)(10.5,3.0) 
\psline{-}(11.0,0.5)(11.0,3.0) \psline{-}(11.5,0.5)(11.5,3.0) \psline{-}(12.0,0.5)(12.0,3.0) 
\psline{-}(12.5,0.5)(12.5,3.0) \psline{-}(13.0,0.5)(13.0,3.0) 
}




\rput(0.25,0.5){{\scriptsize $1$}}
\rput(0.25,1.0){{\scriptsize $2$}} \rput(0.25,1.5){{\scriptsize $3$}}
\rput(0.25,2.0){{\scriptsize $4$}} \rput(0.25,2.5){{\scriptsize $5$}}
\rput(0.25,3.0){{\scriptsize $6$}}

\rput(14.2,2.8){{\scriptsize$r'=3$}} 
\rput(14.2,1.8){{\scriptsize$r'=2$}} 
\rput(14.2,0.8){{\scriptsize$r'=1$}} 
\rput(1.0,0.25){{\scriptsize $1$}} \rput(2.0,0.25){{\scriptsize $3$}}
\rput(3.0,0.25){{\scriptsize $5$}} \rput(4.0,0.25){{\scriptsize $7$}}
\rput(5.0,0.25){{\scriptsize $9$}} \rput(6.0,0.25){{\scriptsize $11$}}
\rput(7.0,0.25){{\scriptsize $13$}} \rput(8.0,0.25){{\scriptsize $15$}}
\rput(9.0,0.25){{\scriptsize $17$}} \rput(10.0,0.25){{\scriptsize $19$}}
\rput(11.0,0.25){{\scriptsize $21$}} \rput(12.0,0.25){{\scriptsize $23$}}
\rput(13.0,0.25){{\scriptsize $25$}}



\psset{linestyle=solid}

\psline{-}(.5,1.5)(1.0,1.0) \psline{-}(1.0,1.0)(1.5,0.5)
\psline{-}(1.5,0.5)(2.0,1.0) \psline{-}(2.0,1.0)(2.5,1.5)
\psline{-}(2.5,1.5)(3.0,2.0) \psline{-}(3.0,2.0)(3.5,1.5)
\psline{-}(3.5,1.5)(4.0,2.0) \psline{-}(4.0,2.0)(4.5,2.5)
\psline{-}(4.5,2.5)(5.0,3.0) \psline{-}(5.0,3.0)(5.5,2.5)
\psline{-}(5.5,2.5)(6.0,2.0) \psline{-}(6.0,2.0)(6.5,1.5)
\psline{-}(6.5,1.5)(7.0,2.0) \psline{-}(7.0,2.0)(7.5,2.5)
\psline{-}(7.5,2.5)(8.0,2.0) \psline{-}(8.0,2.0)(8.5,1.5)
\psline{-}(8.5,1.5)(9.0,1.0)\psline{-}(9.0,1.0)(9.5,1.5)
\psline{-}(9.5,1.5)(10.0,2.0)\psline{-}(10.0,2.0)(10.5,1.5)
\psline{-}(10.5,1.5)(11.0,2.0)\psline{-}(11.0,2.0)(11.5,1.5)
\psline{-}(11.5,1.5)(12.0,2.0)\psline{-}(12.0,2.0)(12.5,1.5)
\psline{-}(12.5,1.5)(13.0,2.0)\psline{-}(13.0,2.0)(13.5,1.5)


\psset{fillcolor=white}
\psset{dotsize=5pt}\psset{dotstyle=o}
\psdots(2.5,1.5)(4.5,2.5)(7.5,2.5)(9.5,1.5)
\psset{dotsize=5pt}\psset{dotstyle=*}
\psdots(1,1)(6.0,2.)(8,2.)(9,1)

\end{pspicture}
\end{center}
\end{figure}


\begin{figure}[ht]
\caption{{\footnotesize A RSOS$(2,9)$ path representing a state in the $(1,5)=(1,4)$ module.
The dots indicate the scoring vertices (of weight $0,3,2,4,4,6,5,7,7,8,8,11$, adding to 62). By flattening the gray band and folding the lower part of the rectangle onto the upper one, it is mapped to the path of Fig. \ref{fig7}. The length of the terminating zigzag pattern in the gray band does not affect the corresponding B$_{3,0}$ path, where $a$ is fixed by $a=k-s=4-4=0$, choosing the value $s= 4$ to fit in  the range $1\leq s\leq k$.}}
\vskip-1.4cm
\label{fig22}
\begin{center}
\begin{pspicture}(1,0)(13.0,5)




{
\psset{linestyle=none}
\psset{fillstyle=solid}
\psset{fillcolor=lightgray}
\psframe(0.5,1.5)(13., 2.0)
}

\psset{linestyle=solid}
\psline{-}(0.5,0.)(13.,0.)
\psline{-}(0.5,3.5)(13.,3.5)

\psline{-}(0.5,0.0)(0.5,3.5)
\psline{-}(13.,0.0)(13.,3.5)

{
\psset{linewidth=0.25pt,linestyle=dashed, dash=2.5pt 1.5pt,linecolor=gray}

\psline{-}(0.5,1.0)(13.,1.0)
\psline{-}(0.5,1.5)(13,1.5)
\psline{-}(0.5,2.0)(13,2.0)
\psline{-}(0.5,2.5)(13,2.5)
\psline{-}(0.5,.5)(13.,0.5)
\psline{-}(0.5,3.)(13.,3)

\psline{-}(1.0,0)(1.0,3.5) \psline{-}(1.5,0)(1.5,3.5) 
\psline{-}(2.0,0)(2.0,3.5) \psline{-}(2.5,0)(2.5,3.5) \psline{-}(3.0,0)(3.0,3.5) 
\psline{-}(3.5,0)(3.5,3.5) \psline{-}(4.0,0)(4.0,3.5) \psline{-}(4.5,0)(4.5,3.5) 
\psline{-}(5.0,0)(5.0,3.5) \psline{-}(5.5,0)(5.5,3.5) \psline{-}(6.0,0)(6.0,3.5) 
\psline{-}(6.5,0)(6.5,3.5) \psline{-}(7.0,0)(7.0,3.5) \psline{-}(7.5,0)(7.5,3.5) 
\psline{-}(8.0,0)(8.0,3.5) \psline{-}(8.5,0)(8.5,3.5) \psline{-}(9.0,0)(9.0,3.5) 
\psline{-}(9.5,0)(9.5,3.5) \psline{-}(10.0,0)(10.0,3.5) \psline{-}(10.5,0)(10.5,3.5) 
\psline{-}(11.0,0)(11.0,3.5) \psline{-}(11.5,0)(11.5,3.5) \psline{-}(12.0,0)(12.0,3.5) 
\psline{-}(12.5,0)(12.5,3.5) 
}





\rput(1.0,-0.25){{\scriptsize $1$}} \rput(2.0,-0.25){{\scriptsize $3$}}
\rput(3.0,-0.25){{\scriptsize $5$}} \rput(4.0,-0.25){{\scriptsize $7$}}
\rput(5.0,-0.25){{\scriptsize $9$}} \rput(6.0,-0.25){{\scriptsize $11$}}
\rput(7.0,-0.25){{\scriptsize $13$}} \rput(8.0,-0.25){{\scriptsize $15$}}
\rput(9.0,-0.25){{\scriptsize $17$}} \rput(10.0,-0.25){{\scriptsize $19$}}
\rput(11.0,-0.25){{\scriptsize $21$}} \rput(12.0,-0.25){{\scriptsize $23$}}
\rput(13.0,-0.25){{\scriptsize $25$}}


\rput(0.25,0.){{\scriptsize $1$}}
\rput(0.25,.5){{\scriptsize $2$}} \rput(0.25,1.){{\scriptsize $3$}}
\rput(0.25,1.5){{\scriptsize $4$}} \rput(0.25,2.){{\scriptsize $5$}}
\rput(0.25,2.5){{\scriptsize $6$}}
\rput(0.25,3){{\scriptsize $7$}}
\rput(0.25,3.5){{\scriptsize $8$}}



\psset{linestyle=solid}

\psline{-}(0.5,2.)(2,.5) \psline{-}(2,.5)(3,1.5)
\psline{-}(3,1.5)(3.5,1.0) \psline{-}(3.5,1.0)(5.5,3.)
\psline{-}(5.5,3)(6,2.5) \psline{-}(6.0,2.5)(6.5,3)
\psline{-}(6.5,3)(7.5,2.0) \psline{-}(7.5,2.0)(8,2.5)
\psline{-}(8,2.5)(8.5,2.0) \psline{-}(8.5,2.0)(10,3.5)
\psline{-}(10,3.5)(12.0,1.5)\psline{-}(12.0,1.5)(12.5,2)
\psline{-}(12.5,2)(13.0,1.5)

\psset{fillcolor=white}
\psset{dotsize=5pt}\psset{dotstyle=o}
\psdots(3.0,1.5)(4,1.5)(5.5,3.)(6.5,3)(8,2.5)(10,3.5)
\psset{dotsize=5pt}\psset{dotstyle=*}
\psdots(2.,.5)(3.5,1)(6.,2.5)(7.5,2.)(8.5,2.0)(11.5,2.0)

\end{pspicture}
\end{center}
\end{figure}
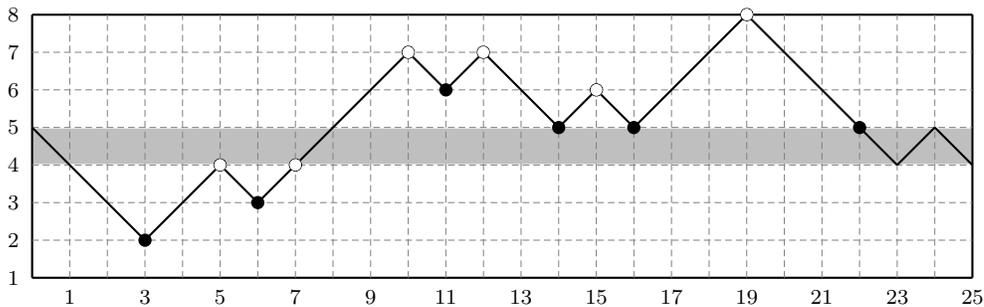


Another weight function for RSOS paths, denoted by $\ws$, is presented in \cite{FLPW}. In the terminology of these references, a vertex is either scoring (hence the label sc), with a non-zero weight $\ws_i=u_i$ or $\ws_i= v_i$, with $u_i$ and $v_i$ given by
 \begin{equation}\label{defuv}
 u_i = \frac12 (i-\ell_i+\ell_0) \;, \qquad v_i = \frac12(i+\ell_i-\ell_0),
\end{equation}  or non-scoring, meaning  that it does not contribute to the weight. The scoring vertices of weight $u_i$ are those where the path enters a gray band from below and the local maxima in white bands, while scoring vertices of weight $v_i$ are those where the path enters a gray band from above and the local minima in white bands. The weight is thus
 \begin{equation}\label{defuws}
 \ws= \sum_{\substack{i=1\\ \text{$i$ is scoring}}}^{L-1} \ws_i. 
\end{equation}  
The scoring vertices for the path of Figs \ref{fig21}-\ref{fig22} are indicated by white dots when their weight is $u_i$ and by black dots 
if it is $v_i$.
This weight prescription is absolute: there is no need to subtract the weight of the ground state with the same boundaries:
\begin{equation}
\ws=h-h_{r,s}.
\end{equation}
It  is clear from this weighting that an infinite path with finite conformal dimension in the sector $(r,s)$ has an infinite tail confined in the $r$-th gray band (cf. Figs \ref{fig21}-\ref{fig22}).

The generating functions for all the RSOS$(p',p)$ paths have been obtained in \cite{Wel} using the recursive approach initiated in \cite{FLPW} (see also \cite{FW}) and some tools first introduced  in \cite{BMlmp,BMS}.

\subsection{Generalized Bressoud paths and the $\M(p',p)$ minimal  models with $p\geq 2p'-1$}\label{Sgb}

In \cite{JMpar}, the RSOS$(p',p)$ paths, for $p\geq 2p'-1$, are transformed  to new paths, denoted as B$(p',p)$, that  generalize naturally the Bressoud paths. The transformation is as follows: 

\begin{enumerate}
\item flatten  the $p'-1$ gray bands (SE and NE edges within the gray bands become H edges in the new path); 
 
 \item fold the part of the strip below the first gray band 
 onto the region just  above it;
\item reset the starting height to 0.

\end{enumerate}

\n The new paths are defined in the strip $ x\geq 0$ and $0\leq y\leq y_{{\rm max}}= p-p'- \l \lf {p}/{p'}\r \rf$.
H edges are allowed
 at all height $y(r')$, with $1\leq r'\leq p'-1$ given by:
 \begin{equation}\label{Hedge}
y(r')= \l \lf \frac{r'p}{p'}\r \rf -\l \lf \frac{p}{p'}\r \rf -  r'+1.
\end{equation}
A path pertaining to the module $(r,s)$ terminates at the height  $y(r)$.
Finally, for the transformation to be reversible,  two  gray bands cannot be adjacent;  this is ensured by the condition $p\geq 2p'-1$.

The generalized Bressoud paths corresponding the RSOS paths of Figs \ref{fig21} and \ref{fig22} are given respectively in Fig. \ref{fig23} and in Fig. \ref{fig7}. Note that the B$(2,2k+1)$ paths are precisely the standard  Bressoud paths $\B$.


Quite remarkably, the weight of these new paths can be computed from the sum of the $x$-position of their peaks and half the $x$-position of their half-peaks, where a half-peak is  a vertex in-between (NE,H) or (H,SE) edges (their $x$-position is underlined in Fig. \ref{fig22}). More precisely, with
\begin{equation} \label{wbre} w= \sum_{x\geq1} w(x)\qquad {\rm where} \qquad w(x)=
\begin{cases}
x & \text{if $x$ is the position of a peak} ,\\
\frac{x}2 & \text{if $x$ is the position of a half-peak} ,\\
 0 &\text{otherwise}\;, \end{cases}
\end{equation}
the weight $w$ of a path, relative to the ground-state configuration with given  end points, reproduce the relative conformal dimension, that is,
   \begin{equation}\label{whw}
 w-w_{\text{gs}(r,s)} = h-h_{r,s}.
\end{equation}
($w$ is a direct generalization of the weight expression for the $\B$ paths and for this reason, we have kept the same notation.)
This new form is obtained by pairing the scoring vertices in a suitable way, which results into a particle interpretation for the B$(p',p)$ path \cite{JMpar}.
Note that since the relation between the B$(p',p)$ and RSOS$(p',p)$ paths is one-to-one, the form (\ref{wbre}) can be viewed as a third  expression for the weight function of the RSOS$(p',p)$ paths when $p\geq 2p'-1$.
  
\begin{figure}[ht]
\caption{{\footnotesize The B(4,7) path corresponding to the RSOS(4,7) path of Fig. \ref{fig3}. Here, H edges are allowed at all values of $y\leq 2$. The horizontal positions that are underlined indicate the positions of the half-peaks. The weight is thus $14+(4+8+10+16+18)/2$.}}
\label{fig23}
\begin{center}
\begin{pspicture}(1.5,-0.5)(13,1.5)

{\psset{yunit=0.55cm,xunit=0.55cm,linewidth=.8pt}


\psline{-}(0,0)(0,2)
\psline{->}(0,0)(26.5,0)

\psline(0,1)(0.2,1)
\psline(0,2)(0.2,2)

\psline(1,0)(1,.2)     \psline(2,0)(2,.2)        \psline(3,0)(3,.2)
\psline(4,0)(4,.2)     \psline(5,0)(5,.2)        \psline(6,0)(6,.2)
\psline(7,0)(7,.2)     \psline(8,0)(8,.2)        \psline(9,0)(9,.2)
\psline(10,0)(10,.2)     \psline(11,0)(11,.2)        \psline(12,0)(12,.2)
\psline(13,0)(13,.2)     \psline(14,0)(14,.2)        \psline(15,0)(15,.2)
\psline(16,0)(16,.2)     \psline(17,0)(17,.2)        \psline(18,0)(18,.2)
\psline(19,0)(19,.2)     \psline(20,0)(20,.2)        \psline(21,0)(21,.2)
\psline(22,0)(22,.2)    \psline(23,0)(23,.2)    \psline(24,0)(24,.2)    
\psline(25,0)(25,.2)\psline(26,0)(26,.2)    


\rput(-.5,0){\scriptsize 0}
\rput(-.5,1){\scriptsize 1}
\rput(-.5,2){\scriptsize 2}

\rput(4,-0.5){\scriptsize \underline 4}    \rput(8,-0.5){\scriptsize  \underline 8}
\rput(10,-0.5){\scriptsize  \underline {10}}
\rput(14,-0.5){\scriptsize 14}    \rput(16,-0.5){\scriptsize  \underline {16}}
\rput(18,-0.5){\scriptsize  \underline {18}}


\psline(0,1)(1,0)
\psline(3,0)(4,1)
\psline(4,1)(7,1)
\psline(7,1)(8,2)
\psline(8,2)(10,2)
\psline(10,2)(11,1)
\psline(11,1)(13,1)
\psline(13,1)(14,2)
\psline(14,2)(15,1)
\psline(15,1)(16,1)
\psline(16,1)(17,0)
\psline(17,0)(18,1)
\psline(18,1)(26,1)


\psset{dotsize=2pt}\psset{dotstyle=*}

}

\end{pspicture}
\end{center}
\end{figure}
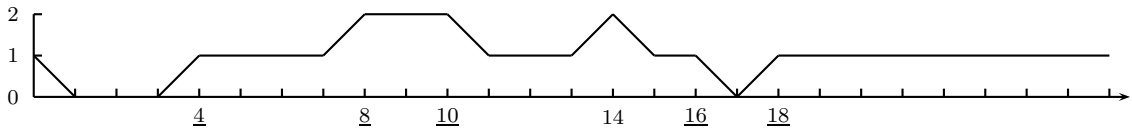

\section{Miscellaneous concluding remarks}
\subsection{Summary of the various path-partition links}

 The different relations presented here between the two types of paths and partitions are summarized in Fig. \ref{fig30}, which  amplifies the diagram of Fig. \ref{fig0} by the addition of a more precise characterization of  the various links, and the relation between the parameters ($a$, $i$ and $\ell$) specifying the `boundary conditions'. The diagram renders particularly visible the central role played by the exchange relation.

\begin{figure}[ht]
\caption{{\footnotesize The links established between restricted partitions, defined by eqs (\ref{difk}) and (\ref{bd}), the multiple partitions, defined in eqs (\ref{mul}), (\ref{difone}), and (\ref{bon}), and the two types of paths. The term `cluster' refers to the clustering procedure described in Section \ref{Srptomp}, and the  `unfolding' stands the inverse operation described in Section \ref{Smptorp}. Similarly, the projection designates the transformation $\Nk\rw\N$ given in (\ref{mula}), while the lift is its inverse in (\ref{mulb}). The `exchange' refers to the reordering of the clusters by means of the  exchange relation (\ref{com}).
It is involved in all the links but the Burge correspondence.}}
\label{fig30}
\begin{center}
\begin{pspicture}(-4,-2.5)(13,4.5)


\rput[tr](0,0){\rnode{A}{\psframebox{B$_{k-1,a}$ paths}}}
\rput[tr](6,4){\rnode{B}{\psframebox{Restricted Partitions }}}
\rput[tr](11,0){\rnode{C}{\psframebox{ABF$_{k,\ell}$ paths}}}
\rput[tr](6,0){\rnode{E}{\psframebox{Multiple ~Partitions~~}}}
\ncline[nodesep=3pt]{<->}{A}{E}
\ncline[nodesep=3pt]{<->}{E}{C}
\ncline[nodesep=3pt]{<->}{E}{B}
\ncline[nodesep=3pt]{<->}{A}{B}

\rput(0.5,1.7){\scriptsize{Burge}}
\rput(2.8,1.7){\scriptsize{{\color{gray!70!black}$a=k-i$}}}

\rput(7.3,0.5){\scriptsize{{\color{gray!70!black}$a=\text{max}\, (\ell-1,0)$}}}

\rput(1,-1){\scriptsize{Exchange}}
\rput(7.3,-1){\scriptsize{Projection ($\shortleftarrow)$}}
\rput(6.88,-1.5){\scriptsize{Lift ($\shortrightarrow$)}}
\rput(6.95,-2){\scriptsize{Exchange}}
\rput(5.5,2.5){\scriptsize{Cluster $(\shortdownarrow)$}} 
\rput(5.7,2){\scriptsize{Unfolding $(\shortuparrow)$}}
\rput(5.45,1.5){\scriptsize{Exchange}}

\end{pspicture}
\end{center}
\end{figure}
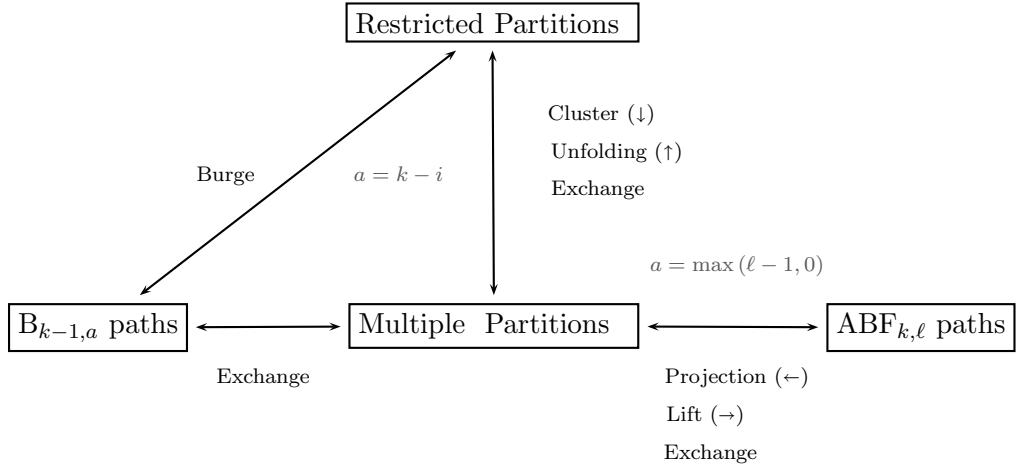

\subsection{Outlook: other parafermionic models}
As indicated in starting, the $\Z_k$ parafermionic theories have important simplifying features. The structurally determining property  for obtaining its basis is the cyclic symmetry which entails the abelian (or single-channel) nature of the operator product algebra. It remains to be seen whether this allows for an analogous  analysis of its various extensions.
In that vein, note that apart from the graded parafermionic model (cf. Section \ref{Sgp}), there is no other model in the class specified by the general coset $\widehat{g}_k/\widehat{u}(1)^r$ \cite{Gep}\footnote{This particular coset construction was first studied in the mathematics literature, albeit not from the conformal point of view -- see e.g. \cite{Lep} for early references.} (where $r$ is the rank of $g$), for which the basis of states has been obtained -- although, rather amazingly,  the general  form of their fermionic characters has been conjectured in \cite{KNS}.

Apart from the Gepner-type coset models, there are further classes of parafermionic models such as those  considered in App. A of \cite{ZFa}, 
for which neither the quasi-particle bases nor the fermionic characters are known. These models are characterized by a parameter $\gamma$, specifying the dimension of the parafermionic fields $\psi_\rho$, namely $\gamma \rho (1-\rho/k)$ 
(the usual $\Z_k$ parafermions corresponding to the case $\gamma=1$). The models with $\gamma=2$ are briefly considered in \cite{ZFa} and studied in more details in \cite{DJS}, while those with $\gamma=3/2$ (with $k$ even) are constructed in \cite{JMbeta}.
There, these so -called $\Z_k^{(\gamma)}$ parafermionic theories  were conjectured to be related to the ${\cal W}_k(k+1,k+2\gamma)$ minimal 
models.\footnote{In that vein, the Jack polynomials associated with the partitions satisfying $\la_j\geq \la_{j+k}+2\gamma$ at coupling $\beta=-(2\gamma-1)/(k+1)$ were conjectured in \cite{Fetal} to be related to the very same $\cal W$ models.}


\subsection{Outlook: minimal models}

As indicated in Section \ref{Sm2p}, the basis of states for the $\M(2,2k+1)$ minimal models is formulated in terms of restricted  partitions. Let us briefly comment on the other  known partition-type bases of states for minimal models. 

A basis has been constructed  for the $\M(3,p)$ models in \cite{JM3p} considered from the point of view of an extended chiral algebra generated by the $\phi_{2,1}$ modes. The same basis is derived in \cite{FJM} using a  vertex algebra obtained by the tensor product of the $\M(3,p)$ model times a free boson.
A different $\M(3,p)$ basis, involving both the $\phi_{2,1}$ and the Virasoro modes, is displayed in \cite{JM3pa} -- which appears as a sort of multiple-partition version of the former basis expressed in terms of special partitions.
 Note that when $p'=3$, the OPE of $\phi_{2,1}$ with itself gives the identity family, meaning in particular that it has a single channel.\footnote{The extension of these constructions to all minimal models, where the role of $\phi_{2,1}$ is replaced by the simple current $\phi_{p'-1,1}$ is initiated in \cite{MR}  (the issue of bases is not addressed directly there but the absence of singular vectors in such a quasi-particle set-up is demonstrated on general grounds).}

Recently, in a set of three astonishing papers \cite{F(12a),F(12b),F(13)}, $\phi$-type bases have been obtained for all minimal models.
 In particular, the $\phi_{2,1}$-basis is proved for $p>2p'$ \cite{F(12a)} and conjectured in the other cases \cite{F(12b)}, while in \cite{F(13)}, a $\phi_{1,3}$-basis is conjectured for $p>2p'$. It should be stressed that these bases are formulated in terms of the modes of a field whose OPE with itself has more than one channel. This, by itself, prevents the derivation of generalized commutation relations.
 
 On the other hand, as indicated in Section \ref{SFB}, RSOS paths do represent a basis of states, from which the fermionic forms can be derived \cite{Wel}. These paths turn out to be related to partitions subject to special conditions: these are the prescribed hook differences (which generalize the conditions on  successive ranks referred to in the introduction) \cite{ABBBFV}. However, these types of partitions do not appear to have a natural operator interpretation. It is still an open problem to interpret the RSOS path basis in a conformal-field-theory formalism. The transformation of RSOS paths into generalized Bressoud ones indicated in Section \ref{Sgb} appears to be a step in that direction given that the particle content of these latter paths become then manifest and that it matches that of ($\phi_{1,3}$-perturbed) off-critical models, via the restricted sine-Gordon spectrum \cite{LeC,Smi}.

\vskip0.3cm
\noindent {\bf ACKNOWLEDGMENTS}

I thank P. Jacob for his collaboration on most of the works reviewed here. This  work is supported  by NSERC.


\begin{thebibliography}{99}
\addcontentsline{toc}{section}{References}


 \bibitem{AgB}
 A.K. Agarwal and D. Bressoud, {\it Latttice paths and
 hypergeometric series}, Pacific J. Math. {\bf 136} (1989) 209-228.


\bibitem{AnS}
G.E. Andrews, {\it Sieves in the theory of partitions}, Amer. J. Math. {\bf 94} (1972) 1214-1230.


\bibitem{An}
G.E. Andrews, {\it An
analytic generalization of the Rogers-Ramanujan identities for odd moduli}, Proc. Nat. Acad. Sci. USA {\bf 71} (1974)
4082-4085.

 \bibitem{Andr}
 G.E. Andrews, {\it The theory of
partitions}, Cambridge Univ. Press, Cambridge, (1984).


 
\bibitem{ABF}
G. E. Andrews, R. J. Baxter and P. J. Forrester,
\textit{Eight-vertex SOS model and generalized Rogers--Ramanujan-type
identities},
J. Stat. Phys. \textbf{35} (1984) 193--266.

\bibitem{AnB}
G.E. Andrews and D. Bressoud, {\it On  the Burge correspondence between partitions and binary words}. Rocky Mtn. J.  Math. {\bf 24} (1980) 225-233.


\bibitem{ABBBFV}
G. E. Andrews, R. J. Baxter, D.M. Bressoud, W.H. Burge, P. J. Forrester and G.Viennot, {\it Partitions with prescribed hook differences}, Europ. J. Comb. {\bf 8} (1987) 341-350. 








\bibitem{BFJM}
L. B\'egin, J.-F. Fortin, P. Jacob and P. Mathieu,  {\it Fermionic characters
for graded parafermions}, Nucl. Phys {\bf B659} (2003) 365-386.


\bibitem{BPZ}
A. Belavin, A. Polyakov, and A. Zamolodchikov,
\newblock {\it Infinite Conformal Symmetry in Two-Dimensional Quantum Field
  Theory},
 Nucl. Phys., {\bf B241} (1984) 333--380.


\bibitem{Ber}
A. Berkovich, {\it Fermionic counting of RSOS-states and Virasoro character formulas for the unitary minimal series $M(\nu,\nu+1)$. Exact results},  Nucl. Phys. {\bf B431} (1994) 315-348.

\bibitem{BMlmp}
 A. Berkovich and B. M. McCoy, {\it Continued fractions and fermionic representations for characters of $\M(p,p')$ minimal models}, Lett. Math. Phys. {\bf 37} (1996) 49-66.
 
\bibitem{BMS}
A, Berkovich, B. M. McCoy, A. Schilling, {\it  Rogers-Schur-Ramanujan type identities for the $\M(p,p')$ minimal models of conformal field theory},
Commun. Math. Phys. {\bf 191} (1998) 325-395


\bibitem{BP}
A. ÊBerkovich and P. Paule, {\it Lattice paths, $q$-multinomials and two variants of the Andrews-Gordon identities},  Ramanujan J. {\bf 5} (2002) 409--425.

\bibitem{BH}
B. A. Bernevig and F. D. M. Haldane, {\it Fractional quantum Hall states and Jack polynomials}
 Phys. Rev. Lett. {\bf 100} (2008)  246802-06.


\bibitem{BreL}
D. Bressoud, {\it Lattice paths and Rogers-Ramanujan identities}, in
{\it Number theory, Madras 1987}, ed. K. Alladi, Lecture Notes in
Mathematics {\bf 1395} (1987) 140-172.


 \bibitem{Bu}
 W.H. Burge, {\it A correspondence between partitions related to generalizations of the Ramanujan-Rogers identities}, Discrete Math. {\bf 34} (1981) 9-15.






\bibitem{CRS}
J. M. Camino, A. V. Ramallo and
J. M. Sanchez de Santos, {\it Graded parafermions}, Nucl. Phys. {\bf B530} (1998) 715-741.
%

 
\bibitem{DN}
G. Delfino and G. Niccoli,
{\it Isomorphism of critical and off-critical operator spaces in two-dimensional quantum field theory}, Nucl. Phys. {\bf B799} (2008) 364-378;
G. Delfino, {\it On the space of quantum fields in massive two-dimensional theories},
 Nucl. Phys. {\bf B807} (2009) 455-470.
 
\bibitem{Kyoto}
 E. Date, M. Jimbo, A. Kuniba, T. Miwa and M. Okado, {\it Exactly solvable SOS models: local height probabilities and theta function identities}, Nucl. Phys. {\bf B290} (1987) 231-273.
 
 \bibitem{DJS}
V. S. Dotsenko, J. L. Jacobsen and R. Santachiara, 
{\it Parafermionic theory with the symmetry $\Z_5$},
 Nucl. Phys. {\bf B656} (2003) 259-324;
{\it Parafermionic theory with the symmetry $\Z_N$, for $N$ even},
 Nucl. Phys. {\bf B664} (2003) 477-511;
{\it Parafermionic theory with the symmetry $\Z_N$, for $N$ even},
 Nucl. Phys. {\bf B679} (2004) 464-494.




 \bibitem{CFT}
P. Di Francesco, P. Mathieu, and D. S\'{e}n\'{e}chal,
{\em {Conformal Field Theory}},
Springer-Verlag, New York (1997). 
 
 
%

\bibitem{FF82}
B.L. Feigin and D.B. Fuchs, {\it Skew-symmetric differential operators
on the line and Verma modules over the Virasoro algebra}, Funct. Anal. and
Appl. {\bf 16} {(1982)} 114-136; {\it Verma modules over the Virasoro algebra}, Funct. Anal. and
Appl. {\bf 17} {(1983)} 241-243.



\bibitem{FNO}
B.L. Feigin, T. Nakanishi  and H. Ooguri,   
{\it The annihilating ideal of minimal modeks},
Int. J. Mod. Phys.
{\bf A7} Suppl. {\bf 1A} (1992) 217-238.

\bibitem{Fetal}
B. Feigin, M. Jimbo, T. Miwa and E. Mukhin, {\it
A differential ideal of symmetric polynomials spanned by Jack polynomials at $\beta=-(r-1)/(k+1)$}, 
Int. Math. Res. Not. {\bf 23} (2002) 1223--1237.



\bibitem{FJM}
B. Feigin, M. Jimbo, and T. Miwa.
\newblock {\it Vertex operator algebra arising from the minimal series $\M \left( 3
  , p \right)$ and monomial basis},
Prog. Math. Phys. {\bf 23} (2002) 179--204.




\bibitem{F(12a)}
B. Feigin, M. Jimbo, T. Miwa, E. Mukhin and Y. Takayama,  {\it A monomial basis for the Virasoro minimal series $\M \left( p , p'
  \right)$: the case $1 < p'/p < 2$}.
\newblock {\em Comm. Math. Phys.}, {\bf 257} (2005) 395--423, 2005;.


\bibitem{F(12b)}
B. Feigin, M. Jimbo, T. Miwa, E. Mukhin and Y. Takayama,  {\it Set of rigged paths with Virasoro characters}, Ramanujan J. {\bf 15} (2008) 123--145. 

\bibitem{F(13)}
B. Feigin, E. Feigin, M. Jimbo, T. Miwa, Y. Takeyama
{\it A $\phi_{1,3}$-filtration of the Virasoro minimal series $\M(p,p')$ with $1<p'/p< 2$},
Publ. Res. Inst. Math. Sci. {\bf 44} (2008) 213--257

 \bibitem{FP}
 G.  Feverati and  P. A. Pearce, {\it Critical RSOS and minimal models I: Paths fermionic algebras and Virasoro modules}, Nucl. Phys. {\bf B663 } (2003) 409-442.


  
 
 \bibitem{FWa}
 O. Foda and T. Welsh, {\it  Melzer's identities revisited},  Contemp. Math. {\bf 248} (1999)  207-234. 
 
 

\bibitem{FLPW}
O. Foda,  K.S. M. Lee, Y. Pugai  and T. A. Welsh, {\it Path generating transforms}, in {\it q-Series from a contemporary perspective},  Contemp. Math. {\bf 254} (2000) 157--186.


 
 
 
\bibitem{FW}
O. Foda and T. A. Welsh, {\it 
On the combinatorics of Forrester-Baxter models},
Prog. Comb. {\bf 191} (2000) 49-103. 


\bibitem{FB}
P. J. Forrester and R. J. Baxter, 
\textit{Further exact solutions of the eight-vertex SOS model 
and generalizations of the Rogers-Ramanujan identities},
J. Stat. Phys. \textbf{38} (1985) 435--472. 


\bibitem{FJM.R}
J.-F. Fortin, P. Jacob and P. Mathieu, {\it Jagged partitions}, Ramanujan J. {\bf 10} (2005) 215-235.


\bibitem{FJM.E}
J.-F. Fortin, P. Jacob and P. Mathieu, {\it Generating function for $K$-restricted jagged partitions}, Electronic J. Comb.
  {\bf 12} (2005) No 1, R12 (17 p.).


\bibitem{FMO}
J.-F. Fortin,  P. Mathieu and  S. O. Warnaar, {\it Characters of graded parafermion conformal field theory}, Adv. Theor. Math. Phys. {\bf 11} (2007) 945-989.

\bibitem{Geo}
G. Georgiev, 
J. Pure Appl. Algebra {\bf 112} (1996) 247; 
  {\it  Combinatorial constructions of modules for infinite-dimensional Lie algebras, II. Parafermionic space}, q-alg/9504024.


\bibitem{Gep}
{D. Gepner}, {\it New conformal  field theories
associated with Lie algebras and their partition
functions}, Nucl. Phys {\bf B290} (1987) 10-24.



\bibitem{Huse}
D.A. Huse, {\it Exact exponents for infinitely many new multicritical points},
 Phys. Rev. {\bf B30} (1984) 3908-3915.


\bibitem{JMa}
P. Jacob and P. Mathieu,  {\it Parafermionic character formula}, Nucl. Phys.
{\bf B587} (2000) 514-542.


\bibitem{JMb}
P. Jacob and P. Mathieu, {\it Parafermionic quasi-particle basis and fermionic-type characters}
 Nucl. Phys. {\bf B620} (2002) 351-379.


\bibitem{JM}
P. Jacob and P. Mathieu, {\it Graded parafermions: standard and quasi-particle bases}, Nucl. Phys.
{\bf B630} (2002) 433-452.

 \bibitem{JM.A}
  P. Jacob and P. Mathieu, {\it Parafermionic derivation of the  Andrews-type multiple sums}, J. Phys. A: Math. Gen. {\bf 38} (2005) 8225-8238.
 
 \bibitem{JMbeta}
 P. Jacob and P. Mathieu, {\it  The $Z_k^{su(2),3/2}$ parafermions},
 Phys. Lett. {\bf B627} (2005) 224-232.
 
\bibitem{JM3p}
P. Jacob and P. Mathieu, {\it A quasi-particle description  of the $\M(3,p)$ models}, Nucl. Phys. {\bf B733} (2006) 205-232;.


\bibitem{JM3pa}
P. Jacob and P. Mathieu,
\newblock {\it Embedding of bases: from the $\M \left( 2 , 2 \kappa + 1 \right)$ to
  the $\M \left( 3 , 4 \kappa + 2 - \delta \right)$ models},
Phys. Lett. {\bf B635} (2006) 350--354.



\bibitem{JMpath}
P. Jacob and P. Mathieu, {\it Paths for  $\z_k$ parafermionic models}, Lett. Math. Phys. {\bf 81} (2007) 211-226.


\bibitem{JSTAT}
P. Jacob and P. Mathieu,
 {\it New path description for the  ${\cal M} (k+1,2k+3)$ models and the dual $\z_k$ graded parafermions}, J. Stat. Mec. (2007) P11005, 43 pages.

 
\bibitem{Mult}
P. Jacob and P. Mathieu, {\it  Multiple partitions, lattice paths and a Burge-Bressoud-type  correspondence}, Discrete Math. {\bf  309}
 (2009) 878-886.


\bibitem{JMan}
P. Jacob and P. Mathieu, {\it Jagged partitions and lattice paths}, math.CO/0605551, Ann. Comb., to appear.


\bibitem{JMpar}
P. Jacob and P. Mathieu,
 {\it Particles in RSOS paths}, J. Phys. A: Math. Theor. 42 (2009) 122001-122015.



\bibitem{Kac}
{V.G. Kac}, {\it Contravariant form for infinite dimensional
Lie algebras and superalgebras}, {Lecture notes in physics vol. {\bf 94}, (1979)}
441.

\bibitem{KKMMa}
R. Kedem, T.R. Klassen, B. M. McCoy and E. Melzer, {\it Fermionic quasi-particle representations for characters of ${(G^{(1)})_1  \times (G^{(1)})_1 / (G^{(1)})_2}$}, 
Phys. Lett. {\bf B304} (1993) 263-270

\bibitem{KKMMb}
R. Kedem, T.R. Klassen, B. M. McCoy and E. Melzer, {\it Fermionic sum representations for conformal feld theory characters},  Phys. Lett. {\bf B307} (1993) 68-76.

\bibitem{KMM} 
 R. Kedem, B. M. McCoy and E. Melzer, {\it The sums of Rogers, Schur and Ramanujan and the Bose-Fermi  correspondence in $1+1$-dimensional quantum field theory }, in {\it Recent progress in statistical mechanics and quantum field theory}, ed. by P. Bouwknegt et al, World Scientific (1995) 195-219. 

\bibitem{KNS}
A. Kuniba, T. Nakanishi and J. Suzuki
{\it Characters in Conformal Field Theories from Thermodynamic Bethe Ansatz}
 Mod. Phys. Lett. {\bf A8} (1993) 1649-1660.

\bibitem{LV}
               {L. Lapointe and L. Vinet}, \emph{Exact operator solution of the {C}alogero-{S}utherland
model},
           Commun.\ Math.\ Phys.\
             {\bf 178} (1996) 425-452.


 \bibitem{LeC}
A. LeClair, 
{\it  Restricted sine-Gordon theory and the minimal conformal models}, 
Phys. Lett. {\bf B230} (1989) 103-107.


\bibitem{LP}
J. Lepowsky and M. Primc,
{\em {Structure of the standard Modules for the affine Lie algebra
  $A_1^{\left( 1 \right)}$}}, volume~46 of {\em Contemporary Mathematics}.
American Mathematical Society, Providence, 1985.

\bibitem{LW}
J. Lepowsky and R.L. Wilson,
{\it  A new family of algebras underlying the Rogers-Ramanujan identities and generalizations}, Proc. Nat. Sci. USA {\bf 78} (1981) 7254-7258.

\bibitem{Lep}
J. Lepowsky, {\it Some developments in vertex operator algebra theory, old and new}, arXiv:0706.4072.

\bibitem{MR} P. Mathieu and  D. Ridout {\it The aextended algebra of the minimal models}, Nucl. Phys. {\bf B776} (2007) 365-404.

\bibitem{Mac}
               {I.~G.~ Macdonald},
                \emph{Symmetric functions and {H}all polynomials},
2nd ed., The Clarendon Press/Oxford University Press
                (1995).




\bibitem{Mel}
E. Melzer, {\it Fermionic character sums and the corner transfer matrix}, Int. J. Mod. Phys. {\bf A9} (1994) 1115-1136.
 
\bibitem{RR}
N. Read and E. Rezayi, {\it  
 Beyond paired quantum Hall states: parafermions and incompressible states in the first excited Landau level},
 Phys. Rev. {\bf B59} (1999) 8084-8092.

\bibitem{Nak}
T. Nakanishi, {\it Nonunitary minimal models and RSOS models}, Nucl. Phys. {\bf B334} (1980) 745-766.


\bibitem{Rig}
H. Riggs, {\it Solvable lattice models with minimal and nonunitary critical behavior in two-dimensions},  Nucl. Phys. {\bf B326} (1989) 673-688.

\bibitem{Roc}
{A. Rocha-Caridi}, {\it Vacuum vector representations of the Virasoro
algebra}, in {\it Vertex Operators in Mathematics and Physics}, ed.  J. Lepowsky,
et al,
Publ. Math. Sciences Res. Inst. {\bf  3}, Springer-Verlag,
(1985) 451-473.




\bibitem{Smi}
F.A. Smirnov, 
{\it The perturbated $c< 1$ conformal field theories as reductions of sine-Gordon model}
 Int. J. Mod. Phys. {\bf A4} (1989) 4213-4220;
{\it Reductions of the sine-Gordon model as a perturbation of minimal models of conformal field theory}
 Nucl.Phys. {\bf B337} (1990) 156-180.

 

\bibitem{Stan}
               R.~P.~Stanley,
               \emph{Some combinatorial properties of Jack symmetric
functions}, Adv.\ Math.\ {\bf77} (1988) 76-115.





\bibitem{CMS}
B.~Sutherland,
\emph{Exact results for a quantum many body problem in
one-dimension}, Phys.\ Rev.\  {\bf A4} (1971) 2019-2021 ; \emph{Exact
results for a quantum many body problem in one-dimension. II},
Phys.\ Rev.\ {\bf A5} (1972) 1372-1376.




 \bibitem{OleJS}
 S. O. Warnaar,
\textit{Fermionic solution of the Andrews--Baxter--Forrester model. I.
Unification of CTM and TBA methods},
J. Stat. Phys. \textbf{82} (1996) 657--685.

\bibitem{OleJSb}
S. O. Warnaar,
\textit{Fermionic solution of the Andrews--Baxter--Forrester model. II.
Proof of Melzer's polynomial identities},
J. Stat. Phys. \textbf{84} (1996), 49--83. 
 
 
 \bibitem{W97}
S.O.  Warnaar
{\it The Andrews-Gordon identities and $q$-multinomial coefficients},
Comm. Math. Phys. {\bf 184}  (1997) 203-232.

 \bibitem{Wel}
T. Welsh, {\it Fermionic expressions for minimal model Virasoro characters},  
Memoirs of the American Mathematical Society, vol. {827}, AMS, RI (2005).
\bibitem{ZFa}
A.B.
Zamolodchikov and V.A. Fateev,  {\it Nonlocal (Parafermion) Currents in Two-Dimensional Conformal Quantum
  Field Theory and Self-Dual Critical Points in $\Z_N$-Symmetrical Statistical
  Systems}, Sov. Phys. JETP {\bf 43} (1985) 215-225.



 
\bibitem{Zam}
A.B. Zamolodchikov, {\it Integrable field theory from conformal field theory}
Adv. Stud. Pure Math. {\bf 19} (1989) 641-674.

 
  \end{thebibliography}
\end{document}